%% file: v2/main_v2.tex
\documentclass[aip,reprint,floatfix,raggedbottom]{revtex4-1}

\input packages.tex

\input plotstyles.tex
\input macros.tex
\input layout_setup.tex

\begin{document}

%\preprint{}

\title[Vlasov--Maxwell equilibria and the $E\times B$ cancellation]{Conservative Numerical Construction of Vlasov--Maxwell Equilibria and an Exact Cancellation of $E\times B$ Flow by an Electrostatic Pressure-Gradient Flow} %Title of paper



%\author{}
%\thanks{}
%\affiliation{}

\author{Benjamin J. Sturdevant}
\affiliation{Rensselaer Polytechnic Institute}
\email{sturdb@rpi.edu}
\author{Hong Qin}%
\affiliation{University of Wisconsin-Madison}


\date{\today}

\begin{abstract}
\par A class of Vlasov--Maxwell equilibria with pressure gradients is
considered under symmetry assumptions that reduce the equilibrium
equations to a nonlinear system of ordinary differential equations.
Such equilibria provide self-consistent kinetic states with
nonuniform thermodynamic profiles and potentially strong equilibrium
electric fields, useful for fully kinetic stability
studies of gradient-driven instabilities in the presence of
equilibrium flows. The reduced equilibrium problem is posed as a
nonlinear boundary-value problem for numerical solution, and a
discretization is developed that preserves an associated energy
invariant exactly at the discrete
level. The resulting method provides a systematic means of computing
nontrivial equilibria without relying on scale separation, and an
alternative to constructing only approximately self-consistent
initial states.
\par A key property of the species flows, identified within this model
for any distribution function of the constants of motion, is an exact
cancellation between the $E\times B$ flow and an electrostatically
generated component of the pressure-gradient-driven flow. This follows
directly from a structural property of the species force balance that
arises from this type of symmetry reduction, and is verified
numerically.
\par Additional numerical results demonstrate the conservative properties of the
discretization. Representative equilibrium configurations, relevant to
future kinetic instability studies, are also presented. An asymptotic
approximation for the interior equilibrium, derived in the
well-magnetized, quasi-neutral regime, provides an independent check
on these results. The present setting is deliberately simple,
providing a controlled framework for constructing and analyzing
self-consistent kinetic equilibria and a starting point for extensions
to more general symmetry-reduced configurations.
\end{abstract}

\pacs{}% insert suggested PACS numbers in braces on next line

\maketitle %\maketitle must follow title, authors, abstract and \pacs

%\section{}
%\label{}
%\subsection{}
%\subsubsection{}

\input v2/section_01.tex
\input v2/section_02.tex
\input v2/section_03.tex
\input v2/section_04.tex
\input v2/section_05.tex
\input v2/section_06.tex
\input v2/section_07.tex
\input v2/section_08.tex

% If in two-column mode, this environment will change to single-column format so that long equations can be displayed. 
%\begin{widetext}
%$$\mbox{put long equation here}$$
%\end{widetext}

% Figures should be put into the text as floats. 
% Use the graphics or graphicx packages (distributed with LaTeX2e).
% Put the label that you will use with \ref{} command in the braces of the \label{} command.
%
% \begin{figure}
% \end{figure}

% \begin{table}
% \end{tabular}
% \end{table}

\appendix
\input v2/appendix_a.tex
\input v2/appendix_b.tex
\clearpage

% \begin{acknowledgments}
% Put your acknowledgments here.
% \end{acknowledgments}

\nocite{*}
\bibliography{bibliography}% Produces the bibliography via BibTeX.

\end{document}

%% file: packages.tex
\usepackage{amssymb}
\usepackage{amsmath}
\usepackage[margin=.75in]{geometry}        %%% not in template
\usepackage {graphicx}
\usepackage[export]{adjustbox}
\usepackage{array}
\usepackage{tikz}
\usetikzlibrary{decorations.markings}
\usepackage{pgfplots}
\pgfplotsset{
    compat=1.18,
    every axis/.append style={
        major grid style={line width=0.1pt, gray!30},
    },
    every axis plot post/.append style={line join=round, line cap=round},
}
\usepgfplotslibrary{groupplots}
\usepackage{xcolor}
\usepackage{soul}
\usepackage{setspace}
\usepackage[bb=boondox]{mathalfa}
\usepackage{hyperref}
\usepackage[section]{placeins}

%% file: plotstyles.tex
\def\bapEmpty{}

\def\BoxW{0.376\textwidth}
\def\BoxH{0.320\textwidth}
\def\PanelGap{1.8cm}
\def\LeftReserve{1.75cm}
\def\RightReserve{0.78cm}
\def\TopReserve{0.65cm}
\def\BottomReserve{0.85cm}

\def\PanelAtPos{}

\usetikzlibrary{patterns.meta}

\newlength{\bapWidthLeft}
\newlength{\bapWidthMid}
\newlength{\bapWidthRight}

\pgfkeys{
  /1dpanel/.is family,
  /1dpanel/.cd,
  default/.style = {
    xleftmin/.initial=-12,
    xleftmax/.initial=-10,
    xmidmin/.initial=-6,
    xmidmax/.initial=6,
    xrightmin/.initial=10,
    xrightmax/.initial=12,
    xtickleft/.initial={},
    xtickmid/.initial={-6,-4,-2,0,2,4,6},
    xtickright/.initial={},
    totalwidth/.initial=0.36\textwidth,
    height/.initial=0.38\textwidth,
    gap/.initial=0.5cm,
    hatchdistance/.initial=3pt,
    hatchcolor/.initial=gray!28,
    ymin/.initial=0,
    ymax/.initial=1,
    ytick/.initial={},
    ylabel/.initial={},
    xlabel/.initial={$x/L_0$},
    title/.initial={},
    data/.initial={},
    datadir/.initial={},
    extradata/.initial={},
    legendname/.initial={},
    legendcolumns/.initial={},
    extraouter/.initial={},
    extramid/.initial={},
  },
}

\newcommand{\OneDBrokenAxisPanel}[1]{%
  \pgfkeys{/1dpanel/.cd, default, #1}%
  \pgfmathsetmacro{\bapXLeftMin}{\pgfkeysvalueof{/1dpanel/xleftmin}}%
  \pgfmathsetmacro{\bapXLeftMax}{\pgfkeysvalueof{/1dpanel/xleftmax}}%
  \pgfmathsetmacro{\bapXMidMin}{\pgfkeysvalueof{/1dpanel/xmidmin}}%
  \pgfmathsetmacro{\bapXMidMax}{\pgfkeysvalueof{/1dpanel/xmidmax}}%
  \pgfmathsetmacro{\bapXRightMin}{\pgfkeysvalueof{/1dpanel/xrightmin}}%
  \pgfmathsetmacro{\bapXRightMax}{\pgfkeysvalueof{/1dpanel/xrightmax}}%
  \pgfmathsetmacro{\bapSpanLeft}{\bapXLeftMax-\bapXLeftMin}%
  \pgfmathsetmacro{\bapSpanMid}{\bapXMidMax-\bapXMidMin}%
  \pgfmathsetmacro{\bapSpanRight}{\bapXRightMax-\bapXRightMin}%
  \pgfmathsetmacro{\bapSpanTotal}{\bapSpanLeft+\bapSpanMid+\bapSpanRight}%
  \pgfmathsetlength{\bapWidthLeft}{\pgfkeysvalueof{/1dpanel/totalwidth}*\bapSpanLeft/\bapSpanTotal}%
  \pgfmathsetlength{\bapWidthMid}{\pgfkeysvalueof{/1dpanel/totalwidth}*\bapSpanMid/\bapSpanTotal}%
  \pgfmathsetlength{\bapWidthRight}{\pgfkeysvalueof{/1dpanel/totalwidth}*\bapSpanRight/\bapSpanTotal}%
  \edef\bapXTickLeft{\pgfkeysvalueof{/1dpanel/xtickleft}}%
  \ifx\bapXTickLeft\bapEmpty
    \edef\bapXTickLeft{\pgfkeysvalueof{/1dpanel/xleftmin},\pgfkeysvalueof{/1dpanel/xleftmax}}%
  \fi
  \edef\bapXTickRight{\pgfkeysvalueof{/1dpanel/xtickright}}%
  \ifx\bapXTickRight\bapEmpty
    \edef\bapXTickRight{\pgfkeysvalueof{/1dpanel/xrightmin},\pgfkeysvalueof{/1dpanel/xrightmax}}%
  \fi
  \def\bapLegendName{\pgfkeysvalueof{/1dpanel/legendname}}%
  \def\bapDataPath{figures/data/\pgfkeysvalueof{/1dpanel/datadir}/\pgfkeysvalueof{/1dpanel/data}.tex}%
  \begin{tikzpicture}
  \begin{groupplot}[
      group style={group size=3 by 1, horizontal sep=\pgfkeysvalueof{/1dpanel/gap}},
      height=\pgfkeysvalueof{/1dpanel/height},
      scale only axis,
      tick label style={font=\footnotesize},
      grid=major,
      scaled y ticks=true,
      ymin=\pgfkeysvalueof{/1dpanel/ymin},
      ymax=\pgfkeysvalueof{/1dpanel/ymax},
      ytick={\pgfkeysvalueof{/1dpanel/ytick}},
  ]
  \nextgroupplot[
      width=\bapWidthLeft,
      xmin=\bapXLeftMin,xmax=\bapXLeftMax,
      xtick={\bapXTickLeft},
      ylabel={\pgfkeysvalueof{/1dpanel/ylabel}},
      axis y line*=left,
      \pgfkeysvalueof{/1dpanel/extraouter},
  ]
  {\renewcommand{\addlegendentry}[1]{}\input{\bapDataPath}\pgfkeysvalueof{/1dpanel/extradata}}%
  \nextgroupplot[
      width=\bapWidthMid,
      xmin=\bapXMidMin,xmax=\bapXMidMax,
      xtick={\pgfkeysvalueof{/1dpanel/xtickmid}},
      xlabel={\pgfkeysvalueof{/1dpanel/xlabel}},
      yticklabels=\empty,
      axis y line=left,
      y axis line style={draw=none},
      ytick style={draw=none},
      title={\pgfkeysvalueof{/1dpanel/title}},
      \pgfkeysvalueof{/1dpanel/extramid},
      \ifx\bapLegendName\bapEmpty\else
        legend to name=\bapLegendName,
        legend style={legend columns=\pgfkeysvalueof{/1dpanel/legendcolumns},font=\footnotesize,column sep=1ex},
        legend cell align=left,
      \fi
  ]
  \input{\bapDataPath}
  \pgfkeysvalueof{/1dpanel/extradata}
  \nextgroupplot[
      width=\bapWidthRight,
      xmin=\bapXRightMin,xmax=\bapXRightMax,
      xtick={\bapXTickRight},
      yticklabels=\empty,
      axis y line*=right,
      \pgfkeysvalueof{/1dpanel/extraouter},
  ]
  {\renewcommand{\addlegendentry}[1]{}\input{\bapDataPath}\pgfkeysvalueof{/1dpanel/extradata}}%
  \end{groupplot}
  \fill[pattern={Hatch[angle=45,distance=\pgfkeysvalueof{/1dpanel/hatchdistance},line width=0.3pt]},pattern color=\pgfkeysvalueof{/1dpanel/hatchcolor}] (group c1r1.north east) rectangle (group c2r1.south west);
  \fill[pattern={Hatch[angle=45,distance=\pgfkeysvalueof{/1dpanel/hatchdistance},line width=0.3pt]},pattern color=\pgfkeysvalueof{/1dpanel/hatchcolor}] (group c2r1.north east) rectangle (group c3r1.south west);
  \draw[black,dashed] (group c1r1.north east) -- (group c2r1.north west);
  \draw[black,dashed] (group c1r1.south east) -- (group c2r1.south west);
  \draw[black,dashed] (group c2r1.north east) -- (group c3r1.north west);
  \draw[black,dashed] (group c2r1.south east) -- (group c3r1.south west);
  \end{tikzpicture}%
}

\pgfkeys{
  /1dplain/.is family,
  /1dplain/.cd,
  default/.style = {
    xmin/.initial=-12,
    xmax/.initial=12,
    xtick/.initial={-12,-10,-8,-6,-4,-2,0,2,4,6,8,10,12},
    ymin/.initial=0,
    ymax/.initial=1,
    ytick/.initial={},
    ylabel/.initial={},
    xlabel/.initial={$x/L_0$},
    title/.initial={},
    data/.initial={},
    datadir/.initial={},
    extradata/.initial={},
    extra/.initial={},
    width/.initial=\BoxW,
    height/.initial=\BoxH,
  },
}

\newcommand{\OneDPlainPanel}[1]{%
  \pgfkeys{/1dplain/.cd, default, #1}%
  \ifx\PanelAtPos\bapEmpty
    \begin{tikzpicture}
    \begin{axis}[
        width=\pgfkeysvalueof{/1dplain/width},
        height=\pgfkeysvalueof{/1dplain/height},
        scale only axis,
        tick label style={font=\footnotesize},
        yticklabel style={xshift=-4pt},
        grid=major,
        yminorgrids=true,
        minor y tick num=1,
        minor grid style={line width=0.1pt, gray!30},
        scaled y ticks=true,
        xmin=\pgfkeysvalueof{/1dplain/xmin},
        xmax=\pgfkeysvalueof{/1dplain/xmax},
        xtick={\pgfkeysvalueof{/1dplain/xtick}},
        ymin=\pgfkeysvalueof{/1dplain/ymin},
        ymax=\pgfkeysvalueof{/1dplain/ymax},
        ytick={\pgfkeysvalueof{/1dplain/ytick}},
        xlabel={\pgfkeysvalueof{/1dplain/xlabel}},
        ylabel={\pgfkeysvalueof{/1dplain/ylabel}},
        title={\pgfkeysvalueof{/1dplain/title}},
        \pgfkeysvalueof{/1dplain/extra},
    ]
    \input{figures/data/\pgfkeysvalueof{/1dplain/datadir}/\pgfkeysvalueof{/1dplain/data}.tex}
    \pgfkeysvalueof{/1dplain/extradata}
    \end{axis}
    \end{tikzpicture}%
  \else
    \begin{axis}[
        at/.expand once=\PanelAtPos, anchor=south west,
        width=\pgfkeysvalueof{/1dplain/width},
        height=\pgfkeysvalueof{/1dplain/height},
        scale only axis,
        tick label style={font=\footnotesize},
        yticklabel style={xshift=-4pt},
        grid=major,
        yminorgrids=true,
        minor y tick num=1,
        minor grid style={line width=0.1pt, gray!30},
        scaled y ticks=true,
        xmin=\pgfkeysvalueof{/1dplain/xmin},
        xmax=\pgfkeysvalueof{/1dplain/xmax},
        xtick={\pgfkeysvalueof{/1dplain/xtick}},
        ymin=\pgfkeysvalueof{/1dplain/ymin},
        ymax=\pgfkeysvalueof{/1dplain/ymax},
        ytick={\pgfkeysvalueof{/1dplain/ytick}},
        xlabel={\pgfkeysvalueof{/1dplain/xlabel}},
        ylabel={\pgfkeysvalueof{/1dplain/ylabel}},
        title={\pgfkeysvalueof{/1dplain/title}},
        \pgfkeysvalueof{/1dplain/extra},
    ]
    \input{figures/data/\pgfkeysvalueof{/1dplain/datadir}/\pgfkeysvalueof{/1dplain/data}.tex}
    \pgfkeysvalueof{/1dplain/extradata}
    \end{axis}
  \fi
}

\pgfmathdeclarefunction{lncosh}{1}{%
  \pgfmathparse{abs(#1) - ln(2) + ln(1+exp(-2*abs(#1)))}%
}
\pgfkeys{
  /1dcurv/.is family,
  /1dcurv/.cd,
  default/.style = {
    rawxmin/.initial=-12,
    rawxmax/.initial=12,
    xbreak/.initial=6,
    sharpness/.initial=3.0,
    compress/.initial=0.3,
    xtickpairs/.initial={%
      -12/-12,-11/,-10/,-9/,-8/,-7/,-6/-6,-5/,-4/-4,-3/,-2/-2,-1/,0/0,%
      1/,2/2,3/,4/4,5/,6/6,7/,8/,9/,10/,11/,12/12%
    },
    ymin/.initial=0,
    ymax/.initial=1,
    ytick/.initial={},
    ylabel/.initial={},
    xlabel/.initial={$x/L_0$},
    title/.initial={},
    data/.initial={},
    datadir/.initial={},
    extradata/.initial={},
    legendname/.initial={},
    legendcolumns/.initial={},
    extra/.initial={},
    width/.initial=\BoxW,
    height/.initial=\BoxH,
  },
}
\newcommand{\OneDCurvilinearPanel}[1]{%
  \pgfkeys{/1dcurv/.cd, default, #1}%
  \def\curvLegendName{\pgfkeysvalueof{/1dcurv/legendname}}%
  \pgfmathsetmacro{\curvBreak}{\pgfkeysvalueof{/1dcurv/xbreak}}%
  \pgfmathsetmacro{\curvK}{\pgfkeysvalueof{/1dcurv/sharpness}}%
  \pgfmathsetmacro{\curvC}{\pgfkeysvalueof{/1dcurv/compress}}%
  \pgfmathsetmacro{\curvXMin}{%
    \curvC*\pgfkeysvalueof{/1dcurv/rawxmin}
    + (1-\curvC)/(2*\curvK)*(%
        lncosh(\curvK*(\pgfkeysvalueof{/1dcurv/rawxmin}+\curvBreak))
        -lncosh(\curvK*(\pgfkeysvalueof{/1dcurv/rawxmin}-\curvBreak)))%
  }%
  \pgfmathsetmacro{\curvXMax}{%
    \curvC*\pgfkeysvalueof{/1dcurv/rawxmax}
    + (1-\curvC)/(2*\curvK)*(%
        lncosh(\curvK*(\pgfkeysvalueof{/1dcurv/rawxmax}+\curvBreak))
        -lncosh(\curvK*(\pgfkeysvalueof{/1dcurv/rawxmax}-\curvBreak)))%
  }%
  \def\curvWarpedTickList{}%
  \def\curvRawTickList{}%
  \edef\curvXTickPairs{\pgfkeysvalueof{/1dcurv/xtickpairs}}%
  \foreach \curvV/\curvLbl in \curvXTickPairs {%
    \pgfmathsetmacro{\curvWarpedV}{%
      \curvC*\curvV + (1-\curvC)/(2*\curvK)*(%
        lncosh(\curvK*(\curvV+\curvBreak))-lncosh(\curvK*(\curvV-\curvBreak)))%
    }%
    \ifx\curvWarpedTickList\bapEmpty
      \xdef\curvWarpedTickList{\curvWarpedV}%
      \xdef\curvRawTickList{\curvLbl}%
    \else
      \xdef\curvWarpedTickList{\curvWarpedTickList,\curvWarpedV}%
      \xdef\curvRawTickList{\curvRawTickList,\curvLbl}%
    \fi
  }%
  \ifx\PanelAtPos\bapEmpty
    \begin{tikzpicture}
    \begin{axis}[
        x filter/.code={%
          \pgfmathparse{%
            \curvC*\pgfmathresult
            + (1-\curvC)/(2*\curvK)*(%
                lncosh(\curvK*(\pgfmathresult+\curvBreak))
                -lncosh(\curvK*(\pgfmathresult-\curvBreak)))%
          }%
        },
        width=\pgfkeysvalueof{/1dcurv/width},
        height=\pgfkeysvalueof{/1dcurv/height},
        scale only axis,
        tick label style={font=\footnotesize},
        yticklabel style={xshift=-4pt},
        grid=major,
        scaled y ticks=true,
        xmin=\curvXMin, xmax=\curvXMax,
        xtick/.expand once=\curvWarpedTickList,
        xticklabels/.expand once=\curvRawTickList,
        ymin=\pgfkeysvalueof{/1dcurv/ymin}, ymax=\pgfkeysvalueof{/1dcurv/ymax},
        ytick={\pgfkeysvalueof{/1dcurv/ytick}},
        xlabel={\pgfkeysvalueof{/1dcurv/xlabel}},
        ylabel={\pgfkeysvalueof{/1dcurv/ylabel}},
        title={\pgfkeysvalueof{/1dcurv/title}},
        \pgfkeysvalueof{/1dcurv/extra},
        \ifx\curvLegendName\bapEmpty\else
          legend to name=\curvLegendName,
          legend style={legend columns=\pgfkeysvalueof{/1dcurv/legendcolumns},font=\footnotesize,column sep=1ex},
          legend cell align=left,
        \fi
    ]
    \input{figures/data/\pgfkeysvalueof{/1dcurv/datadir}/\pgfkeysvalueof{/1dcurv/data}.tex}
    \pgfkeysvalueof{/1dcurv/extradata}
    \end{axis}
    \end{tikzpicture}%
  \else
    \begin{axis}[
        at/.expand once=\PanelAtPos, anchor=south west,
        x filter/.code={%
          \pgfmathparse{%
            \curvC*\pgfmathresult
            + (1-\curvC)/(2*\curvK)*(%
                lncosh(\curvK*(\pgfmathresult+\curvBreak))
                -lncosh(\curvK*(\pgfmathresult-\curvBreak)))%
          }%
        },
        width=\pgfkeysvalueof{/1dcurv/width},
        height=\pgfkeysvalueof{/1dcurv/height},
        scale only axis,
        tick label style={font=\footnotesize},
        yticklabel style={xshift=-4pt},
        grid=major,
        scaled y ticks=true,
        xmin=\curvXMin, xmax=\curvXMax,
        xtick/.expand once=\curvWarpedTickList,
        xticklabels/.expand once=\curvRawTickList,
        ymin=\pgfkeysvalueof{/1dcurv/ymin}, ymax=\pgfkeysvalueof{/1dcurv/ymax},
        ytick={\pgfkeysvalueof{/1dcurv/ytick}},
        xlabel={\pgfkeysvalueof{/1dcurv/xlabel}},
        ylabel={\pgfkeysvalueof{/1dcurv/ylabel}},
        title={\pgfkeysvalueof{/1dcurv/title}},
        \pgfkeysvalueof{/1dcurv/extra},
        \ifx\curvLegendName\bapEmpty\else
          legend to name=\curvLegendName,
          legend style={legend columns=\pgfkeysvalueof{/1dcurv/legendcolumns},font=\footnotesize,column sep=1ex},
          legend cell align=left,
        \fi
    ]
    \input{figures/data/\pgfkeysvalueof{/1dcurv/datadir}/\pgfkeysvalueof{/1dcurv/data}.tex}
    \pgfkeysvalueof{/1dcurv/extradata}
    \end{axis}
  \fi
}

\newcommand{\OneDProfilePair}[3]{%
  \noindent\makebox[\linewidth]{%
  \begin{tikzpicture}
  \useasboundingbox (-\LeftReserve,-\BottomReserve)
    rectangle ({2*\BoxW+\PanelGap+\RightReserve},{\BoxH+\TopReserve});
  \def\PanelAtPos{(0,0)}%
  #1%
  \def\PanelAtPos{(\BoxW+\PanelGap,0)}%
  #2%
  \end{tikzpicture}%
  }%
  \edef\bapPairLegendName{#3}%
  \ifx\bapPairLegendName\bapEmpty\else
    \par\vspace{0.5em}%
    \begin{center}{\NoHyper\ref{#3}\endNoHyper}\end{center}%
  \fi
}

\newcommand{\LegendSwatch}[2]{\tikz[baseline=-0.5ex]{\draw[#1,thick,#2] (0,0) -- (0.45cm,0);}}

\newwrite\CropBoxFile
\newcommand{\ExportTrimBox}[3]{%
  \immediate\openout\CropBoxFile=#1_trim.tex
  \immediate\write\CropBoxFile{\string\includegraphics[trim=0pt \the\dimexpr\paperheight-(#3)\relax\space\the\dimexpr\paperwidth-(#2)\relax\space 0pt,clip]{figures/compiled/#1.pdf}}%
  \immediate\closeout\CropBoxFile
}

%% file: macros.tex
\DeclareMathAlphabet{\pazocal}{OMS}{zplm}{m}{n}

\newcommand{\refeq}[1]{Eq.(\ref{#1})}
\newcommand{\refeqs}[2]{Eqs.(\ref{#1})--(\ref{#2})}

\newcommand{\refequations}[2]{Equations (\ref{#1})--(\ref{#2})}
\newcommand{\reffig}[1]{Fig.(\ref{#1})}
\newcommand{\reffigs}[2]{Figs.(\ref{#1})--(\ref{#2})}

\newcommand{\beq}{\begin{equation}}
\newcommand{\eeq}{\end{equation}}

\newcommand{\bdm}{\begin{displaymath}}
\newcommand{\edm}{\end{displaymath}}

\newcommand{\beqa}{\begin{eqnarray}}
\newcommand{\eeqa}{\end{eqnarray}}

\newcommand{\beqas}{\begin{eqnarray*}}
\newcommand{\eeqas}{\end{eqnarray*}}

\newcommand{\barr}{\begin{array}}
\newcommand{\earr}{\end{array}}

\newcommand{\bit}{\begin{itemize}}
\newcommand{\eit}{\end{itemize}}

\newcommand{\Sp}[1]{{\pazocal{#1}}}
\newcommand{\sA}{\Sp{A}}

\newcommand{\sH}{\Sp{H}}

\newcommand{\sL}{\Sp{L}}

\newcommand{\sN}{\Sp{N}}

\newcommand{\sP}{\Sp{P}}
\newcommand{\sQ}{\Sp{Q}}

\newcommand{\sT}{\Sp{T}}

\newcommand{\sV}{\Sp{V}}
\newcommand{\sW}{\Sp{W}}
\newcommand{\sX}{\Sp{X}}

\newcommand{\V}[1]{{\bf #1}}

\newcommand{\vu}{\V{u}}
\newcommand{\vv}{\V{v}}

\newcommand{\vx}{\V{x}}

\newcommand{\vA}{\V{A}}
\newcommand{\vB}{\V{B}}

\newcommand{\vE}{\V{E}}

\newcommand{\vJ}{\V{J}}

\newcommand{\vP}{\V{P}}

\newcommand{\vV}{\V{V}}

\newcommand{\vzero}{\V{0}}

\newcommand{\pd}[2]{\frac{\partial #1}{\partial #2}}
\newcommand{\ppd}[2]{\frac{\partial^2 #1}{\partial #2^2}}
\newcommand{\pmd}[3]{\frac{\partial^2 #1}{\partial #2 \partial #3}}

\newcommand{\dd}[0]{\, \mathrm{d}}

%% file: layout_setup.tex
\makeatletter
\newcommand{\captionof}[2]{%
  \refstepcounter{#1}%
  \par
  \vskip\abovecaptionskip
  \begingroup
    \small\rmfamily
    \samepage
    \flushing
    \let\footnote\@footnotemark@gobble
    \@make@capt@title{\csname fnum@#1\endcsname}{#2}\par
  \endgroup
  \vskip\belowcaptionskip
}
\makeatother

\makeatletter
\def\widefigure@toprule{%
  \noindent\hbox to\linewidth{%
    \color{black!30}%
    \leaders\hrule height0.4\p@\hfil
  }%
}
\def\widefigure@botrule{%
  \noindent\hbox to\linewidth{%
    \color{black!30}%
    \leaders\hrule height0.4\p@\hfil
  }%
}
\makeatother

\makeatletter
\newbox\widefigure@box
\newenvironment{widefigure}{%
  \par\ignorespaces
  \setbox\widefigure@box\vbox\bgroup
  \hsize\textwidth
  \linewidth\hsize
  \parindent\z@
  \widefigure@toprule
  \vskip8\p@
}{%
  \vskip6\p@
  \widefigure@botrule
  \par
  \egroup
  \onecolumngrid
  \vskip26\p@
  \box\widefigure@box
  \vskip16\p@
  \twocolumngrid\global\@ignoretrue
  \@endpetrue
}
\makeatother

%% file: v2/section_01.tex
\section{\label{sec:introduction}Introduction}

Plasma stability and transport analysis assumes a time-independent
solution of the governing equations. For collisionless plasmas, analytical
kinetic stability theory has long relied on Vlasov--Maxwell equilibrium solutions 
as base states for such analyses
\cite{Harris1962,Krall1963,Channell1976,Mynick1979}.
In linear and nonlinear kinetic simulations, equilibrium states play an equally important role. 
In $\delta f$ methods, a zeroth-order time-independent
state $f_0$ is prescribed and only the departure from it, $\delta f$, is evolved
\cite{ParkerLee1993}. If $f_0$ is not an exact equilibrium solution, the
zeroth-order terms in the Vlasov equation do not cancel in the resulting equation 
for $\delta f$. The residual either acts as a source that can drive, possibly large,
transient dynamics or, more commonly, is dropped entirely from the
formulation, effectively modifying the underlying equations being solved.
In methods which evolve the full distribution function, 
any departure of the initial state from equilibrium evolves with, and 
cannot be separated from, the dynamics under study. This can lead to 
simulations dominated by the relaxation of the initial state rather 
than the intended physics.

In practice, time-independent base states for magnetized plasma simulations 
are commonly constructed from fluid information rather than at the kinetic level.
The magnetic field is generally taken from an MHD (Grad--Shafranov)
equilibrium \cite{GradRubin1958,Shafranov1958} with prescribed density and 
temperature profiles as flux functions. Each species is then
represented by a local Maxwellian distribution function matching those profiles. When the
gyroradius, and in toroidal geometry the drift-orbit width, is small
compared with the gradient scale lengths, such a state departs from an
exact equilibrium at first order in that small ratio, which may be
adequate for certain simulations of well-magnetized plasmas.
When orbit widths are comparable to the gradient scale, however, there
is no longer reason to expect such a construction to be close to a steady-state
solution. Such initial states relax, often exciting large transients, and may
settle into states other than the one intended, complicating the
interpretation of simulation results. Examples include substantial electric fields from charge separation
between species with different gyroradii in particle simulations of
current sheets and tangential layers initialized from local Maxwellians
\cite{Belmont2012,Dorville2015,Allanson2017}, and lower- and upper-hybrid
waves and boundary sheaths in continuum Vlasov simulations of magnetized
shear layers initialized from two-fluid or approximate kinetic equilibria
\cite{Vogman2019,Vogman2020}. In full-$f$ gyrokinetic
simulations of ion-temperature-gradient (ITG) turbulence, a local
Maxwellian with profiles constant on flux surfaces has been shown to
generate a large radial electric field on short time scales, whose sheared
$E\times B$ flow can suppress the instability under study
\cite{DifPradalier2008}. In global gyrokinetic simulations of ITG
turbulence, initializing from a distribution that is not a kinetic
equilibrium affects the zonal flows and turbulent transport that develop,
with spurious zonal-flow oscillations reported in realistic aspect-ratio
configurations \cite{Idomura2003,Angelino2006}. In total-$f$ gyrokinetic
simulations of the tokamak edge, a local Maxwellian constructed from
flux-surface-averaged profiles excites large-amplitude geodesic acoustic
modes that can persist into the early turbulence phase \cite{Ku2026}.
The present work is motivated by these same concerns in the fully kinetic
setting, in which the full Lorentz-force gyromotion is resolved rather than
gyro-averaged.

Regimes in which orbit widths and gradient scales are not well separated 
are of growing practical interest. Development of fully kinetic simulation
capabilities has been motivated in part by the H-mode pedestal region of magnetic 
confinement devices, where gradients are steep and the scale separation 
underlying typical gyrokinetic ordering assumptions is not so well satisfied
\cite{ChenParker2009,Schnack2013,Sturdevant2016,Sturdevant2017,Miecnikowski2018,Hu2018}.
In recent years, a range of fully kinetic particle and continuum methods
has been developed with an emphasis on conservation, structure
preservation, and scalability
\cite{ChenChaconBarnes2011,ChenChacon2015,Kraus2017,Juno2018,Kormann2019,RicketsonChacon2020,XiaoQin2021,WangQin2021,Vogman2021,ChenChacon2023,Finn2023,Finn2026},
and these stand to benefit from initial states that are exact
Vlasov--Maxwell equilibria.
Steep-gradient regions such as the pedestal are also where equilibrium electric fields are of fundamental importance:
sheared $E\times B$ flow is understood to suppress turbulence and
drive the formation of transport barriers \cite{Burrell1997}. Exact
kinetic equilibria with self-consistent electric
fields are therefore valuable starting points for kinetic studies of
these regions. The generation and effects of equilibrium electric
fields within the present model are explored in this paper.

In gyrokinetic simulation, where these issues have been studied
extensively, defining an equilibrium is itself subtle \cite{Qin2000}.
A common approach is a canonical Maxwellian, a distribution function of the
constants of motion \cite{Idomura2003,Angelino2006}. In a torus, the curvature and
$\nabla B$ drifts separate ions and electrons vertically; a canonical
Maxwellian carries a parallel flow that compensates this charge
separation, so that the spurious radial electric field produced by a local
Maxwellian does not develop \cite{DifPradalier2008}. A background
electrostatic potential can also be incorporated in such distributions
\cite{McMillan2011}. Constructing canonical Maxwellian distribution functions
to reproduce prescribed profiles together with a self-consistent radial
electric field remains difficult, particularly in the steep-gradient edge
\cite{Ku2026,Li2025}. Although the present construction does not address
the gyrokinetic setting directly, it follows the same principle of building
the equilibrium from the constants of motion, here for the Vlasov--Maxwell
case and with the electric field determined self-consistently.

In the present work, each species' distribution function is taken to depend 
only on the constants of motion, following Jeans' theorem \cite{Jeans1915} and
already treated in early textbooks \cite{Longmire1963}, so that
Maxwell's equations reduce to a pair of nonlinear ordinary differential
equations for the electrostatic and vector potentials, $\phi$ and
$A_y$. The charge and current densities are then derivatives of a
single stress tensor component, $\rho=-\partial P_{xx}/\partial\phi$ and
$J_y=\partial P_{xx}/\partial A_y$, a classical reduction to the motion
of a pseudo-particle in a pseudo-potential
\cite{Grad1961,Bertotti1963,Kan1972,Channell1976,Mynick1979,Harrison2009pop},
used extensively to model current sheets and boundary layers observed
by spacecraft \cite{Roth1996,Mottez2003,Belmont2012,Dorville2015,Allanson2017}.
Most constructions built on this reduction choose distribution
functions with closed-form moments
\cite{Harris1962,Kan1972,Channell1976,Harrison2009prl,Neukirch2009,Allanson2015,Allanson2017},
integrate the field equations as initial-value problems
\cite{Roth1996,Kaltsas2023,Kaltsas2025} rather than solving a
self-consistent nonlinear boundary-value problem as is done here, and
either set $\phi=0$ \cite{Harris1962,Channell1976,Neukirch2009} or
eliminate it in favor of a quasi-neutral $\phi(A)$ obtained by
solving $\rho(\phi,A)=0$
\cite{Sestero1964,Sestero1966,Mynick1979,Roth1996,Schindler2002,Tasso2014,Kuiroukidis2015,Kaltsas2023,Kaltsas2024,Kaltsas2025}.
Outside this reduction, Vlasov--Poisson equilibria in a fixed,
uniform magnetic field have been constructed numerically: customizable
two-species equilibria, built from distribution functions of the constants
of motion with a uniform temperature, are obtained by solving the nonlinear
Poisson equation as a boundary-value problem with numerically evaluated
moments, and are used to initialize continuum kinetic simulations of
Kelvin--Helmholtz and lower-hybrid drift instabilities
\cite{Vogman2019,Vogman2020,Vogman2021}. There, the magnetic field is
prescribed rather than determined self-consistently.

Here, the full system for $\phi$ and $A_y$ is posed as a nonlinear
boundary-value problem for an arbitrary distribution function of the constants of motion,
with the full Poisson equation retained and the magnetic field determined
self-consistently, and solved with a
discretization that conserves the pseudo-particle Hamiltonian exactly.
The moments are evaluated by quadrature, so the distribution function
is not restricted to forms with closed-form moments. As a concrete
illustration of the method, a canonical bi-Maxwellian family is
introduced, with density and temperature profiles prescribed as
functions of a guiding-center-like first integral; it reduces to the
familiar local-Maxwellian prescription when orbits are small but
remains an exact equilibrium when they are not. A procedure
for loading particles consistent with the equilibrium in
particle-in-cell simulations is given, and an asymptotic analysis of
the interior in the well-magnetized, quasi-neutral limit provides an
independent check on the numerical solutions.

Within this model, and in terms of the well-magnetized, quasi-neutral
ordering of Appendix~\ref{sec:asymptotic_approach}, equilibrium
electric fields arise at zeroth order when the species are not
charge-balanced in their canonical density profiles, and at first
order due to finite-Larmor-radius (FLR) separation between species
and finite-$\beta$ effects; Sec.~\ref{sec:equilibrium_efield}
examines these mechanisms in detail and identifies a key property of
the species flow: an exact cancellation between the $E\times B$ flow and
part of the pressure-gradient-driven flow. In the model, each species' perpendicular flow may be described as the sum of
the $E\times B$ flow and a pressure-gradient-driven flow computed from a component
of the species stress tensor. Because the species stress depends on position only 
through the potentials, the latter splits by the chain rule into an electrostatically and a magnetostatically
generated part, resulting in a decomposition of the form
\begin{equation*}
V_{s,\perp}=V_E+V_{s,*\phi}+V_{s,*A_y}.
\label{eq:intro_flow}
\end{equation*}
In the present model the electrostatically generated part cancels the
$E\times B$ flow \emph{exactly}, for each species separately and for
any distribution function of the constants of motion, as a direct
consequence of the species identity $\pd{P_{s,xx}}{\phi}=-\rho_s$. 
In other words, 
\begin{equation*}
V_{s,*\phi}=-V_E,
\label{eq:intro_cancellation}
\end{equation*}
leaving the total perpendicular flow
of each species to come only from the magnetostatically generated
part, $V_{s,*A_y}$, which leads to the classic diamagnetic current
commonly associated with the pressure-gradient-driven flow as a whole.
A large self-consistent electric field therefore need not produce a
correspondingly large species flow. Configurations illustrating such a result 
are constructed and verified numerically. This cancellation follows from a structural property of the species
force balance that arises from this type of symmetry reduction, and
does not depend on consistency with Maxwell's equations or whether $\phi$ 
is determined by the Poisson equation or by quasi-neutrality. A
single-particle picture of the cancellation, related to the classic
distinction between particle drifts and fluid flows \cite{Spitzer1952},
is given in Appendix~\ref{sec:appendix_particle_picture}.

Elements of this type of cancellation can be recognized in earlier
work. In the analysis of thin current sheets, where quasi-neutrality is
used from the outset to impose $\phi=\phi(A)$, a chain-rule expansion of
the species pressure is used to identify the electron Hall current
\cite{Schindler2002}. A balance between $E\times B$ and diamagnetic flows
has been observed numerically \cite{Shoucri2004}, and later reported
directly in a similar study comparing the radial electric field to the
ion pressure gradient \cite{Shoucri2012}. At the textbook level, a fluid
treatment shows the $E\times B$ and diamagnetic flows of isothermal
electrons to be equal and opposite, as a consequence of the specific
(Boltzmann) functional form assumed for the electron density \cite{Chen_textbook}.

The remainder of the paper is organized as follows. Section~\ref{sec:VM_system}
introduces the time-independent Vlasov--Maxwell system and the symmetry
assumptions used to simplify it. Section~\ref{sec:reduction_to_odes} reduces
the resulting system to a pair of coupled nonlinear ordinary differential
equations for the electrostatic and vector potentials, together with their
Lagrangian and Hamiltonian structure and dimensionless form.
Section~\ref{sec:numerical_approach} presents a conservative spatial
discretization of this boundary-value problem and the nonlinear solver used
to obtain equilibrium solutions. Section~\ref{sec:pseudo_bimaxwellian_family}
introduces a concrete canonical bi-Maxwellian distribution family, including its
numerical evaluation and an associated particle-loading procedure for
particle-in-cell simulations. Section~\ref{sec:numerical_verification}
verifies the numerical method against discrete Hamiltonian conservation,
spatial and quadrature convergence, and the asymptotic description derived in
Appendix~\ref{sec:asymptotic_approach}, and characterizes a representative
resolved equilibrium, including its fields, densities, and species flows.
Section~\ref{sec:equilibrium_efield}
examines the generation of equilibrium electric fields and the $E\times B$
cancellation property described above. Conclusions are given in
Section~\ref{sec:conclusions}.

%% file: v2/section_02.tex
\section{The time-independent Vlasov--Maxwell system and symmetry assumptions \label{sec:VM_system}}

Time-independent solutions of the Vlasov--Maxwell system, i.e. solutions that satisfy this system under the constraint that partial derivatives with respect to time of all quantities are zero, are considered. Each particle species $s$ is described by a distribution function $f_s(\vx,\vv): \mathbb{R}^6 \rightarrow \mathbb{R}^{+}$ satisfying a time-independent Vlasov equation
\begin{align}
\vv \cdot \nabla f_s + \frac{q_s}{m_s} \left( \vE + \vv \times \vB \right) \cdot \nabla_\vv f_s = 0. \label{eq:vlasov}
\end{align}
Here, $q_s$ and $m_s$ denote the charge and mass of species $s$, respectively. The electric and magnetic fields, $\vE$ and $\vB$, are determined by the time-independent Maxwell's equations
\begin{align}
\epsilon_0 \nabla \cdot \vE &= \rho \label{eq:ti_gauss} \\
\frac{1}{\mu_0} \nabla \times \vB &= \vJ \label{eq:ti_ampere} \\
\nabla \cdot \vB &= 0 \label{eq:ti_gauss_mag} \\
\nabla \times \vE &= \vzero, \label{eq:ti_faraday} 
\end{align}
where the charge density, $\rho$, and the current density, $\vJ$, consist of contributions from each species $s$. In particular, 
\begin{equation}
\rho(\vx) = \sum_s \rho_s \quad \mathrm{and} \quad \vJ(\vx) = \sum_s \vJ_s,
\end{equation}
where $\rho_s$ and $\vJ_s$ are the charge density and current density of species $s$, respectively, and are computed from the species distribution functions $f_s$ according to
\begin{align}
\rho_s(\vx) &= q_s \int_{\mathbb{R}^3} f_s(\vx,\vv) d^3 v \\
\vJ_s(\vx) & = q_s \int_{\mathbb{R}^3} \vv f_s(\vx,\vv) d^3 v.
\end{align}
\refequations{eq:ti_gauss_mag}{eq:ti_faraday} allow the electric and magnetic fields to be expressed in terms of a scalar and vector potential respectively as $\vE = -\nabla \phi$ and $\vB = \nabla \times \vA$. Hence \refeqs{eq:ti_gauss}{eq:ti_ampere} are equivalent to the following two equations:
\begin{align}
-\epsilon_0 \nabla \cdot \nabla \phi &= \rho \label{eq:poisson} \\
\frac{1}{\mu_0} \nabla \times \nabla \times \vA &= \vJ \label{eq:ampere}.
\end{align}
Here, an important constraint on the solutions is noted: the force-balance equation for each species, which is readily obtained by multiplying \refeq{eq:vlasov} by $m_s \vv$ and integrating over velocity space. This can be written as
\begin{align}
\nabla \cdot \vP_s - \rho_s \vE - \vJ_s \times \vB = \vzero,\,
\end{align}
where $\vP_s$ is the species stress tensor. The total force-balance equation is then obtained by summing over $s$:
\begin{align}
\nabla \cdot \vP - \rho \vE - \vJ \times \vB = \vzero,\, \label{eq:general_force_balance}
\end{align}
with $\vP = \sum_s \vP_s$.
\par To render the task tractable, a number of symmetry assumptions are imposed on the solutions. The following ansatz is adopted for $\vE$ and $\vB$:
\begin{align}
\vE &= E_x \hat{x}, \qquad E_x \equiv -\pd{\phi}{x}, \label{eq:field_ansatz} \\
\vB &= B_z \hat{z}, \qquad B_z \equiv \pd{A_y}{x},
\end{align}
where $\phi$ and $A_y$ are functions of $x$ only, and $E_x$, $B_z$ denote the corresponding (only nonzero) field components. Additionally, solutions are sought where $f_s$ is uniform in the $y$ and $z$ coordinates,
\begin{equation}
\pd{f_s}{y} = \pd{f_s}{z} = 0, \label{eq:fs_assumptions_uniform}
\end{equation}
and have vanishing first moments in $v_x$ and $v_z$,
\begin{equation}
\int_{\mathbb{R}^3} v_x f_s d^3 v = \int_{\mathbb{R}^3} v_z f_s d^3 v = 0. \label{eq:fs_assumptions_moments}
\end{equation}
These assumptions lead to the following simpler system of equations:
\begin{multline}
v_x \pd{f_s}{x}
- \frac{q_s}{m_s} \pd{\phi}{x} \pd{f_s}{v_x}
\\
+ \frac{q_s}{m_s} \pd{A_y}{x}
\left(
v_y \pd{f_s}{v_x}
- v_x \pd{f_s}{v_y}
\right)
= 0,
\label{eq:simp_vlasov}
\end{multline}
for each $s$, and
\begin{align}
- \epsilon_0 \ppd{\phi}{x} &= \rho \label{eq:simp_poisson} \\
- \frac{1}{\mu_0} \ppd{A_y}{x} &= J_y. \label{eq:simp_ampere}
\end{align}
The force-balance equations for each species take the simplified form
\begin{align}
\pd{P_{s,xx}}{x} + \rho_s \pd{\phi}{x} - J_{s,y} \pd{A_y}{x} = 0, \label{eq:simp_species_force_balance}
\end{align}
where $P_{s,xx}$ is the stress tensor component of species $s$, corresponding to the $m_s v_x^2$ moment of $f_s$. Summation over $s$ leads to
\begin{align}
\pd{P_{xx}}{x} + \rho \pd{\phi}{x} - J_{y} \pd{A_y}{x} = 0. \label{eq:simp_force_balance}
\end{align}
The existence of a class of functions, $f_s$, which satisfy \refeq{eq:simp_vlasov} and the aforementioned assumptions in \refeqs{eq:fs_assumptions_uniform}{eq:fs_assumptions_moments} will subsequently be demonstrated. The remaining challenge will then be to ensure consistency with \refeqs{eq:simp_poisson}{eq:simp_ampere}.

%% file: v2/section_03.tex
\section{Reduction to a System of Nonlinear Ordinary Differential Equations \label{sec:reduction_to_odes}}

In this section, the problem of obtaining solutions to the system \refeqs{eq:simp_vlasov}{eq:simp_ampere}, subject to \refeqs{eq:fs_assumptions_uniform}{eq:fs_assumptions_moments}, is reduced to that of solving a system of two coupled nonlinear second-order ordinary differential equations (ODEs) for $\phi$ and $A_y$. First, a general form for solutions of \refeq{eq:simp_vlasov} is derived for prescribed but arbitrary $\phi$ and $A_y$. Some care is required in choosing the class of functions satisfying \refeq{eq:simp_vlasov} so that they decay sufficiently rapidly for the moments appearing in \refeq{eq:fs_assumptions_moments} and \refeqs{eq:simp_poisson}{eq:simp_ampere} to be well-defined. From this general solution, the moments of $f_s$ appearing in \refeqs{eq:simp_poisson}{eq:simp_ampere} are expressed as nonlinear functions of $\phi$ and $A_y$, resulting in the ODE system. Lagrangian and Hamiltonian descriptions of the system are then given. The section concludes with a nondimensionalization of the system to clarify the relevant physical scales and the magnitudes of the fundamental dimensionless parameters.

\subsection{A General Vlasov Solution for Prescribed Potentials}

For prescribed $\phi$ and $A_y$, \refeq{eq:simp_vlasov} is a first-order, linear, homogeneous partial differential equation (PDE) in the three independent variables $x, v_x$, and $v_y$. Any dependence of $f_s$ on $v_z$ is parametric in the context of this equation, while \refeq{eq:fs_assumptions_uniform} requires $f_s$ to be independent of $y$ and $z$. Solutions to PDEs of this class are constant along characteristic curves, which can be described by a maximal set of functionally independent first integrals. Because \refeq{eq:simp_vlasov} has three independent variables, its characteristic curves are determined by two functionally independent first integrals. These may be taken as
\begin{align}
X_s(A_y;\vv) &= \frac{A_y}{B_0} + \frac{m_s}{q_s B_0} v_y \label{eq:gcenter} \\
W_{s, \perp}(\phi;\vv) &= \frac{m_s}{2} \left( v_x^2 + v_y^2 \right) + q_s \phi. \label{eq:perp_energy}
\end{align}
Here, $X_s$ is a scaled canonical momentum conjugate to $y$ with dimensions of length, while $W_{s, \perp}$ gives the contribution to the particle energy associated with motion perpendicular to the magnetic field. In \refeq{eq:gcenter}, $B_0$ is a reference magnetic-field strength characterizing, for example, an externally applied field.
Consequently, for each fixed $v_z$, a general $C^1$ solution of \refeq{eq:simp_vlasov} can be expressed as an arbitrary function of $X_s$ and $W_{s,\perp}$; see, for example, \cite{Zachmanoglou} or \cite{PrasadRavindran} for a rigorous treatment of this result for first-order PDEs.

To also enforce the vanishing $v_z$ moment in \refeq{eq:fs_assumptions_moments}, the dependence on $v_z$ is taken to be even. A convenient choice is to express this dependence through
\begin{align}
W_{s, \parallel}(\vv) &= \frac{m_s}{2} v_z^2, \label{eq:par_energy}
\end{align}
which gives the contribution to the particle energy associated with motion parallel to the magnetic field. The solutions considered here therefore take the form
\begin{multline}
\label{eq:general_F}
f_s(x,\vv) = \\ 
F_s\left(X_s(A_y(x);\vv),W_{s, \perp}(\phi(x);\vv),W_{s, \parallel}(\vv)\right),
\end{multline}
with $F_s$ taken to be $C^1$ in its three arguments. Provided these functions decay sufficiently rapidly in velocity space for the required moments to exist, they also satisfy \refeqs{eq:fs_assumptions_uniform}{eq:fs_assumptions_moments}, since the coordinates $y$ and $z$ are absent from $X_s$, $W_{s, \perp}$, and $W_{s, \parallel}$, while the dependence on $v_x$ and $v_z$ is even.

\subsection{Coupling to Maxwell's Equations}

Consistency with \refeqs{eq:simp_poisson}{eq:simp_ampere} is obtained by taking the required moments of $f_s$ and substituting them into Maxwell's equations. This results in two coupled nonlinear autonomous second-order ODEs whose solution determines the potentials $\phi$ and $A_y$ needed to construct self-consistent solutions of \refeqs{eq:simp_vlasov}{eq:simp_ampere}.
\begin{align}
-\epsilon_0 \ppd{\phi}{x} &= \rho \left(\phi,A_y\right) \label{eq:phi_ode} \\
- \frac{1}{\mu_0} \ppd{A_y}{x} &= J_y \left(\phi,A_y\right). \label{eq:A_ode}
\end{align}
In particular, the charge density and the $y$-component of the current density depend on position only through $\phi$ and $A_y$, as a direct consequence of the spatial dependence of $f_s$ entering through the first integrals \refeqs{eq:gcenter}{eq:par_energy}.

\subsection{Lagrangian and Hamiltonian Descriptions}
The ODE system \refeqs{eq:phi_ode}{eq:A_ode} admits Lagrangian and Hamiltonian formulations. Consider the species stress tensor component $P_{s,xx}$, corresponding to the $m_s v_x^2$ moment of $f_s$. Like $\rho_s$ and $J_{s,y}$, $P_{s,xx}$ depends on position only through $\phi$ and $A_y$. Applying the chain rule to the species force-balance equation \refeq{eq:simp_species_force_balance} gives
\begin{equation}
\left( \pd{P_{s,xx}}{\phi} + \rho_s \right) \pd{\phi}{x} + \left( \pd{P_{s,xx}}{A_y} - J_{s,y} \right) \pd{A_y}{x} = 0.
\end{equation}
The species force-balance equation is simply a result of the Vlasov
equation, valid for any independently prescribed $\phi$ and $A_y$,
regardless of how they are determined in the end. Their values and
derivatives at any point can therefore be assigned independently, so
the relation above, being linear in $\pd{\phi}{x}$ and $\pd{A_y}{x}$,
forces each coefficient to vanish separately. It therefore follows that
\begin{align}
\pd{P_{s,xx}}{\phi} &= -\rho_s, \label{eq:species_phi_identity} \\
\pd{P_{s,xx}}{A_y} &= J_{s,y}. \label{eq:species_A_identity}
\end{align}
Alternatively, these identities follow by differentiating the
$m_s v_x^2$ moment of \refeq{eq:general_F} with respect to $\phi$ and
$A_y$, rewriting the resulting derivatives of $F_s$ in terms of
velocity derivatives of $f_s$ via the chain rule, and integrating by
parts in velocity space.
Summing over species recovers the same identities for the total stress, $\pd{P_{xx}}{\phi}=-\rho$ and $\pd{P_{xx}}{A_y}=J_y$.

This pair of identities, or its magnetostatic half alone, is well established. The magnetostatic half alone appears in \cite{Grad1961,Krall1963,Channell1976,Mottez2003,Allanson2015,Allanson2017,Wilson2017,Kaltsas2024}, while the full pair, including the electrostatic half, appears in \cite{Bertotti1963,Kan1972,Mynick1979,Harrison2009pop,Wilson2011,Neukirch2009,Schindler2002,Kaltsas2023}. Most of these treat the total, species-summed stress; \cite{Schindler2002} is a notable exception, working, as is done here, at the species level.
The magnetostatic half of this identity, $\partial P_{xx}/\partial A_y=J_y$, generalizes the classic requirement that pressure be a flux function in MHD equilibrium, as in the Grad--Shafranov equation \cite{GradRubin1958,Shafranov1958}, to the case of a second potential $\phi$; the two coincide when $P_{xx}$ has no $\phi$-dependence.
Using these identities, the ODE system can be written as
\begin{align}
\epsilon_0 \ppd{\phi}{x} &= \pd{P_{xx}}{\phi} \\
-\frac{1}{\mu_0} \ppd{A_y}{x} &= \pd{P_{xx}}{A_y},
\end{align}
which are recognized as the Euler-Lagrange equations generated by the Lagrangian
\begin{align}
L = \frac{\epsilon_0}{2} \left(\pd{\phi}{x}\right)^2 - \frac{1}{2 \mu_0} \left(\pd{A_y}{x}\right)^2 + P_{xx}(\phi,A_y). \label{eq:lagrangian}
\end{align}
The corresponding Hamiltonian is
\begin{align}
H = \frac{\epsilon_0}{2} \left(\pd{\phi}{x}\right)^2 - \frac{1}{2 \mu_0} \left(\pd{A_y}{x}\right)^2 - P_{xx}(\phi,A_y), \label{eq:hamiltonian}
\end{align}
where $x$ plays the role of the independent variable in the corresponding dynamical system. Since the Lagrangian has no explicit dependence on $x$, the Hamiltonian is conserved:
\begin{align}
\pd{H}{x} = 0.
\end{align}
The same conservation law can also be obtained by eliminating $\rho$ and $J_y$ from \refeq{eq:simp_force_balance} using \refeq{eq:simp_poisson} and \refeq{eq:simp_ampere}, respectively, and integrating once with respect to $x$.

\subsection{Dimensionless Form}

It is convenient to introduce dimensionless variables. Characteristic scales for density, temperature, and spatial variation are denoted by $N_0$, $T_0$, and $L_0$, respectively. Using these scales together with the previously introduced reference magnetic-field strength $B_0$, define the dimensionless potentials and spatial coordinate by
\begin{equation}
\sV = \frac{e\phi}{T_0},
\qquad
\sA = \frac{A_y}{B_0 L_0},
\qquad
\xi = \frac{x}{L_0}.
\end{equation}
The stress tensor component $P_{xx}$ is similarly normalized as
\begin{equation}
\sP = \frac{P_{xx}}{N_0T_0}.
\end{equation}
Two dimensionless parameters arise naturally in the resulting ODE system. The first is the normalized Debye length,
\begin{equation}
\lambda = \frac{\epsilon_0 T_0}{L_0 e^2 N_0},
\end{equation}
and the second is the characteristic plasma beta
\begin{equation}
\beta = \frac{2\mu_0N_0T_0}{B_0^2}.
\end{equation}
In terms of these variables, the ODE system \refeqs{eq:phi_ode}{eq:A_ode} becomes
\begin{align}
\lambda^2 \ppd{\sV}{\xi}
&= \pd{\sP}{\sV}, \label{eq:dl_phi_ode} \\
-\frac{2}{\beta}\ppd{\sA}{\xi}
&= \pd{\sP}{\sA}. \label{eq:dl_A_ode}
\end{align}

The corresponding dimensionless Lagrangian is
\begin{equation}
\sL =
\frac{1}{2}\lambda^2
\left(\pd{\sV}{\xi}\right)^2
-\frac{1}{\beta}
\left(\pd{\sA}{\xi}\right)^2
+\sP(\sV,\sA),
\label{eq:dl_lagrangian}
\end{equation}
and the dimensionless Hamiltonian is
\begin{equation}
\sH =
\frac{1}{2}\lambda^2
\left(\pd{\sV}{\xi}\right)^2
-\frac{1}{\beta}
\left(\pd{\sA}{\xi}\right)^2
-\sP(\sV,\sA).
\label{eq:dl_hamiltonian}
\end{equation}
As in the dimensional formulation, $\sH$ is conserved.

For evaluating the stress from the distribution functions, it is also useful to introduce dimensionless velocity coordinates and first integrals. Define a characteristic velocity for each species by
\begin{equation}
v_{0,s} = \sqrt{\frac{T_0}{m_s}},
\end{equation}
and the corresponding species-dependent dimensionless velocity coordinates by 
\begin{equation}
\vu_s = \vv / v_{0,s}.
\end{equation}
The corresponding signed gyroradius scale and signed normalized
gyroradius parameter are
\begin{align}
\ell_{\mathrm{gy},s}
&=
\frac{m_s v_{0,s}}{q_s B_0},
\notag\\
\varrho_s
&=
\frac{\ell_{\mathrm{gy},s}}{L_0}.
\end{align}
The normalized gyroradius parameters are collected into the vector
\begin{equation}
\boldsymbol{\varrho} = \left(\varrho_1,...,\varrho_{N_s} \right)^T.
\end{equation}
Because the stress generally depends on the signed normalized
gyroradius parameters of all species, this dependence may be made
explicit when useful by writing
\begin{equation}
\sP = \sP(\sV,\sA;\boldsymbol{\varrho}).
\end{equation}

Finally, the dimensionless first integrals are given by
\begin{align}
\sX_s &= \frac{X_s}{L_0} = \sA + \varrho_s u_{s,y} \label{eq:dl_gcenter} \\
\sW_{s,\perp} &= \frac{W_{s,\perp}}{T_0} = \frac{1}{2} \left( u_{s,x}^2 + u_{s,y}^2 \right) + \frac{q_s}{e} \sV \label{eq:dl_perp_energy} \\
\sW_{s,\parallel} &= \frac{W_{s,\parallel}}{T_0} = \frac{1}{2} u_{s,z}^2. \label{eq:dl_par_energy}
\end{align}

Both dimensional and dimensionless forms are used in what follows, as a matter of style: dimensionless
variables for numerical and asymptotic results, where they keep track
of fewer parameters and make relative sizes more apparent, and
dimensional variables for physics results, where they are more
familiar.

\subsection{Boundary Conditions} \label{sec:bcs}
To fully determine the potentials $\sV$ and $\sA$, additional conditions must be specified, for example, initial or boundary values. Here, the system is posed as a boundary-value problem, since this is the more natural formulation for equilibria intended for use in simulations. An open interval $\xi \in \left( \xi_L, \xi_R \right)$ is assumed over which the ODE system is satisfied, with values of $\sV$ and $\sA$ specified at the endpoints. For $\sA$, the boundary conditions are taken to be
\begin{equation}
\sA(\xi_L) = \xi_L,
\qquad
\sA(\xi_R) = \xi_R,
\end{equation}
which recover the solution $\sA = \xi$ in the limit $\beta \rightarrow 0$. In dimensional variables, this corresponds to a uniform externally applied magnetic field of strength $B_0$ in the $\hat{z}$ direction. For $\sV$, homogeneous Dirichlet boundary conditions are considered,
\begin{equation}
\sV(\xi_L) = \sV(\xi_R) = 0.
\end{equation}

%% file: v2/section_04.tex
\section{Numerical Approach \label{sec:numerical_approach}}

Here, the numerical approach is presented, beginning with the spatial discretization for a selected velocity-space integration method. An example of a particular choice of integration method is given for the problem considered in the following section. Since the ODE system is posed as a boundary-value problem, the spatial discretization results in a nonlinear algebraic system for the nodal values of the potentials. The discretization is constructed so that a discrete Hamiltonian is exactly conserved when this nonlinear system is satisfied. An algebraic solver, such as the quasi-Newton method presented in the following section, may be used to obtain the solution. Details of the solver design in general will depend on the particular form and numerical evaluation of $\sP$.

\subsection{Spatial Discretization}

Consider a uniform grid on the interval $\left[\xi_L,\xi_R\right]$, with nodes given by
\begin{equation}
\xi_j = \xi_L + j\Delta\xi,
\quad
j=0,\ldots,N,
\quad
\Delta\xi = \frac{\xi_R-\xi_L}{N}.
\label{eq:spatial_grid}
\end{equation}
The nodal approximations to the potentials are denoted by $\sV_j$ and $\sA_j$. Their boundary values are prescribed according to Section~\ref{sec:bcs}, leaving the values at $j=1,\ldots,N-1$ as unknowns.

For a quantity $\sQ$ defined at the grid nodes, define its cell-centered average by
\begin{equation}
\sQ_{j+1/2} = \frac{\sQ_j+\sQ_{j+1}}{2}.
\label{eq:cell_centered_average}
\end{equation}
The centered difference operators at cell centers and nodes are then defined by
\begin{equation}
D_{1,j+\frac{1}{2}} \sQ = \frac{\sQ_{j+1} - \sQ_{j}}{\Delta \xi},
\label{eq:cell_center_difference_operator}
\end{equation}
and
\begin{equation}
D_{1,j} \sQ = \frac{\sQ_{j+\frac{1}{2}} - \sQ_{j-\frac{1}{2}}}{\Delta \xi},
\label{eq:difference_operators}
\end{equation}
respectively. The centered second-difference operator at node $j$ is defined by
\begin{equation}
D_{2,j} \sQ = \frac{\sQ_{j+1} - 2 \sQ_j + \sQ_{j-1}}{\Delta \xi^2}.
\end{equation}
The cell-centered potential values used below are therefore $\sV_{j+\frac{1}{2}}$ and $\sA_{j+\frac{1}{2}}$, obtained from the preceding averaging operation.

Let $\widehat{\sP}$ denote a numerical approximation to the dimensionless stress appearing in \refeqs{eq:dl_phi_ode}{eq:dl_A_ode}, obtained using a selected velocity-space integration method, e.g., Gaussian quadrature or Monte Carlo-based integration. The numerical stress is evaluated only using cell-centered potential values. To compactly represent averages of these evaluations, let $\widehat{\sP}_{(\alpha,\beta)}$ denote the numerical stress associated with the $\sV$ and $\sA$ locations specified by the first and second indices, respectively. A half-integer index denotes direct evaluation at the corresponding cell-centered potential, whereas an integer index denotes a symmetric average over the two adjacent cell-centered evaluations. Thus,
\begin{equation}
\widehat{\sP}_{(j+\frac{1}{2},j+\frac{1}{2})}
=
\widehat{\sP}
\left(
\sV_{j+\frac{1}{2}},
\sA_{j+\frac{1}{2}}
\right),
\label{eq:cell_centered_stress}
\end{equation}
while, for example,
\begin{equation}
\widehat{\sP}_{(j+\frac{1}{2},j)}
=
\frac{
\widehat{\sP}_{(j+\frac{1}{2},j+\frac{1}{2})}
+
\widehat{\sP}_{(j+\frac{1}{2},j-\frac{1}{2})}
}{2}.
\label{eq:averaged_stress}
\end{equation}

The discrete partial-derivative operators are then defined by
\begin{align}
\overline{\partial}_{\sV,j}\widehat{\sP}
&=
\frac{\widehat{\sP}_{(j+\frac{1}{2},j)}-\widehat{\sP}_{(j-\frac{1}{2},j)}}{\Delta \xi D_{1,j} \sV},
\label{eq:discrete_partial_V}
\\
\overline{\partial}_{\sA,j}\widehat{\sP}
&=
\frac{\widehat{\sP}_{(j,j+\frac{1}{2})}-\widehat{\sP}_{(j,j-\frac{1}{2})}}{\Delta \xi D_{1,j} \sA},
\label{eq:discrete_partial_A}
\end{align}
and the equilibrium equations are discretized at each interior node as
\begin{align}
\lambda^2 D_{2,j} \sV &= \overline{\partial}_{\sV,j}\widehat{\sP},
\label{eq:discrete_V_equilibrium}
\\
-\frac{2}{\beta} D_{2,j} \sA &= \overline{\partial}_{\sA,j}\widehat{\sP}.
\label{eq:discrete_A_equilibrium}
\end{align}
Thus, the field terms use the standard centered second-order approximation, while the stress terms use discrete partial derivatives constructed to preserve a discrete Hamiltonian.

\subsection{Discrete Hamiltonian Conservation}

The discrete partial derivatives \refeqs{eq:discrete_partial_V}{eq:discrete_partial_A} satisfy an exact discrete chain rule
\begin{equation}
D_{1,j} \widehat{\sP} = D_{1,j} \sV \ 
\overline{\partial}_{\sV,j}\widehat{\sP}
+ D_{1,j} \sA \ 
\overline{\partial}_{\sA,j}\widehat{\sP},
\label{eq:discrete_chain_rule}
\end{equation}
where
\begin{equation}
D_{1,j} \widehat{\sP} = \frac{\widehat{\sP}_{(j+
\frac{1}{2},j+
\frac{1}{2})}-\widehat{\sP}_{(j-
\frac{1}{2},j-
\frac{1}{2})}}{\Delta \xi}.
\end{equation}
This identity follows directly from \refeqs{eq:discrete_partial_V}{eq:discrete_partial_A}, together with the $\widehat{\sP}_{(\alpha,\beta)}$ definition for an integer index.

Define the cell-centered discrete Hamiltonian by
\begin{multline}
\sH_{j+1/2} = \frac{1}{2}\lambda^2 \left(D_{1,j+\frac{1}{2}} \sV \right)^2 - \frac{1}{\beta} \left(D_{1,j+\frac{1}{2}} \sA \right)^2 \\ - \widehat{\sP}_{(j+\frac{1}{2},j+\frac{1}{2})},
\label{eq:discrete_hamiltonian}
\end{multline}
and consider the difference between adjacent cell centers $\sH_{j+\frac{1}{2}}-\sH_{j-\frac{1}{2}}$. For any nodally defined quantity $\sQ$, the difference operators satisfy the identity
\begin{multline}
\frac{1}{2} \frac{\left(D_{1,j+
\frac{1}{2}} \sQ \right)^2 - \left(D_{1,j-
\frac{1}{2}} \sQ \right)^2}{\Delta \xi} \\ = \left(D_{1,j} \sQ\right) \left( D_{2,j} \sQ\right).
\end{multline}
This follows directly by factoring the difference of the squares and using the definitions of $D_{1,j}$ and $D_{2,j}$. Applying this identity to the two field terms, together with the discrete chain rules, gives
\begin{equation}
\sH_{j+1/2} - \sH_{j-1/2}
= \Delta \xi \left( D_{1,j} \sV \ R_j^{\sV} + D_{1,j} \sA \ R_j^{\sA} \right),
\label{eq:hamiltonian_residual_identity}
\end{equation}
where the nonlinear residuals at node $j$ are defined by
\begin{align}
R_j^{\sV}
&=
\lambda^2 D_{2,j} \sV - \overline{\partial}_{\sV,j}\widehat{\sP},
\label{eq:residual_V}
\\
R_j^{\sA}
&=
-\frac{2}{\beta} D_{2,j} \sA - \overline{\partial}_{\sA,j}\widehat{\sP}.
\label{eq:residual_A}
\end{align}
Consequently, any solution of \refeqs{eq:discrete_V_equilibrium}{eq:discrete_A_equilibrium} corresponds to vanishing residuals, resulting in a discrete Hamiltonian that is exactly constant over all cell centers.

\subsection{Numerical Integration of Stress and Nonlinear Solution}

The spatial discretization must be supplemented by a choice of velocity-space integration method for evaluating the stress. An appropriate choice will generally depend on the particular form of $F_s$ in \refeq{eq:general_F} and on the intended application. When the form of $F_s$ permits a lower-dimensional evaluation, that reduction should be exploited.

A sample-based evaluation may be useful, for example, in a particle-in-cell setting. In such an approach, a set of reference samples may be drawn once, for example uniformly on a unit cube, and held fixed throughout the nonlinear solve. The velocity-space samples and associated weights are then obtained from mappings that depend on the local potentials. Provided these mappings depend smoothly on the potentials, the resulting finite-sample approximation to the stress also varies smoothly with the nonlinear unknowns, rather than introducing new sampling noise between nonlinear iterations. An example of a related mapping used to load particles from a solved equilibrium is given later and could be adapted for such a sample-based stress evaluation.

Direct quadrature may instead be natural, particularly in a continuum implementation. In that setting, the stress may be evaluated using the same, or a closely related, velocity-space quadrature as the kinetic code. For the example considered in the following section, the velocity-space integration reduces to a one-dimensional integral that is evaluated using Gauss--Hermite quadrature.

Once a velocity-space integration procedure has been specified, the numerical stress and its derivatives define a closed nonlinear algebraic system for the interior values of the potentials. The solution of this system is considered next.

For a quantity $\sQ$ defined at the nodes, introduce the vector of interior nodal values
\begin{equation}
\mathbf{Q}
=
\begin{bmatrix}
\sQ_1 & \cdots & \sQ_{N-1}
\end{bmatrix}^{T},
\end{equation}
and let $\mathbf{D}_2$ denote the $(N-1)\times(N-1)$ tridiagonal matrix
\begin{equation}
\mathbf{D}_2
=
\frac{1}{\Delta\xi^2}
\begin{bmatrix}
-2 & 1 & & \\
1 & -2 & 1 & \\
& \ddots & \ddots & \ddots \\
& & 1 & -2
\end{bmatrix}.
\end{equation}
The contribution of the prescribed boundary values to the second difference is
\begin{equation}
\mathbf{b}^\sQ
=
\frac{1}{\Delta\xi^2}
\left(
\sQ_0\mathbf{e}_1
+
\sQ_N\mathbf{e}_{N-1}
\right),
\end{equation}
where $\mathbf{e}_j$ denotes the $j$th coordinate vector in $\mathbb{R}^{N-1}$. The second differences at the interior nodes may therefore be assembled as
\begin{equation}
\begin{bmatrix}
D_{2,1}\sQ\\
\vdots\\
D_{2,N-1}\sQ
\end{bmatrix}
=
\mathbf{D}_2\mathbf{Q}+\mathbf{b}^\sQ.
\end{equation}
Similarly, define the vectors of numerical stress derivatives by
\begin{equation}
\overline{\boldsymbol{\partial}}_{\sQ}\widehat{\sP}
=
\begin{bmatrix}
\overline{\partial}_{\sQ,1}\widehat{\sP}\\
\vdots\\
\overline{\partial}_{\sQ,N-1}\widehat{\sP}
\end{bmatrix}.
\end{equation}
The nonlinear residuals can then be written as
\begin{equation}
\mathbf{R}^{\sV}
=
\lambda^2
\left[
\mathbf{D}_2\mathbf{V}+\mathbf{b}^\sV
\right]
-
\overline{\boldsymbol{\partial}}_{\sV}\widehat{\sP},
\end{equation}
and
\begin{equation}
\mathbf{R}^{\sA}
=
-\frac{2}{\beta}
\left[
\mathbf{D}_2\mathbf{A}+\mathbf{b}^\sA
\right]
-
\overline{\boldsymbol{\partial}}_{\sA}\widehat{\sP}.
\end{equation}
The discrete solution satisfies
\begin{equation}
\mathbf{R}^{\sV}=\mathbf{0},
\qquad
\mathbf{R}^{\sA}=\mathbf{0}.
\end{equation}
A Newton linearization of this system gives
\begin{equation}
\begin{bmatrix}
\lambda^2\mathbf{D}_2-\mathbf{K}_{\sV\sV}
&
-\mathbf{K}_{\sV\sA}
\\
-\mathbf{K}_{\sA\sV}
&
-\dfrac{2}{\beta}\mathbf{D}_2-\mathbf{K}_{\sA\sA}
\end{bmatrix}
\begin{bmatrix}
\delta\mathbf{V}\\
\delta\mathbf{A}
\end{bmatrix}
=
\begin{bmatrix}
-\mathbf{R}^{\sV}\\
-\mathbf{R}^{\sA}
\end{bmatrix},
\end{equation}
where the stress-dependent Jacobian matrices are
\begin{equation}
\mathbf{K}_{\sV\sV}
=
\frac{\partial}{\partial\mathbf{V}}
\left(
\overline{\boldsymbol{\partial}}_{\sV}\widehat{\sP}
\right),
\qquad
\mathbf{K}_{\sV\sA}
=
\frac{\partial}{\partial\mathbf{A}}
\left(
\overline{\boldsymbol{\partial}}_{\sV}\widehat{\sP}
\right),
\end{equation}
and
\begin{equation}
\mathbf{K}_{\sA\sV}
=
\frac{\partial}{\partial\mathbf{V}}
\left(
\overline{\boldsymbol{\partial}}_{\sA}\widehat{\sP}
\right),
\qquad
\mathbf{K}_{\sA\sA}
=
\frac{\partial}{\partial\mathbf{A}}
\left(
\overline{\boldsymbol{\partial}}_{\sA}\widehat{\sP}
\right).
\end{equation}
Each nodal residual depends only on the potential values at that node and its two nearest neighbors, so each matrix $\mathbf{K}_{\alpha\beta}$ is tridiagonal. At the first and last interior nodes, the prescribed boundary values entering the stress evaluations are held fixed when forming these derivatives. The corresponding first and last rows therefore contain derivatives only with respect to interior unknowns. Thus, all four blocks of the full Newton Jacobian are tridiagonal. The linearized system is solved for the corrections $\delta \vV$ and $\delta \vA$ to the current iterate, which are then used to update the interior nodal values of the potentials.

For the problem considered in the following section, a simpler quasi-Newton approximation was found to be sufficient. The matrices $\mathbf{K}_{\sV\sA}$ and $\mathbf{K}_{\sA\sV}$ are neglected, so that each quasi-Newton iteration requires only two independent tridiagonal linear solves; $\mathbf{K}_{\sA\sA}$ is neglected as well. In place of $\mathbf{K}_{\sV\sV}$, an analytic approximation is used: the continuum curvature
\begin{equation}
K_j \equiv \left.\frac{\partial^2\widehat{\sP}}{\partial\sV^2}\right|_{\sV_j,\sA_j},
\end{equation}
evaluated directly at the nodal potentials, is spread onto row $j$ with weights $\tfrac14,\tfrac12,\tfrac14$ on the sub-, main-, and super-diagonal. This weighting is not arbitrary: $\overline{\partial}_{\sV,j}\widehat{\sP}$ is itself built from the cell-centered values $\sV_{j\pm1/2}=\tfrac12(\sV_j+\sV_{j\pm1})$. For a smooth, well-resolved solution it therefore behaves as the average of $\partial\widehat{\sP}/\partial\sV$ at those two neighboring cell centers, with corrections appearing only at second order in the grid spacing $\Delta\xi$. Differentiating through this cell-centered averaging, and evaluating the curvature once per node rather than separately at each cell center, yields exactly this tridiagonal row. This analytic curvature needs only one velocity-space quadrature evaluation per node. The exact $\mathbf{K}_{\sV\sV}$, by contrast, would require differentiating $\overline{\partial}_{\sV,j}\widehat{\sP}$ with respect to each of the three stencil values separately, each needing its own quadrature evaluation; the approximation used here trades exactness in the Jacobian for efficiency, and as with any quasi-Newton method, the resulting inexactness affects only the convergence rate, not the correctness of a converged solution. The resulting approximate Jacobian is
\begin{equation}
\widetilde{\mathbf{J}}
=
\begin{bmatrix}
\lambda^2\mathbf{D}_2-\widetilde{\mathbf{K}}_{\sV\sV}
&
\mathbf{0}
\\[2mm]
\mathbf{0}
&
-\dfrac{2}{\beta}\mathbf{D}_2
\end{bmatrix},
\end{equation}
with $\widetilde{\mathbf{K}}_{\sV\sV}$ the tridiagonal matrix just described. This approximation is used here for efficiency in both the construction and inversion of the Jacobian, and was sufficient for the numerical example considered below. Additional Jacobian terms, including the exact $\mathbf{K}_{\sV\sV}$ and the off-diagonal blocks, can be retained when needed to improve the convergence or robustness of the nonlinear solve, or in problems where $\sV$ and $\sA$ are strongly coupled.

%% file: v2/section_05.tex
\section{A Canonical Bi-Maxwellian Equilibrium Family}
\label{sec:pseudo_bimaxwellian_family}

In this section, the construction is made concrete by introducing a canonical bi-Maxwellian
distribution. This is a natural starting point: Maxwellians and
bi-Maxwellians are standard distributions considered in plasma
physics. Here, the usual density and temperature parameters are
replaced by profiles of the first integral $X_s$, which reduces to the
spatial coordinate $x$ in the well-magnetized limit. These profiles can
therefore impose nonuniform structure while still satisfying the time-independent
Vlasov equation. For this choice, all but one velocity integration in the
species stress can be performed analytically, leaving a one-dimensional
quadrature. The same distribution function also lends itself to a
simplified particle-loading procedure for particle-in-cell (PIC) simulations. This
section is intended to show how this particular family of distribution
functions can be implemented end to end, and to provide a template for
other choices.

\subsection{Canonical Bi-Maxwellian Distribution}

Consider a distribution in first-integral space of the factored form
\begin{multline}
F_s(X_s,W_{\perp s},W_{\parallel s}) \\
=
N_s(X_s)
F_{\perp s}(X_s,W_{\perp s})
F_{\parallel s}(X_s,W_{\parallel s}),
\label{eq:pseudo_bimaxwellian}
\end{multline}
where the perpendicular and parallel factors are
\begin{equation}
F_{\perp s}(X_s,W_{\perp s})
=
\frac{m_s}{2\pi T_{\perp s}(X_s)}
\exp\left[
-\frac{W_{\perp s}}{T_{\perp s}(X_s)}
\right]
\label{eq:perpendicular_factor}
\end{equation}
and
\begin{equation}
F_{\parallel s}(X_s,W_{\parallel s})
=
\sqrt{
\frac{m_s}{2\pi T_{\parallel s}(X_s)}
}
\exp\left[
-\frac{W_{\parallel s}}{T_{\parallel s}(X_s)}
\right].
\label{eq:parallel_factor}
\end{equation}
$F_s$ uses the first integrals as its arguments. The corresponding
phase-space distribution $f_s(x,\vv)$ is obtained by evaluating those
first integrals through the potentials $\phi(x)$ and $A_y(x)$ and the
velocity coordinates $\vv$, as in \refeq{eq:general_F}; see
\refeqs{eq:gcenter}{eq:par_energy}.

The functions $N_s$, $T_{\perp s}$, and $T_{\parallel s}$ are
referred to as the canonical (prescribed) density and canonical
(prescribed) temperatures. Although they enter $F_s$ in the same
manner as the density and temperatures of a bi-Maxwellian, they are
functions of the first integral $X_s$ and are not, in general, equal
to the corresponding velocity moments of $f_s$ at fixed $x$, referred
to below as the true (resulting) density and temperature.

The parallel factor is normalized for every $X_s$:
\begin{equation}
\int_{-\infty}^{\infty}
F_{\parallel s}
\left(
X_s,W_{\parallel s}(v_z)
\right)
\,dv_z
=
1.
\label{eq:parallel_factor_normalization}
\end{equation}
Consequently, $T_{\parallel s}(X_s)$ does not enter the velocity
moments used to determine the equilibrium. No particular parallel
canonical temperature profile is therefore required at this stage, and
$T_{\parallel s}(X_s)$ may be chosen independently when constructing
the full distribution function.

It is useful to record the change of variables at the level of the
velocity-space measure:
\begin{equation}
\frac{1}{N_0}
f_s(L_0\xi,\vv)
\dd^3v
=
\frac{v_{0,s}^3}{N_0}
f_s(L_0\xi,v_{0,s}\vu_s)
\dd^3u_s.
\label{eq:dimensionless_phase_space_measure}
\end{equation}
Since $m_s v_{0,s}^2=T_0$, the species contribution to the
dimensionless stress is the dimensionless second moment
\begin{equation}
\sP_s(\sV,\sA)
=
\int_{\mathbb{R}^3}
u_{s,x}^2
\frac{v_{0,s}^3}{N_0}
f_s(L_0\xi,v_{0,s}\vu_s)
\dd^3u_s.
\label{eq:dimensionless_species_stress_moment}
\end{equation}
Substituting the factored distribution into
\refeq{eq:dimensionless_species_stress_moment}, using the
dimensionless first integrals \refeqs{eq:dl_gcenter}{eq:dl_perp_energy},
and applying the normalization \refeq{eq:parallel_factor_normalization},
the $u_{s,z}$ dependence is marginalized out. For fixed $u_{s,y}$, the
remaining $u_{s,x}$ dependence is Gaussian and can also be integrated
analytically. Writing the remaining velocity variable as $u=u_{s,y}$
yields
\begin{multline}
\sP_s(\sV,\sA)
= \int_{-\infty}^{\infty}
\frac{
\sN_s\bigl(\sA+\varrho_s u\bigr)
}
{\sqrt{2\pi}}
\left[
\sT_s\bigl(\sA+\varrho_s u\bigr)
\right]^{1/2}
\\
\times
\exp\left[
-\frac{
u^2/2+(q_s/e)\sV
}
{\sT_s\bigl(\sA+\varrho_s u\bigr)}
\right]
\dd u,
\label{eq:reduced_species_stress}
\end{multline}
where
\begin{equation}
\sN_s(\sX_s)
=
\frac{N_s(L_0 \sX_s)}{N_0},
\quad
\sT_s(\sX_s)
=
\frac{T_{\perp s}(L_0 \sX_s)}{T_0}.
\label{eq:dimensionless_pseudoprofile_definitions}
\end{equation}
The total dimensionless stress is obtained by summing this quantity
over species.

\subsection{Gaussian Quadrature}
\label{sec:gaussian_quadrature}

The one-dimensional integral in \refeq{eq:reduced_species_stress} can
be evaluated using Gauss--Hermite quadrature by exposing a Gaussian
weight in the remaining velocity coordinate. For a fixed evaluation
point $(\sV,\sA)$ and species $s$, the value $\sT_s(\sA)$ is used as
the reference temperature, distinct from the profile values evaluated
at the shifted first-integral coordinate. Set
\begin{equation}
u=\sqrt{2\sT_s(\sA)}\,z,
\qquad
\sX_s(z)
=
\sA+\varrho_s\sqrt{2\sT_s(\sA)}\,z.
\label{eq:gh_change_of_variables}
\end{equation}
The second relation is simply the first integral
\refeq{eq:dl_gcenter} after this change of variables. Define the
quadrature kernel
\begin{multline}
\mathcal{K}_s(z) =
\sqrt{\frac{\sT_s(\sA) \sT_s\bigl(\sX_s(z)\bigr)}{\pi}}
\sN_s\bigl(\sX_s(z)\bigr) \\
\times
\exp\left[
\frac{-(q_s/e)\sV+\left(\sT_s(\sX_s(z))-\sT_s(\sA)\right)z^2}
{\sT_s\bigl(\sX_s(z)\bigr)}
\right].
\label{eq:gh_quadrature_kernel}
\end{multline}
The species stress becomes
\begin{equation}
\sP_s
=
\int_{-\infty}^{\infty}
\mathcal{K}_s(z)
e^{-z^2}
\dd z.
\label{eq:gh_stress_integral}
\end{equation}
If $z_i$ and $w_i$ denote the nodes and weights of an $n$-point
Gauss--Hermite rule for the weight $e^{-z^2}$, then the corresponding
numerical species stress is simply
\begin{equation}
\widehat{\sP}_s
=
\sum_{i=1}^n w_i
\mathcal{K}_s(z_i).
\label{eq:gh_stress_quadrature}
\end{equation}
The total stress is obtained by summing the species contributions.

\subsection{Particle Loading}
\label{sec:particle_loading}

For particle-in-cell (PIC) applications, one must generate particle
samples from the equilibrium distribution.
For given potentials $\sV(\xi)$ and $\sA(\xi)$, whether from a
nonlinear solver iterate or from a completed equilibrium, the
canonical bi-Maxwellian reduces this task to sampling a nonseparable joint PDF in two variables.
Starting from the full phase-space density in
\refeq{eq:dimensionless_phase_space_measure}, the Gaussian factors in
$u_{s,x}$ and $u_{s,z}$ can be marginalized analytically. The result is
an unnormalized marginal density in position $\xi$ and velocity $u_{s,y}$, which are
the variables that remain coupled in the loading procedure. Writing
$u=u_{s,y}$ and
\begin{equation}
\sX_s(\xi,u)
=
\sA(\xi)+\varrho_s u,
\label{eq:loading_first_integral}
\end{equation}
the unnormalized joint density to be sampled from is
\begin{multline}
g_s(\xi,u)
=
\frac{
\sN_s\bigl(\sX_s(\xi,u)\bigr)
}
{
\sqrt{2\pi\sT_s\bigl(\sX_s(\xi,u)\bigr)}
}
\\
\times
\exp\left[
-\frac{
u^2/2+(q_s/e)\sV(\xi)
}
{\sT_s\bigl(\sX_s(\xi,u)\bigr)}
\right].
\label{eq:loading_joint_density}
\end{multline}
The inverse Rosenblatt transformation \cite{Rosenblatt1952} for the full distribution is
then reduced to a grid-based sampling step for the coupled pair
$(\xi,u)$, followed by analytic conditional Gaussian draws for the two
remaining velocity components. A convenient recipe is based on a numerical
inverse Rosenblatt transformation of a tabulated version of the joint density,
requiring two uniformly drawn values $(r_\xi,r_u)\in[0,1]^2$ to produce a
sample $(\xi_p,u_p)$ from the joint distribution. These uniform draws
may be, for example, pseudorandom or from a low-discrepancy sequence. The procedure is as follows:

\begin{enumerate}
\item \emph{Tabulate the coupled density.}
Use the $N+1$ spatial nodes $\xi_i$, $i=0,\ldots,N$, defined in
\refeq{eq:spatial_grid}. For the remaining velocity coordinate, choose
a large symmetric interval $[-u_c,u_c]$, with a cutoff velocity $u_c$ large enough that
the truncated tails are negligible, and set
\begin{equation}
u_j=-u_c+j\Delta u,
\quad
\Delta u=\frac{2u_c}{M},
\quad
0\leq j\leq M.
\end{equation}
This truncates the velocity support and allows the joint density to be
sampled on grid nodes. Tabulate this density as $G_{ij}=g_s(\xi_i,u_j)$.

\item \emph{Build the marginal distribution in $\xi$.}
The marginal PDF and CDF are constructed from linear interpolation and
trapezoidal quadrature.
Linear interpolation is positivity preserving for nonnegative
nodal values, and its analytic integrals agree with trapezoidal
quadrature on grid intervals. The tabulated marginal and normalization factor
are
\begin{align}
m_i^{\mathrm{tab}}
&=
\sum_{j=0}^{M-1}
\frac{\Delta u}{2}
\left(G_{ij}+G_{i,j+1}\right),
\notag\\
\mathcal{Z}^{\mathrm{tab}}
&=
\sum_{i=0}^{N-1}
\frac{\Delta\xi}{2}
\left(m_i^{\mathrm{tab}}+m_{i+1}^{\mathrm{tab}}\right).
\end{align}
The tabulated marginal CDF values are
\begin{align}
F_{\xi,0}^{\mathrm{tab}}
&=0,
\notag\\
F_{\xi,k}^{\mathrm{tab}}
&=
\frac{1}{\mathcal{Z}^{\mathrm{tab}}}
\sum_{i=0}^{k-1}
\frac{\Delta\xi}{2}
\left(m_i^{\mathrm{tab}}+m_{i+1}^{\mathrm{tab}}\right).
\end{align}

\item \emph{Invert the marginal CDF to load position.}
Given a uniformly drawn
$r_\xi\in[0,1]$, locate the grid interval $[\xi_k,\xi_{k+1}]$ whose
CDF values bracket $r_\xi$. On this interval, the linearly
interpolated marginal density is determined by
$m_k^{\mathrm{tab}}$ and $m_{k+1}^{\mathrm{tab}}$. The loaded
position is obtained by inverting the local analytic form of the CDF
within the bracket, found by integrating the linear density interpolant from
$\xi_k$ to an arbitrary point $\xi$ within the interval. Thus $\xi_p\in[\xi_k,\xi_{k+1}]$ is the root of
\begin{multline}
\frac{m_{k+1}^{\mathrm{tab}}-m_k^{\mathrm{tab}}}
{2\Delta\xi}
(\xi_p-\xi_k)^2 \\
+
m_k^{\mathrm{tab}}(\xi_p-\xi_k)
-
\mathcal{Z}^{\mathrm{tab}}
\left(r_\xi-F_{\xi,k}^{\mathrm{tab}}\right)=0.
\label{eq:piecewise_linear_inverse_cdf}
\end{multline}
Because the interpolated marginal CDF is continuous and monotone, the
bracketing step guarantees a root in $[\xi_k,\xi_{k+1}]$ whenever the
selected interval has nonzero probability mass. This root is obtained
by solving the resulting quadratic equation on the selected grid
interval.

\item \emph{Invert the conditional CDF to load velocity.}
At each spatial node, form the tabulated conditional CDF values
\begin{align}
F_{(u|\xi_i),0}^{\mathrm{tab}}
&=0,
\notag\\
F_{(u|\xi_i),l}^{\mathrm{tab}}
&=
\frac{1}{m_i^{\mathrm{tab}}}
\sum_{j=0}^{l-1}
\frac{\Delta u}{2}
\left(G_{ij}+G_{i,j+1}\right).
\end{align}
Interpolate these conditional CDF values between the two grid columns
bracketing $\xi_p$. Given a uniformly drawn $r_u\in[0,1]$, locate the
velocity interval $[u_l,u_{l+1}]$ whose conditional CDF values bracket
$r_u$, and apply the same quadratic inversion with $\xi$ replaced by
$u$ and the endpoint density values replaced by the corresponding
conditional density values. This gives $u_p$.
\end{enumerate}

Once $(\xi_p,u_p)$ has been sampled, set
$\sX_{s,p}=\sX_s(\xi_p,u_p)$. The remaining velocities are sampled as
\begin{equation}
\begin{aligned}
u_{s,x,p}
&=
\sqrt{\sT_s(\sX_{s,p})}\,r_{u_x},
\\
u_{s,z,p}
&=
\sqrt{\sT_{\parallel s}(\sX_{s,p})}\,r_{u_z},
\end{aligned}
\end{equation}
where
$r_{u_x}$ and $r_{u_z}$ are independent standard normal variates and
$\sT_{\parallel s}=T_{\parallel s}/T_0$ is defined analogously to the
perpendicular temperature.

This construction also provides a PIC-style alternative for computing
$\widehat{\sP}$, which may be desirable when the equilibrium solver is to be used within a PIC method. After mapping the uniformly drawn values
through the current potential-dependent distribution, the resulting
particle samples can be gathered to the grid and used to compute the
second moment in the usual PIC manner. If this particle-based estimate
is used inside a Newton iteration, the same set of uniformly drawn
values $(r_\xi,r_u)$ should be held fixed so that the iteration remains
deterministic. As the iterates for $\sV$ and $\sA$ change, the table
$G_{ij}$, the tabulated CDF values, and the Rosenblatt map change;
consequently, the mapped locations $(\xi_p,u_p)$ also change. Since the
particle-based stress must be differentiated inside a Newton solve for both the residual and Jacobian construction, an
additional numerical consideration is smoothness. Because the inverse map is
assembled cell by cell from piecewise polynomial CDFs, derivatives with
respect to the potentials may be nonsmooth when a sample crosses a
grid point or when the bracketing cell changes. Smoother positivity-preserving numerical constructions could be
considered to address this.

\subsection{Prescribed Profiles}

The equilibrium is determined by the dimensionless canonical density
$\sN_s$ and canonical temperature $\sT_s$. In the numerical examples of
the next section, both
are prescribed using the same tanh profile form connecting two
asymptotic values $g^-$ and $g^+$, corresponding to the limits as
$\zeta\to-\infty$ and $\zeta\to+\infty$, respectively, with the
transition centered at $\zeta=0$,
\begin{equation}
\mathsf{S}(\zeta;g^-,g^+)
=
g^-\frac{1-\tanh(\zeta)}{2}
+
g^+\frac{1+\tanh(\zeta)}{2}.
\label{eq:profile_template}
\end{equation}
Then
\begin{align}
\sN_s(\sX_s)
&=
\mathsf{S}\left(
\frac{L_0}{w_{sN}} \sX_s;
\frac{N_s^-}{N_0},
\frac{N_s^+}{N_0}
\right),
\notag\\
\sT_s(\sX_s)
&=
\mathsf{S}\left(
\frac{L_0}{w_{sT}} \sX_s;
\frac{T_s^-}{T_0},
\frac{T_s^+}{T_0}
\right).
\label{eq:dimensionless_pseudoprofiles}
\end{align}
The widths $w_{sN}$ and $w_{sT}$ control the sharpness of the
corresponding transitions.

For an equilibrium intended to reproduce specified physical profiles,
the parameters of a chosen canonical profile model would be determined by
matching moments of the resulting distribution to the desired physical
density and temperature profiles. This is an inverse problem and could,
for example, be posed as a constrained optimization problem. A related
inverse problem has recently been addressed by other means, for
uniform-temperature Vlasov--Poisson equilibria in a fixed magnetic
field \cite{Vogman2019} and, in gyrokinetic simulation (recovering a
canonical, motion-invariant distribution close to a prescribed one,
without re-solving the field equations), by a
projection method \cite{Li2025}. For the
numerical examples in the next section, the canonical profile parameters
are prescribed directly.

\subsection{Nominal Parameters}

Table~\ref{tab:illustrative_parameters} gives the nominal dimensionless
parameter values used for the representative equilibrium and numerical
verification of Sec.~\ref{sec:numerical_verification}. The domain
boundaries are fixed at $\xi_L=-12$ and $\xi_R=12$ for all numerical
examples. Later sections
vary the reference magnetic-field strength
(Sec.~\ref{sec:asymptotic_comparison}) and, in some cases, the electron
and ion canonical density transition widths $w_{eN}$, $w_{iN}$, relative
to this nominal case; each such departure is defined explicitly where
it is used.

The electron and ion canonical density transition widths, $w_{eN}=1.0L_0$
and $w_{iN}=1.2L_0$ in Table~\ref{tab:illustrative_parameters}, control
the relative steepness of the two species' transitions and provide a
controlled way to break charge balance of the canonical densities;
either narrowing or widening $w_{eN}$ relative to $w_{iN}$ is used in
later sections. When the widths are equal, $w_{eN}=w_{iN}$, the
electron and ion canonical density profiles coincide in shape, giving
the charge-weighted cancellation
\begin{equation}
\sum_s \frac{q_s}{e}\sN_s(\sX)=0,
\label{eq:charge_neutral_pseudo_profiles}
\end{equation}
for a common profile argument $\sX$, using the limiting values
$N_e^\pm=N_i^\pm$ from Table~\ref{tab:illustrative_parameters}, which
are fixed independently of the transition widths. Away from equal widths, including the nominal case
where $w_{eN}\neq w_{iN}$, the canonical densities no longer cancel in
this sense. The consequences of this charge balance, or its absence,
for the self-consistent potential and for quasi-neutrality of the
physical densities are examined in Sec.~\ref{sec:equilibrium_efield}.

\begin{table}[htbp]
\centering
\setlength{\arrayrulewidth}{0.5pt}
\setlength{\tabcolsep}{6pt}
\renewcommand{\arraystretch}{1.2}
\begin{tabular}{|c|c|c|}
\hline
\multicolumn{3}{|c|}{Asymptotic plasma parameters} \\
\hline
Parameter & \multicolumn{2}{c|}{Value} \\
\hline
$\beta$
&
\multicolumn{2}{c|}{
$5.0\times10^{-3}$
}
\\
$\lambda$
&
\multicolumn{2}{c|}{
$2.5\times10^{-3}$
}
\\
$\|\boldsymbol{\varrho}\|_\infty$
&
\multicolumn{2}{c|}{
$1.5\times10^{-1}$
}
\\
\hline
\multicolumn{3}{|c|}{Species-specific parameters} \\
\hline
Parameter & $s=e$ & $s=i$ \\
\hline
$m_s/m_e$
&
$1.0$
&
$3.7\times10^3$
\\
$q_s/e$
&
$-1.0$
&
$1.0$
\\
$\ell_{\mathrm{gy},s}/L_0$
&
$-2.5\times10^{-3}$
&
$1.5\times10^{-1}$
\\
$N_s^-/N_0$
&
$1.0$
&
$1.0$
\\
$N_s^+/N_0$
&
$1.0\times10^{-1}$
&
$1.0\times10^{-1}$
\\
$w_{sN}/L_0$
&
$1.0$
&
$1.2$
\\
$T_s^-/T_0$
&
$1.0$
&
$1.0$
\\
$T_s^+/T_0$
&
$5.0\times10^{-2}$
&
$5.0\times10^{-2}$
\\
$w_{sT}/L_0$
&
$1.3$
&
$3.0$
\\
\hline
\end{tabular}\vspace{20pt}
\caption{
Nominal parameter values, used unless stated otherwise in the
numerical examples of Secs.~\ref{sec:numerical_verification}--%
\ref{sec:equilibrium_efield}.
}
\label{tab:illustrative_parameters}
\end{table}

%% file: v2/section_06.tex
\section{Numerical Verification}
\label{sec:numerical_verification}

\par The preceding section gives a complete specification of the
canonical bi-Maxwellian example used in the numerical tests. Here, that
example is used to verify the conservative discretization, examine
convergence with respect to the spatial grid and Gauss--Hermite
quadrature, and compare the computed solutions with the asymptotic
description. The emphasis in this section is on numerical verification
of the construction. The next section uses the same family of examples
to discuss the resulting equilibrium fields and moments in more detail.

\subsection{Test Configuration}

\par Unless stated otherwise, the computations use the nominal parameters
of Table~\ref{tab:illustrative_parameters}. The
nonlinear algebraic system \refeqs{eq:discrete_V_equilibrium}
{eq:discrete_A_equilibrium} is solved for the nodal values of
$\sV$ and $\sA$, and the stress is evaluated by the Gauss--Hermite rule
in \refeq{eq:gh_stress_quadrature}.

\par For reference, define the residual scales
\begin{align}
\mathrm{S}^{\sV} &= \lambda^2\|\mathbf{D}_2\mathbf{V}+\mathbf{b}^{\sV}\|_2
+ \sum_s\|\overline{\boldsymbol{\partial}}_{\sV}\widehat{\sP}_s\|_2, \\
\mathrm{S}^{\sA} &= (2/\beta)\|\mathbf{D}_2\mathbf{A}+\mathbf{b}^{\sA}\|_2
+ \sum_s\|\overline{\boldsymbol{\partial}}_{\sA}\widehat{\sP}_s\|_2.
\end{align}
The nonlinear iteration is terminated when
\refeq{eq:discrete_V_equilibrium} and \refeq{eq:discrete_A_equilibrium}
independently satisfy a mixed absolute-relative tolerance test,
\begin{equation}
\|\mathbf{R}^{\sQ}\|_2 \le \epsilon_{\mathrm{rel}}\,\mathrm{S}^{\sQ}+\epsilon_{\mathrm{abs}}
\quad\text{for } \sQ=\sV,\sA,
\label{eq:verification_residual_norm}
\end{equation}
with $\epsilon_{\mathrm{abs}}=10^{-14}$ fixed throughout. Except where
explicitly varied below when testing Hamiltonian conservation,
$\epsilon_{\mathrm{rel}}=10^{-10}$ is used throughout, small enough
that the observed differences in the convergence tests below are
dominated by discretization or quadrature rather than by the nonlinear
solve.

\subsection{Discrete Hamiltonian Conservation}

\par The discrete Hamiltonian conservation result in
\refeq{eq:hamiltonian_residual_identity} implies that the cell-centered
Hamiltonian \refeq{eq:discrete_hamiltonian} is exactly constant when the
algebraic residuals vanish. The numerical test therefore checks that the
observed variation in $\sH_{j+1/2}$ decreases with the nonlinear
residual, independently of the spatial resolution. The normalized
local variation is reported as
\begin{equation}
\left|
\frac{\sH_{j+1/2}-\langle \sH\rangle}{\langle \sH\rangle}
\right|,
\qquad
\langle \sH\rangle
=
\frac{1}{N}\sum_{j=0}^{N-1}\sH_{j+1/2}.
\label{eq:relative_hamiltonian_variation}
\end{equation}

\input{figures/data/hamiltonian_conservation/hamiltonian_conservation_vars.tex}

\reffig{fig:hamiltonian_conservation} shows this quantity on a
semilogarithmic scale for $N=\hamNa$ and $N=\hamNb$ cells
\refeq{eq:spatial_grid} ($\Delta\xi=5.859\times10^{-3}$ and
$3.750\times10^{-1}$, or $\Delta\xi/\lambda=2.344$ and
$1.500\times10^2$, respectively), while varying the
nonlinear residual tolerance. The Hamiltonian variation decreases as the
nonlinear system is solved more accurately, consistent with the residual
identity. Comparable
conservation is obtained at fixed solver tolerance on both grids,
confirming that the conservation property is not a high-resolution
artifact.

\begin{widefigure}
\centering
\scalebox{0.87}{\input{figures/compiled/hamiltonian_conservation_trim.tex}}
\par\vspace{7pt}
\input{figures/data/hamiltonian_conservation/legend_hamiltonian.tex}
\captionof{figure}{\label{fig:hamiltonian_conservation} Semilog plot of
$\left|\dfrac{\sH_{j+1/2}-\langle\sH\rangle}{\langle\sH\rangle}\right|$ for
the nominal parameters of Table~\ref{tab:illustrative_parameters},
shown for several nonlinear residual
tolerances at two resolutions, $N=\hamNa$ (a) and $N=\hamNb$ (b).}
\end{widefigure}

\subsection{Spatial Resolution and Quadrature Convergence}

\par Spatial convergence is tested by fixing a Gauss--Hermite rule with $N_g=32$
quadrature rule and refining the grid in $\xi$. For the refinement study below, $N$ ranges from 32 to 1024,
so $\Delta\xi$ decreases from $7.5\times10^{-1}$ to
$2.3438\times10^{-2}$. With $\lambda=2.5\times10^{-3}$, this
corresponds to $\Delta\xi/\lambda$ decreasing from $3.000\times10^2$ to
$9.375$. Since the exact solution is not known, a
Richardson-type comparison between successive refinements is used. Let $\sQ_N$
denote either computed potential, $\sV_N$ or $\sA_N$, on a grid with
$N$ cells. For nested grids, define the relative difference
\begin{equation}
\Delta_{\sQ,N}
=
\left[
\frac{
\int_{\xi_L}^{\xi_R}
\left|\sQ_N-\mathcal{I}_N\sQ_{2N}\right|^2\dd\xi}
{\int_{\xi_L}^{\xi_R}\left|\mathcal{I}_N\sQ_{2N}\right|^2\dd\xi}
\right]^{1/2},
\label{eq:spatial_successive_difference}
\end{equation}
where $\mathcal{I}_N\sQ_{2N}$ denotes the solution on the finer grid
evaluated on the coarser grid. In the computations below, the integrals
in \refeq{eq:spatial_successive_difference} are evaluated by trapezoidal
quadrature on the coarser grid. This quantity is not the exact error;
it is a computable measure of the change in the numerical solution when
the grid is refined.

\par If the method is in an asymptotic convergence regime, then
\begin{equation}
\sQ_N=\sQ+C_{\sQ}(\Delta\xi)^p+\mathrm{H.O.T.},
\end{equation}
for some order $p$. In that case, the successive differences satisfy
$\Delta_{\sQ,N}/\Delta_{\sQ,2N}\approx 2^p$, so the observed order is
estimated by
\begin{equation}
p_{\sQ,N}
=
\log_2\left(
\frac{\Delta_{\sQ,N}}{\Delta_{\sQ,2N}}
\right).
\label{eq:richardson_observed_order}
\end{equation}
The centered spatial discretization suggests second-order convergence
once the transition region in the prescribed profiles is resolved. At
coarser resolution, under-resolving the pressure gradient can produce
visible oscillations, particularly in $\sV$. This is a comparatively
benign resolution requirement: the grid must resolve the imposed profile
variation, but it need not resolve the Debye length in the quasi-neutral
regime. Tables~\ref{tab:spatial_richardson_v} and
\ref{tab:spatial_richardson_a} report the successive differences and
observed order for $\sV$ and $\sA$, respectively.

\begin{table}[htbp]
\centering
\setlength{\tabcolsep}{6pt}
\input{figures/data/spatial_convergence/richardson_v_table.tex}\vspace{20pt}
\caption{
Richardson-type spatial-convergence study for $\sV$ with $N_g=32$.
Each row compares the $N$-cell solution with the $2N$-cell solution
restricted to the coarser grid, and $p_{\sV,N}$ is the observed order
from \refeq{eq:richardson_observed_order}.}
\label{tab:spatial_richardson_v}
\end{table}

\begin{table}[htbp]
\centering
\setlength{\tabcolsep}{6pt}
\input{figures/data/spatial_convergence/richardson_a_table.tex}\vspace{20pt}
\caption{
Richardson-type spatial-convergence study for $\sA$ with $N_g=32$.
Each row compares the $N$-cell solution with the $2N$-cell solution
restricted to the coarser grid, and $p_{\sA,N}$ is the observed order
from \refeq{eq:richardson_observed_order}.}
\label{tab:spatial_richardson_a}
\end{table}

\par Quadrature convergence is tested separately by fixing a resolved spatial
grid and varying the number $N_g$ of Gauss--Hermite nodes in
\refeq{eq:gh_stress_quadrature}. The expected behavior is rapid
convergence when the quadrature kernel is well resolved by the
Gauss--Hermite rule. This is most favorable in the well-magnetized
regime, while weaker magnetization can require more quadrature points
because finite-Larmor-radius effects increase the velocity-space
structure of the integrand.

\par Unlike the spatial refinement study, there is no
sequence of nested grids to compare against.
Quadrature convergence is therefore assessed relative to a fixed
high-order reference, computed with $N_g=128$ nodes. Define
\begin{equation}
\Delta_{\sQ,N_g}
=
\left[
\frac{
\int_{\xi_L}^{\xi_R}\left|\sQ_{N_g}-\sQ_{128}\right|^2\dd\xi}
{\int_{\xi_L}^{\xi_R}\left|\sQ_{128}\right|^2\dd\xi}
\right]^{1/2},
\label{eq:quadrature_reference_difference}
\end{equation}
where $\sQ_{N_g}$ is either computed potential, $\sV$ or $\sA$, obtained
with $N_g$ Gauss--Hermite nodes, and $\sQ_{128}$ is the same quantity
obtained with $N_g=128$ nodes. Table~\ref{tab:gauss_hermite_convergence}
reports $\Delta_{\sQ,N_g}$ for a doubling sequence $N_g=1,2,4,8,16$,
together with $N_g=24$ and $N_g=32$. Both quantities drop over roughly
ten orders of magnitude across the doubling sequence, with each
doubling of $N_g$ gaining several additional digits, consistent with
the spectral convergence expected of Gauss--Hermite quadrature for a
smooth integrand. The quadrature order enters the equilibrium solve
only through $\widehat{\sP}_s$ at each vertex, so once the
Gauss--Hermite sum defining $\widehat{\sP}_s$ has converged, the solver
responds to an essentially unchanging input and reproduces the same
solution up to floating-point round-off in that quadrature sum.
Consistent with this, $\Delta_{\sV,N_g}$ and $\Delta_{\sA,N_g}$ settle
to within a factor of a few of $10^{-14}$ by $N_g=24$ and remain at
that level at $N_g=32$ rather than decreasing further, indicating that
the remaining difference from the reference reflects round-off in the
quadrature evaluation of $\widehat{\sP}_s$ rather than further
quadrature truncation error.

\begin{table}[htbp]
\centering
\setlength{\tabcolsep}{14pt}
\input{figures/data/quadrature_convergence/gauss_hermite_table.tex}\vspace{20pt}
\caption{
Gauss--Hermite quadrature-convergence study. $\Delta_{\sQ,N_g}$ is the
relative difference, defined in
\refeq{eq:quadrature_reference_difference}, between the $N_g$-node
solution and a reference solution computed with $N_g=128$ nodes.}
\label{tab:gauss_hermite_convergence}
\end{table}

\subsection{Representative Equilibrium and Particle Loading}

\par After establishing convergence, it is useful to characterize a
representative resolved solution, using the same nominal parameters
of Table~\ref{tab:illustrative_parameters}
as in Sec.~\ref{sec:numerical_verification} above, with $N=1024$ and
$N_g=32$. These diagnostics are not separate convergence tests; they show the
equilibrium state produced by the verified numerical construction,
focusing on the potentials, fields, densities, and species flows
examined further in Sec.~\ref{sec:equilibrium_efield}. All diagnostics
below are reported at the interior nodes $j$.
\reffig{fig:representative_potentials} shows the converged potentials
$\sV$ and $\sA$.

\input{figures/data/representative_equilibrium/representative_equilibrium_vars.tex}

\begin{widefigure}
\centering
\scalebox{0.87}{\input{figures/compiled/representative_potentials_trim.tex}}
\vspace{10pt}
\captionof{figure}{\label{fig:representative_potentials} Converged potentials
$\sV=\frac{e\phi}{T_0}$ and the vector-potential perturbation from its
background value, $\sA-\xi=\frac{\frac{A_y}{B_0}-x}{L_0}$, for the representative
equilibrium of Table~\ref{tab:illustrative_parameters},
$N=1024$, $N_g=32$.}
\end{widefigure}

\begin{samepage}
\par Normalized electric and magnetic fields follow from their
discrete gradients,
\begin{align}
\left( \frac{eL_0 E_x}{T_0} \right)_j &= -D_{1,j}\sV, \\
\left( \frac{B_z}{B_0} \right)_j &= D_{1,j}\sA,
\end{align}
and are shown in \reffig{fig:representative_fields}. The magnetic
field shows the expected pressure-balance behavior: $B_z$ is reduced
below $B_0$ on the high-pressure side and raised above $B_0$ on the
low-pressure side.
\end{samepage}

\begin{widefigure}
\centering
\scalebox{0.87}{\input{figures/compiled/representative_fields_trim.tex}}
\vspace{10pt}
\captionof{figure}{\label{fig:representative_fields} Electric field
$\frac{eE_xL_0}{T_0}=-\frac{\partial\sV}{\partial\xi}$ and magnetic field
$\frac{B_z}{B_0}=\frac{\partial\sA}{\partial\xi}$ for the representative equilibrium,
obtained from the discrete gradient of $\sV$ and $\sA$.}
\end{widefigure}

\par The remaining diagnostics follow from the discrete stress
$\widehat{\sP}_s$, via the species-level identities
\refeq{eq:species_phi_identity} and \refeq{eq:species_A_identity}
used in Sec.~\ref{sec:reduction_to_odes} to derive the Lagrangian
and Hamiltonian formulations. The number density for each species
follows from the discrete partial derivative of the stress with
respect to the potential:
\begin{equation}
\left( \frac{n_s}{N_0} \right)_j = -\frac{e}{q_s}\,\overline{\partial}_{\sV,j}\widehat{\sP}_s.
\end{equation}
In \reffig{fig:representative_density}, the prescribed
canonical densities $N_s$ are compared with the true (resulting), self-consistent densities $n_s$.
The canonical densities do not match by construction, but the true
densities do, as expected in the quasi-neutral limit $\lambda^2\ll1$.

\begin{widefigure}
\centering
\scalebox{0.87}{\input{figures/compiled/representative_density_trim.tex}}
\par\vspace{7pt}
\input{figures/data/representative_equilibrium/legend_density.tex}

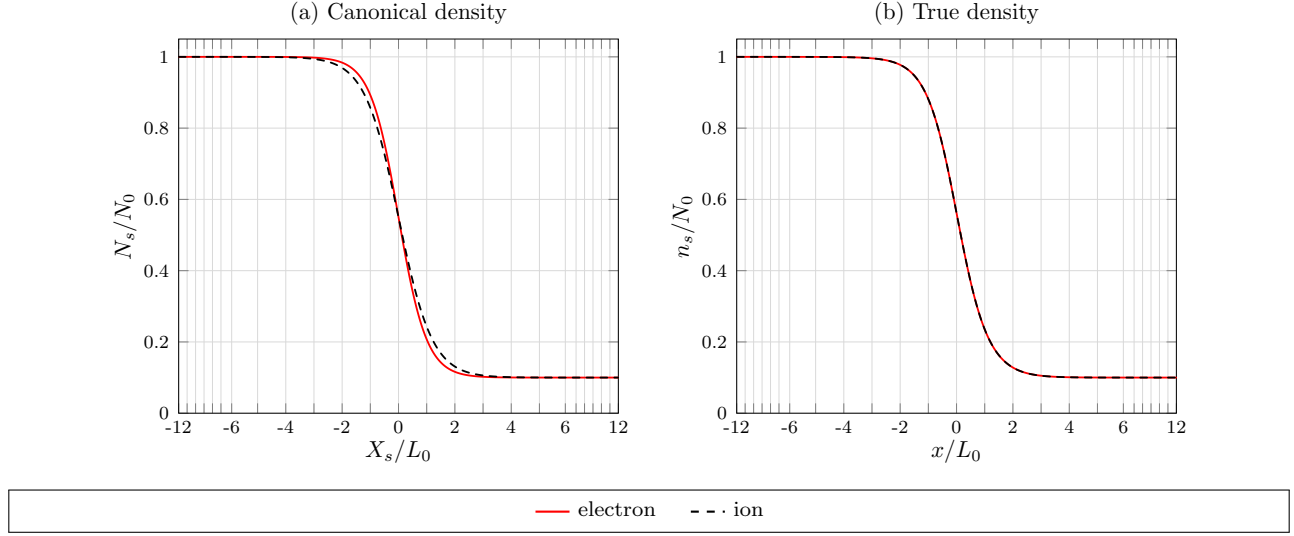
\captionof{figure}{\label{fig:representative_density} Prescribed canonical density
$\frac{N_s}{N_0}$ (a) compared with the true, self-consistent density $\frac{n_s}{N_0}$
(b) for each species. Because the electron and ion canonical density
transition widths differ ($w_{eN}=1.0L_0\neq w_{iN}=1.2L_0$), the
prescribed canonical densities are not charge-balanced; the true density
is nevertheless quasi-neutral up to the small Debye-length correction
$\lambda^2\ll1$, since it is obtained from the self-consistent solve
rather than imposed directly.}
\end{widefigure}

\par The perpendicular flow for each species is obtained by dividing the
$y$-component of the species current density by the species charge
density. In the discrete setting, this ratio is computed directly
from the partial derivatives of the stress with respect to the vector
potential and the potential:
\begin{align}
\left( \frac{V_{s,\perp}}{V_{0,*}} \right)_j &= -\frac{\overline{\partial}_{\sA,j}\widehat{\sP}_s}{\overline{\partial}_{\sV,j}\widehat{\sP}_s}.
\label{eq:discrete_perp_flow}
\end{align}
Here, the characteristic pressure-gradient speed is defined as
\begin{equation}
V_{0,*} \equiv \frac{T_0}{eB_0L_0}.
\label{eq:characteristic_pressure_gradient_speed}
\end{equation}
This also gives a physical interpretation for the normalization used
for $E_x$ above: $eE_xL_0/T_0=E_x/(B_0V_{0,*})$, so $E_x$ is reported
in units of the field that would drive an $E\times B$ flow of
order $V_{0,*}$.

\par Dividing the species force balance
\refeq{eq:simp_species_force_balance} by $\rho_s B_z$ and rearranging
shows the species perpendicular flow $V_{s,\perp}$ to be composed of
the pressure-gradient-driven and $E\times B$ contributions:
\begin{equation}
\underbrace{\frac{J_{s,y}}{\rho_s}}_{V_{s,\perp}}
=
\underbrace{\frac{1}{\rho_s B_z}\frac{dP_{s,xx}}{dx}}_{V_{s,*}}
+
\underbrace{\left(-\frac{E_x}{B_z}\right)}_{V_E},
\label{eq:mean_flow_definition}
\end{equation}
with discrete diagnostics
\begin{align}
\left( \frac{V_{E}}{V_{0,*}} \right)_j &= \frac{D_{1,j}\sV}{D_{1,j}\sA}, \\
\left( \frac{V_{s,*}}{V_{0,*}} \right)_j &= -\frac{D_{1,j}\widehat{\sP}_s}{\overline{\partial}_{\sV,j}\widehat{\sP}_s\,D_{1,j}\sA}.
\end{align}
By the exact discrete chain-rule identity \refeq{eq:discrete_chain_rule},
the sum of these two pieces is exactly equal to $\left(V_{s,\perp}/V_{0,*}\right)_j$ as
computed directly in \refeq{eq:discrete_perp_flow}, so this relation holds at the
level of the discretization, not merely in the continuum limit. A further decomposition
of $V_{s,*}$ into an electrostatic piece and a magnetostatic piece, via the same chain
rule, is presented in Sec.~\ref{sec:flow_invariance}.

\reffig{fig:representative_flow} shows the resulting perpendicular
flow for each species: the pressure-gradient-driven flow $V_{s,*}$ is
large compared with the $E\times B$ flow $V_E$, consistent with
the small magnitude of the normalized $E_x$ shown in
\reffig{fig:representative_fields}. Here, $V_E$ shifts the flow of
both species in the same direction, positive near $\xi=0$.

\begin{widefigure}
\centering
\scalebox{0.87}{\input{figures/compiled/representative_flow_trim.tex}}
\par\vspace{7pt}
\input{figures/data/representative_equilibrium/legend_flow.tex}
\captionof{figure}{\label{fig:representative_flow} Dimensional perpendicular
flow $V_{s,\perp}=\frac{J_{s,y}}{\rho_s}$ \refeq{eq:mean_flow_definition},
in units of the characteristic pressure-gradient speed $V_{0,*}=\frac{T_0}{eB_0L_0}$
\refeq{eq:characteristic_pressure_gradient_speed}, for each species,
decomposed into the pressure-gradient flow $V_{s,*}$ and the
(species-independent) $E \times B$ flow $V_E$, with
$V_{s,\perp}=V_{s,*}+V_E$. Since $V_{0,*}$ does not depend on species,
the two panels are directly comparable.}
\end{widefigure}

\par This subsection is also the natural place to check the particle-loading
procedure from Sec.~\ref{sec:particle_loading}. For the converged
pair $\sV(\xi)$ and $\sA(\xi)$, particles are loaded from the marginal
joint density in $(\xi,u_{s,y})$ using inverse Rosenblatt sampling, as
described there. \reffig{fig:representative_loading} shows, for each
species, a binned realization of $10^6$ loaded particles (color)
compared with contours of the tabulated target density
$g_s(\xi,u_{s,y})$ \refeq{eq:loading_joint_density}. Close agreement
between the particle density and the target contours verifies the
end-to-end use of the equilibrium for PIC initialization, separately
from the accuracy of the field solve itself. The electron phase-space
density resembles a local Maxwellian, since electrons are
well-magnetized (small gyroradius); the ion phase space shows more
structure, since the ion gyroradius is not small compared with the
reference gradient length.

\begin{widefigure}
\centering
\scalebox{0.87}{\input{figures/compiled/representative_loading_trim.tex}}
\vspace{10pt}
\captionof{figure}{\label{fig:representative_loading} Binned realization of
$10^6$ particles loaded by inverse Rosenblatt sampling
(Sec.~\ref{sec:particle_loading}), shown by color (denser regions in
yellow), compared with contours of the tabulated target density
$g_s(\xi,u_{s,y})$ (black lines), for each species.}
\end{widefigure}

\subsection{Comparison with the Asymptotic Description \label{sec:asymptotic_comparison}}

\input{figures/data/asymptotic_verification/asymptotic_verification_vars.tex}

\par As an independent check on the numerical equilibria, this
subsection performs a comparison with the asymptotic approximation derived in
Appendix~\ref{sec:asymptotic_approach} for the well-magnetized,
quasi-neutral regime. To carry out this comparison, a dimensionless
parameter $\theta_{\mathrm{mag}}$ is introduced that multiplies the
reference magnetic field, $B_0 \to \theta_{\mathrm{mag}} B_0$, while
holding the remaining dimensional scales fixed. The value
$\theta_{\mathrm{mag}}=1$ recovers the nominal values of
Table~\ref{tab:illustrative_parameters}. Since the gyroradius scales as
$B_0^{-1}$ and $\beta$ as $B_0^{-2}$, increasing $\theta_{\mathrm{mag}}$
simultaneously decreases the max norm of the signed normalized
gyroradius vector, $\|\boldsymbol{\varrho}\|_\infty$, linearly,
$\ell_{\mathrm{gy},s}\propto\theta_{\mathrm{mag}}^{-1}$, and $\beta$
quadratically, $\beta\propto\theta_{\mathrm{mag}}^{-2}$.

\par The last verification step compares the numerical solution with
the asymptotic expansion as $\theta_{\mathrm{mag}}$ is varied. In the
canonical bi-Maxwellian family, increasing $\theta_{\mathrm{mag}}$ while
holding the remaining parameters fixed decreases both $\epsilon_\beta$ and
$\epsilon_\varrho$ at the rate $\theta_{\mathrm{mag}}^{-2}$. In the regime
where $\epsilon_\lambda$ is small compared with $\epsilon_\varrho$, the potential is
expected to satisfy
\begin{equation}
\frac{\sV-\sV^{(0)}}{\epsilon_\varrho}
\approx
\frac{\epsilon_\beta}{\epsilon_\varrho}\sV^{(\beta)}
+
\sV^{(\varrho)},
\label{eq:asymptotic_phi_verification_form}
\end{equation}
where the right-hand side is independent of
$\theta_{\mathrm{mag}}$. Similarly, the vector potential is expected to
satisfy
\begin{equation}
\frac{\sA-\sA^{(0)}}{\epsilon_\beta}
\approx
\sA^{(\beta)}.
\label{eq:asymptotic_A_verification_form}
\end{equation}
These scaled differences provide a direct check that the numerical
solutions approach the predicted asymptotic terms as the plasma becomes
more strongly magnetized.

\par \reffig{fig:asymptotic_verification} shows these scaled differences for
several values of $\theta_{\mathrm{mag}}$, together with the predicted
right-hand sides. As $\theta_{\mathrm{mag}}$ increases, the numerical
curves collapse onto the prediction, testing the parameter independence in
\refeqs{eq:asymptotic_phi_verification_form}{eq:asymptotic_A_verification_form}.
The ratio $\epsilon_\lambda/\epsilon_\varrho$ remains small over the scan
shown, reaching a maximum of $2.8\times10^{-4}$ at
$\theta_{\mathrm{mag}}=1$, indicating negligible contributions from $\sV^{(\lambda)}$ in the interior. 
Panel (a) also shows, in a light shade, the two terms
$(\epsilon_\beta/\epsilon_\varrho)\sV^{(\beta)}$ and $\sV^{(\varrho)}$
making up the right-hand side of
\refeq{eq:asymptotic_phi_verification_form} individually. Neither term is negligible relative to the other, so
both the finite-beta and FLR separation corrections contribute to the
leading-order prediction.
\refeq{eq:asymptotic_A_verification_form} has no analogous second term,
so panel (b) shows only the prediction itself.

\begin{widefigure}
\centering
\scalebox{0.87}{\input{figures/compiled/asymptotic_verification_trim.tex}}
\par\vspace{7pt}
\input{figures/data/asymptotic_verification/legend_asymptotic.tex}
\captionof{figure}{\label{fig:asymptotic_verification} Scaled potential
differences $(\sV-\sV^{(0)})/\epsilon_\varrho$ (a) and
$(\sA-\sA^{(0)})/\epsilon_\beta$ (b) for a scan in
$\theta_{\mathrm{mag}}$ at the nominal canonical density transition widths
of Table~\ref{tab:illustrative_parameters}, shown dashed, compared
with the asymptotic prediction (solid black) given by the right-hand
sides of
\refeqs{eq:asymptotic_phi_verification_form}{eq:asymptotic_A_verification_form},
evaluated at $\theta_{\mathrm{mag}}=\avMaxTheta$. Panel (a) additionally
shows, in a light shade, the two terms
$(\epsilon_\beta/\epsilon_\varrho)\sV^{(\beta)}$ and $\sV^{(\varrho)}$
making up the prediction individually (also solid, distinguishing them
from the dashed numerical curves).}
\end{widefigure}

%% file: figures/data/hamiltonian_conservation/hamiltonian_conservation_vars.tex
\newcommand{\hamNa}{4096}
\newcommand{\hamNb}{64}
\newcommand{\hamXmin}{-12}
\newcommand{\hamXmax}{12}
\newcommand{\hamLegendColumns}{5}

%% file: figures/compiled/hamiltonian_conservation_trim.tex
\includegraphics[trim=0pt 590.40236pt 110.65811pt 0pt,clip]{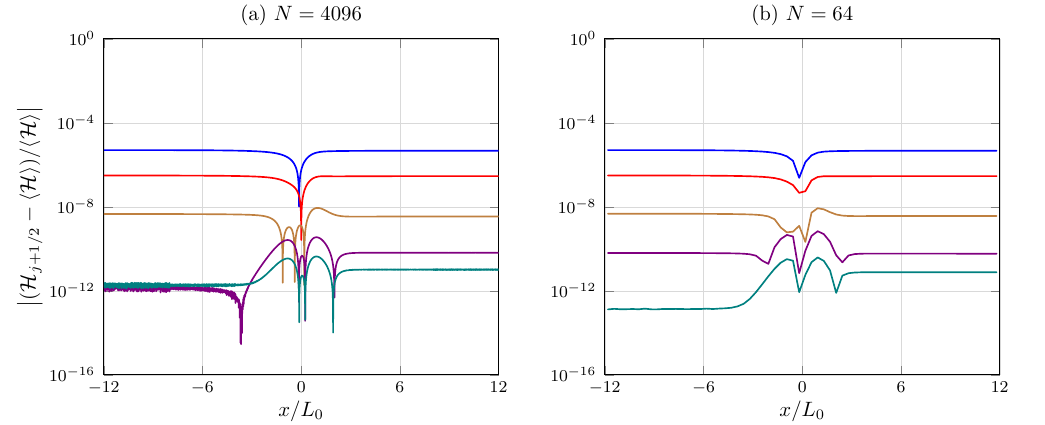}

%% file: figures/data/hamiltonian_conservation/legend_hamiltonian.tex
\begin{center}\fbox{\begin{minipage}{0.94\linewidth}\centering\footnotesize
\begin{tabular}{l@{\hspace{1.5em}}l@{\hspace{1.5em}}l@{\hspace{1.5em}}l@{\hspace{1.5em}}l}
\LegendSwatch{blue}{solid}~$\epsilon_{\mathrm{rel}}=10^{-3}$ & \LegendSwatch{red}{solid}~$\epsilon_{\mathrm{rel}}=10^{-4}$ & \LegendSwatch{brown}{solid}~$\epsilon_{\mathrm{rel}}=10^{-5}$ & \LegendSwatch{violet}{solid}~$\epsilon_{\mathrm{rel}}=10^{-6}$ & \LegendSwatch{teal}{solid}~$\epsilon_{\mathrm{rel}}=10^{-7}$ \\
\end{tabular}
\end{minipage}}\end{center}

%% file: figures/data/spatial_convergence/richardson_v_table.tex
\begin{tabular}{c
r@{\,${}\times{}10$}l
@{\hspace{10pt}}
r@{\,${}\times{}10$}l
c}
\hline\hline
$N$ & \multicolumn{2}{c}{$\Delta\xi$} &
\multicolumn{2}{c}{$\Delta_{\sV,N}$} & $p_{\sV,N}$ \\
\hline
$32$ & $7.500$ & $^{-1}$ & $2.561$ & $^{0}$ & $6.064$ \\
$64$ & $3.750$ & $^{-1}$ & $3.828$ & $^{-2}$ & $2.237$ \\
$128$ & $1.875$ & $^{-1}$ & $8.117$ & $^{-3}$ & $2.029$ \\
$256$ & $9.375$ & $^{-2}$ & $1.989$ & $^{-3}$ & $2.007$ \\
$512$ & $4.688$ & $^{-2}$ & $4.949$ & $^{-4}$ & $2.001$ \\
$1024$ & $2.344$ & $^{-2}$ & $1.237$ & $^{-4}$ & $1.999$ \\
\hline\hline
\end{tabular}

%% file: figures/data/spatial_convergence/richardson_a_table.tex
\begin{tabular}{c
r@{\,${}\times{}10$}l
@{\hspace{10pt}}
r@{\,${}\times{}10$}l
c}
\hline\hline
$N$ & \multicolumn{2}{c}{$\Delta\xi$} &
\multicolumn{2}{c}{$\Delta_{\sA,N}$} & $p_{\sA,N}$ \\
\hline
$32$ & $7.500$ & $^{-1}$ & $5.366$ & $^{-4}$ & $2.014$ \\
$64$ & $3.750$ & $^{-1}$ & $1.329$ & $^{-4}$ & $2.003$ \\
$128$ & $1.875$ & $^{-1}$ & $3.315$ & $^{-5}$ & $2.001$ \\
$256$ & $9.375$ & $^{-2}$ & $8.283$ & $^{-6}$ & $2.000$ \\
$512$ & $4.688$ & $^{-2}$ & $2.070$ & $^{-6}$ & $2.000$ \\
$1024$ & $2.344$ & $^{-2}$ & $5.176$ & $^{-7}$ & $2.000$ \\
\hline\hline
\end{tabular}

%% file: figures/data/quadrature_convergence/gauss_hermite_table.tex
\begin{tabular}{ccc}
\hline\hline
$N_g$ & $\Delta_{\sV,N_g}$ & $\Delta_{\sA,N_g}$ \\
\hline
$1$ & $4.8209\times10^{-2}$ & $1.3866\times10^{-4}$ \\
$2$ & $2.9814\times10^{-2}$ & $8.9653\times10^{-5}$ \\
$4$ & $1.7588\times10^{-4}$ & $2.6019\times10^{-7}$ \\
$8$ & $4.6920\times10^{-7}$ & $9.0966\times10^{-10}$ \\
$16$ & $1.2329\times10^{-12}$ & $2.4132\times10^{-14}$ \\
$24$ & $2.3070\times10^{-14}$ & $6.6124\times10^{-14}$ \\
$32$ & $2.4317\times10^{-14}$ & $4.3398\times10^{-14}$ \\
\hline\hline
\end{tabular}

%% file: figures/data/representative_equilibrium/representative_equilibrium_vars.tex
\newcommand{\reqXmin}{-12}
\newcommand{\reqXmax}{12}
\newcommand{\reqPhiYlim}{0.0254124}
\newcommand{\reqFieldYlim}{0.0422584}
\newcommand{\reqUmin}{-3.5}
\newcommand{\reqUmax}{3.5}
\newcommand{\reqUDataMin}{-4}
\newcommand{\reqUDataMax}{4}

%% file: figures/compiled/representative_potentials_trim.tex
\includegraphics[trim=0pt 590.40236pt 110.65811pt 0pt,clip]{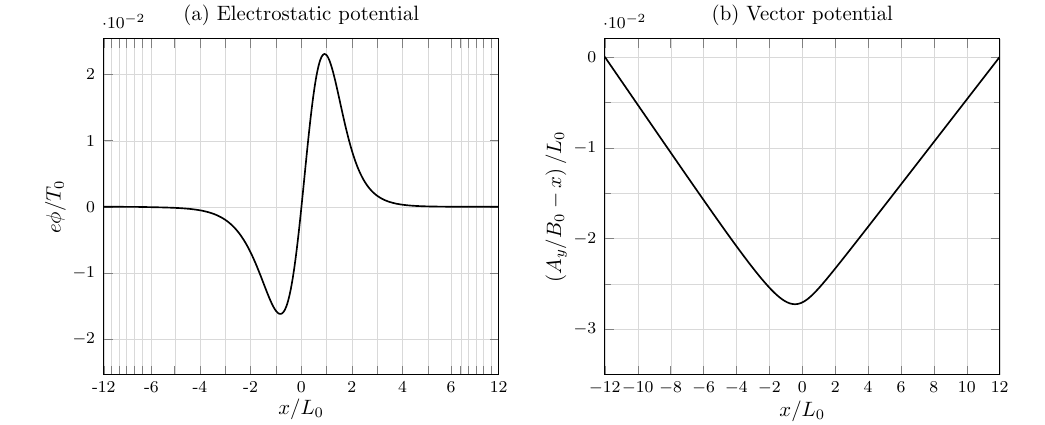}

%% file: figures/compiled/representative_fields_trim.tex
\includegraphics[trim=0pt 590.40236pt 110.65811pt 0pt,clip]{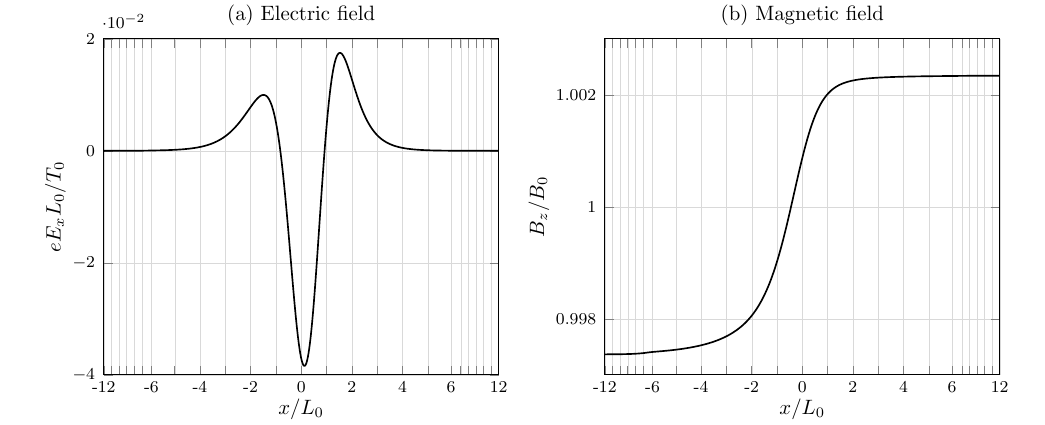}

%% file: figures/compiled/representative_density_trim.tex
\includegraphics[trim=0pt 590.40236pt 110.65811pt 0pt,clip]{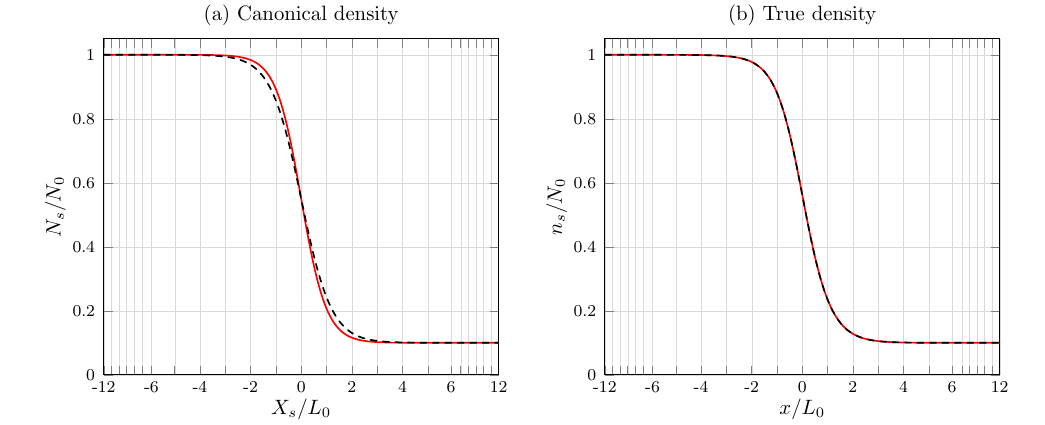}

%% file: figures/data/representative_equilibrium/legend_density.tex
\fbox{\begin{minipage}{0.94\linewidth}\centering\footnotesize
\begin{tabular}{l@{\hspace{1.5em}}l}
\LegendSwatch{red}{solid}~electron & \LegendSwatch{black}{dashed}~ion \\
\end{tabular}
\end{minipage}}\vspace{10pt}\par

%% file: figures/compiled/representative_flow_trim.tex
\includegraphics[trim=0pt 590.40236pt 110.65811pt 0pt,clip]{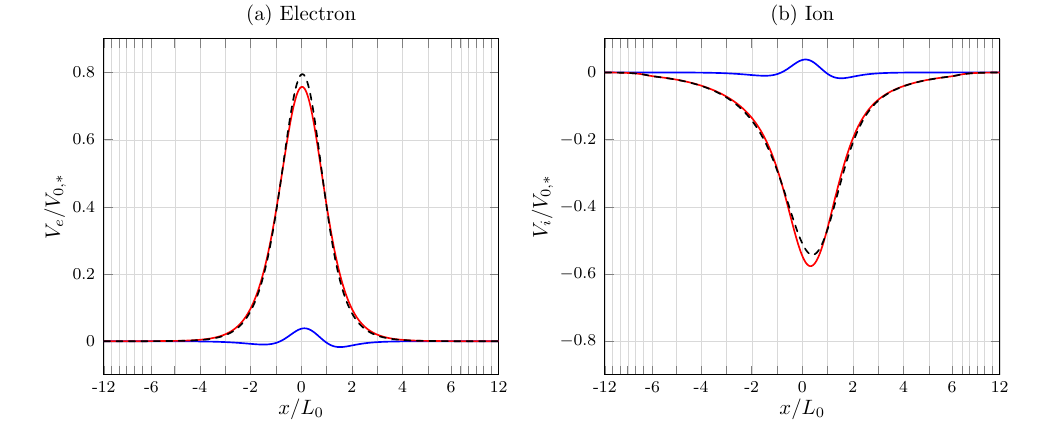}

%% file: figures/data/representative_equilibrium/legend_flow.tex
\fbox{\begin{minipage}{0.94\linewidth}\centering\footnotesize
\begin{tabular}{l@{\hspace{1.5em}}l@{\hspace{1.5em}}l}
\LegendSwatch{blue}{solid}~$V_E$ & \LegendSwatch{red}{solid}~$V_{s,*}$ & \LegendSwatch{black}{dashed}~$V_{s,\perp}$ \\
\end{tabular}
\end{minipage}}\vspace{10pt}\par

%% file: figures/compiled/representative_loading_trim.tex
\includegraphics[trim=0pt 590.40236pt 110.65811pt 0pt,clip]{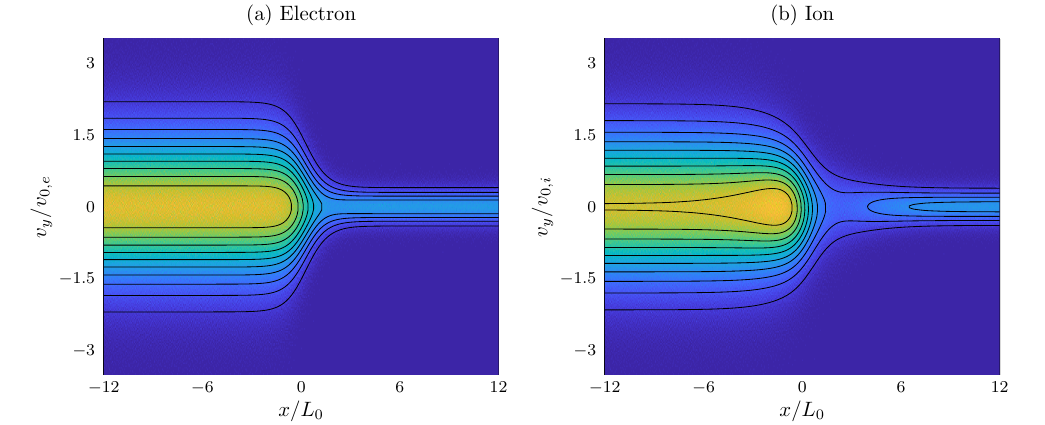}

%% file: figures/data/asymptotic_verification/asymptotic_verification_vars.tex
\newcommand{\avXmin}{-12}
\newcommand{\avXmax}{12}
\newcommand{\avPhiYlim}{0.062}
\newcommand{\avLegendColumns}{4}
\newcommand{\avMaxTheta}{1}

%% file: figures/compiled/asymptotic_verification_trim.tex
\includegraphics[trim=0pt 590.40236pt 110.65811pt 0pt,clip]{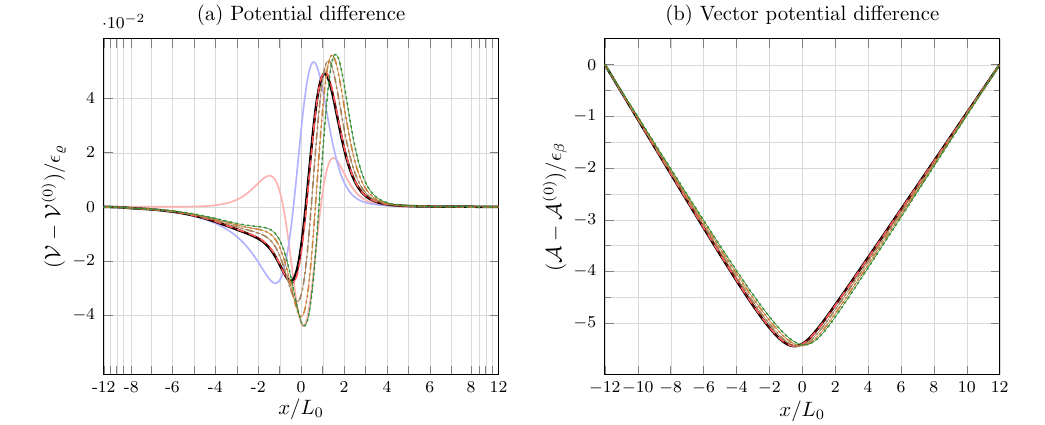}

%% file: figures/data/asymptotic_verification/legend_asymptotic.tex
\begin{center}\fbox{\begin{minipage}{0.94\linewidth}\centering\footnotesize
\begin{tabular}{l@{\hspace{1.5em}}l@{\hspace{1.5em}}l@{\hspace{1.5em}}l}
\LegendSwatch{green!51!black!85!black}{dotted}~$\theta_{\mathrm{mag}}=0.2$ (numerical) & \LegendSwatch{orange!85!black!85!black}{dashdotdotted}~$\theta_{\mathrm{mag}}=0.24$ (numerical) & \LegendSwatch{brown!85!black!85!black}{dashdotted}~$\theta_{\mathrm{mag}}=0.33$ (numerical) & \LegendSwatch{red}{dashed}~$\theta_{\mathrm{mag}}=1$ (numerical) \\
\LegendSwatch{red!30!white}{solid}~$(\epsilon_\beta/\epsilon_\varrho)\sV^{(\beta)}$ & \LegendSwatch{blue!30!white}{solid}~$\sV^{(\varrho)}$ & \LegendSwatch{black}{solid}~$(\epsilon_\beta/\epsilon_\varrho)\sV^{(\beta)}+\sV^{(\varrho)}$ (a) & \LegendSwatch{black}{solid}~$\sA^{(\beta)}$ (b) \\
\end{tabular}
\end{minipage}}\end{center}

%% file: v2/section_07.tex
\section{Equilibrium Electric Fields \label{sec:equilibrium_efield}}

An interesting feature of the present model is the self-consistent description of
equilibrium electric fields. In this section, tests are designed to explore mechanisms by which
such fields may be generated in this model, as well as the collective plasma response to such fields.
The focus here is on the quasi-neutral regime, however, the model is additionally suitable for studying
equilibrium electric fields generated by Debye-scale charge-separation, since the full Poisson equation 
is included.

In the presence of magnetic fields, it is well known that electric fields result in $E \times B$
drift motions in charged particle orbits. Because this drift occurs at the level of
individual particle motions, it corresponds to genuine net particle advection, producing
a collective $E \times B$ flow.
Such flows are understood to have important implications for transport and stability
in magnetized plasma. For example, sheared equilibrium $E \times B$ flows are a characteristic feature
of transport barriers.

Two studies are carried out in this section to explore features of equilibrium electric fields within this model.
In the first study, the generation of such fields due to FLR separation and finite-$\beta$ effects is demonstrated. In the context
of the asymptotic analysis of Appendix~\ref{sec:asymptotic_approach}, these are first-order effects in the ordering parameters.
This is explored for charge-balanced canonical densities in the canonical bi-Maxwellian form for $F_s$, which eliminates
electrostatic potentials at zeroth order. When the charge balance in canonical densities is broken, however, much larger
(zeroth-order) potentials are generated in response, to achieve quasi-neutrality in the true densities.

In the second study, a key prediction of the model is tested: an exact species-level cancellation of the $E \times B$ flow
by an electrostatically generated component of the pressure-gradient-driven flow. To work within the framework of the self-consistent
Vlasov--Maxwell system, a test is designed in which zeroth-order
electric fields large enough to drive an $E\times B$ flow comparable
to the characteristic pressure-gradient speed $V_{0,*}$ are
generated by adjusting the width of the canonical density
profiles of one species while keeping all parameters of the second species fixed, breaking charge balance in the canonical densities. In the
particular case of the canonical bi-Maxwellian distribution model with flat canonical temperature profiles, an even stronger prediction is made:
the invariance of the total flow to both the electric field \textit{and} electrostatic potential. Numerical results demonstrate
the predicted invariance in this setting.

\subsection{Electric Field Generation due to FLR Separation and Finite-$\beta$ Effects \label{sec:efield_flr}}

Applying \refeq{eq:V0_balance_general} to the two-species case with canonical bi-Maxwellians gives:
\begin{equation}
-\sN_e\exp\!\left[\frac{\sV^{(0)}}{\sT_e}\right]
+ \frac{q_i}{e}\,\sN_i\exp\!\left[-\frac{q_i}{e}\,\frac{\sV^{(0)}}{\sT_i}\right] = 0,
\label{eq:V0_balance}
\end{equation}
with $\sN_s$ and $\sT_s$ evaluated at $\sX_s=\sA^{(0)}=\xi$. Hence, the zeroth-order potential is determined by balancing the ion and electron
Boltzmann-like factors. For two species, this may be solved explicitly for the potential:
\begin{equation}
\sV^{(0)} = \frac{\sT_e\sT_i}{\sT_i+\frac{q_i}{e}\sT_e}
\ln\!\left(\frac{q_i}{e}\,\frac{\sN_i}{\sN_e}\right).
\label{eq:V0_explicit}
\end{equation}
Zeroth-order potentials may therefore be generated when canonical densities are charge-separated. When canonical density profiles are chosen to be charge-balanced in a shared
argument, the two-species form of \refeq{eq:charge_neutral_pseudo_profiles} becomes
\begin{equation}
\frac{q_i}{e}\sN_i(\chi) =\sN_e(\chi)
\end{equation}
for common $\chi$. Since both species profiles are evaluated at $\xi$ at zeroth order, the argument of the logarithm in \refeq{eq:V0_explicit} is unity 
and the zeroth-order potential vanishes. An equilibrium electrostatic
potential still may be generated, however, from first order effects in $\epsilon_\beta$ and $\epsilon_\varrho$. It is noted here that electric field generation
due to FLR charge separation effects has been considered previously in the literature
\cite{Hazeltine1993,Shoucri2002,Shoucri2004,Shoucri2012,LeeWhite2017,LeeWhite2019}, and has been suggested
to play a role in $E_r$ wells in the H-mode pedestal.

To examine the charge-balanced canonical density case, the nominal
parameters of Table~\ref{tab:illustrative_parameters} are used, except
the electron canonical density transition width is modified so that
$w_{eN}=w_{iN}=1.2L_0$. As in Sec.~\ref{sec:asymptotic_comparison},
$\theta_{\mathrm{mag}}$ is scanned from $0.20$ to $1.00$, over which
\[
\begin{array}{r@{\;}c@{\;}c@{\;}c@{\;}l}
5.00\times10^{-3} & \le & \beta & \le & 1.25\times10^{-1}, \\[6pt]
1.50\times10^{-1} & \le & |\ell_{\mathrm{gy},i}/L_0| & \le & 7.50\times10^{-1}, \\[6pt]
2.50\times10^{-3} & \le & |\ell_{\mathrm{gy},e}/L_0| & \le & 1.25\times10^{-2}.
\end{array}
\]

The most weakly magnetized case, $\theta_{\mathrm{mag}}=0.20$
($\beta\approx12.5\%$, $|\ell_{\mathrm{gy},i}/L_0|\approx0.75$,
$|\ell_{\mathrm{gy},e}/L_0|\approx1.25\times10^{-2}$), pushes $\beta$ and
$\|\boldsymbol{\varrho}\|_\infty$ large enough that the perturbative
ordering underlying the asymptotic description of
Appendix~\ref{sec:asymptotic_approach} is no longer expected to be
well satisfied. It is included deliberately, to see how the numerical
equilibrium behaves in a regime accessible only to the full
construction.

\reffig{fig:mag_scan_peak_scaling} shows the peak magnitude of $\sV$
and $\sA-\xi$ against $\theta_{\mathrm{mag}}$ on log-log axes, for five
$\theta_{\mathrm{mag}}$ values together with a $\theta_{\mathrm{mag}}^{-2}$
reference line anchored at the most strongly magnetized point, where
the asymptotic estimate is most accurate. Agreement is close across
the full range shown for $\sA-\xi$, and tightens toward the most
strongly magnetized end for $\sV$. This gives a quick check on the
predicted $\theta_{\mathrm{mag}}^{-2}$ scaling, complementing the more
detailed asymptotic comparison in \reffig{fig:asymptotic_verification}.

\input{figures/data/theta_mag_scan/theta_mag_scan_vars.tex}

\begin{widefigure}
\centering
\scalebox{0.87}{\input{figures/compiled/theta_mag_peak_scaling_trim.tex}}
\vspace{10pt}
\captionof{figure}{\label{fig:mag_scan_peak_scaling} Peak magnitude of the
electrostatic potential $\sV$ (a) and vector-potential perturbation
$\sA-\xi$ (b) against $\theta_{\mathrm{mag}}$ (log-log), for five
$\theta_{\mathrm{mag}}$ values spanning the scan, including the three
shown in \reffig{fig:mag_scan_potentials}. Markers are the numerically
computed peak values; the dashed line is a $\theta_{\mathrm{mag}}^{-2}$
reference anchored at the most strongly magnetized point, $\theta_{\mathrm{mag}}=1$.}
\end{widefigure}

\reffigs{fig:mag_scan_potentials}{fig:mag_scan_loading} show the
corresponding potentials, fields, canonical and true densities, species flow decomposed
into pressure-gradient and $E \times B$ flows, and the loaded distribution
function for three of the values selected from the scan in the previous figure.

Both $\sV$ and the vector-potential perturbation $\sA-\xi$ grow in
magnitude as $\theta_{\mathrm{mag}}$ decreases, as seen in the full
profiles in \reffig{fig:mag_scan_potentials}, consistent with
weakening the background field increasing both the FLR separation
between species and $\beta$.

\begin{widefigure}
\centering
\scalebox{0.87}{\input{figures/compiled/theta_mag_potentials_trim.tex}}
\par\vspace{7pt}
\input{figures/data/theta_mag_scan/legend_theta.tex}
\captionof{figure}{\label{fig:mag_scan_potentials} Electrostatic potential $\sV$
(a) and vector potential $\sA$ (b) for the $\theta_{\mathrm{mag}}$ scan.}
\end{widefigure}

Across the scan, the normalized electric field shown in
\reffig{fig:mag_scan_fields} remains small in the sense discussed in
Sec.~\ref{sec:numerical_verification}: the largest magnitude reached
anywhere in the scan is only $eE_xL_0/T_0\approx-3\times10^{-2}$, at
the most weakly magnetized point $\theta_{\mathrm{mag}}=0.20$, well
below the $O(1)$ value that would drive an $E\times B$ flow comparable
to $V_{s,*}$.

The magnetic field $B_z$, also shown in \reffig{fig:mag_scan_fields},
deviates from its background value $B_0$ by about $6\%$ at its most
extreme, at $\theta_{\mathrm{mag}}=0.20$.

\begin{widefigure}
\centering
\scalebox{0.87}{\input{figures/compiled/theta_mag_fields_trim.tex}}
\par\vspace{7pt}
\input{figures/data/theta_mag_scan/legend_theta.tex}
\captionof{figure}{\label{fig:mag_scan_fields} Electric field $E_x$ (a) and
magnetic field $B_z$ (b) for the $\theta_{\mathrm{mag}}$ scan.}
\end{widefigure}

In \reffig{fig:mag_scan_den}, the true density shifts noticeably
toward the low-density side relative to the prescribed canonical density
profile as $\theta_{\mathrm{mag}}$ decreases. This shift can be
explained primarily by the ions: since $\varrho_i\gg\varrho_e$, ion
orbits with $\sX_i$ on the high-density side are thought to reach
appreciably farther into the low-density region as $\varrho_i$ grows,
while the much smaller electron gyroradius would produce little shift
on its own. In
the quasi-neutral regime $\lambda^2\ll1$ used throughout, however, the
true electron density is constrained to closely track the true ion
density, so the electron profile shifts along with the ions despite
the electron orbits themselves remaining small.

\begin{widefigure}
\centering
\scalebox{0.87}{\input{figures/compiled/theta_mag_density_trim.tex}}
\par\vspace{7pt}
\input{figures/data/theta_mag_scan/legend_density_theta.tex}
\captionof{figure}{\label{fig:mag_scan_den} Prescribed canonical density $N_s/N_0$
(a) compared with the true, self-consistent density $n_s/N_0$ (b) for
each species, for the $\theta_{\mathrm{mag}}$ scan. Electron is solid
and a lighter shade, ion is dashed and drawn above the electron curve of
the same $\theta_{\mathrm{mag}}$ value, since the two nearly coincide
once $n_s$ is quasi-neutral. Since the electron and ion canonical density
transition widths are equal here, the canonical densities are
charge-balanced
\refeq{eq:charge_neutral_pseudo_profiles}; the true densities are
quasi-neutral for the same reason as in
\reffig{fig:representative_density}.}
\end{widefigure}

The perpendicular flow in \reffig{fig:mag_scan_flow} shows a
corresponding shift: the electron total flow $V_{e,\perp}$ stays
roughly constant in magnitude while its peak shifts toward the
low-density side as $\theta_{\mathrm{mag}}$ decreases, whereas the ion
total flow $V_{i,\perp}$ both shifts in the same direction and
decreases in magnitude. This shift tracks the shift of the peak
density gradient noted above in \reffig{fig:mag_scan_den}, which can
be explained by the pressure-gradient-driven flow $V_{s,*}$ peaking near
where the pressure gradient itself is largest. The $E\times B$
flow $V_E$ grows
correspondingly larger as $\theta_{\mathrm{mag}}$ decreases, consistent
with the growing $E_x$ noted above in \reffig{fig:mag_scan_fields}.

\begin{widefigure}
\centering
\scalebox{0.87}{\input{figures/compiled/theta_mag_flow_trim.tex}}
\par\vspace{7pt}
\input{figures/data/theta_mag_scan/legend_flow_theta.tex}
\captionof{figure}{\label{fig:mag_scan_flow} Perpendicular flow
$V_{s,\perp}$ \refeq{eq:mean_flow_definition} for each species,
decomposed into the pressure-gradient flow $V_{s,*}$ (dashed) and the
$E \times B$ flow $V_E$ (dotted) as in \reffig{fig:representative_flow},
with the total $V_{s,\perp}$ (solid) drawn on top, for the
$\theta_{\mathrm{mag}}$ scan. Color identifies the $\theta_{\mathrm{mag}}$
value.}
\end{widefigure}

\reffig{fig:mag_scan_loading} shows the loaded distribution function
across the $\theta_{\mathrm{mag}}$ scan, following the same
particle-loading procedure as \reffig{fig:representative_loading}. As
magnetization weakens, particle orbits widen. The particles
contributing to a given point $\xi$ are then drawn from a wider range
of $\sX_s$, where the canonical density and temperature vary. This
mixing is most visible for the ions, whose larger orbits produce a
tilt in the target contours, strongest near $\xi=0$ where the
canonical density gradient is steepest. From the first integral
\refeq{eq:dl_gcenter}, contours of constant $\sX_s$ satisfy
$u_{s,y}=(\sX_s-\sA(\xi))/\varrho_s$, so their slope scales as
$1/\varrho_s$ and grows more tilted as $\varrho_s$ increases with
decreasing $\theta_{\mathrm{mag}}$.

\begin{widefigure}
\centering
\scalebox{0.87}{\input{figures/compiled/theta_mag_loading_trim.tex}}
\vspace{10pt}
\captionof{figure}{\label{fig:mag_scan_loading} Binned realization of $10^6$
particles loaded by inverse Rosenblatt sampling
(Sec.~\ref{sec:particle_loading}), shown by color (denser regions in
yellow), compared with contours of the tabulated target density
$g_s(\xi,u_{s,y})$ (black lines), for each species across the
$\theta_{\mathrm{mag}}$ scan.}
\end{widefigure}
\newpage

\subsection{An exact cancellation of the $E \times B$ flow \label{sec:flow_invariance}}

The form of the pressure-gradient-driven flow, along with the moment identities
\refeqs{eq:species_phi_identity}{eq:species_A_identity}, makes an interesting
prediction for this symmetry-reduced model: the $E \times B$ flow is exactly cancelled by an electrostatically
generated component of the pressure-gradient-driven flow. This is easy to see
starting from the definition of $V_{s,*}$ in \refeq{eq:mean_flow_definition}.
The derivative with respect to $x$ can be expanded by the chain rule into
electrostatically and magnetostatically generated components:
\begin{equation}
V_{s,*} = \underbrace{\frac{1}{\rho_sB_z}\pd{P_{s,xx}}{\phi}\frac{d\phi}{dx}}_{V_{s,*\phi}}
+ \underbrace{\frac{1}{\rho_sB_z}\pd{P_{s,xx}}{A_y}\frac{dA_y}{dx}}_{V_{s,*A_y}}.
\label{eq:pressure_gradient_split}
\end{equation}
Using \refeq{eq:species_phi_identity}, together with
\refeq{eq:field_ansatz} and $V_E=-E_x/B_z$,
\begin{equation}
V_{s,*\phi} = \frac{-\rho_s}{\rho_sB_z}\frac{d\phi}{dx} = \frac{E_x}{B_z} = -V_E,
\label{eq:electrostatic_flow_exact}
\end{equation}
exactly, for any species and for any form chosen for $F_s$ in \refeq{eq:general_F}.

The remaining magnetostatically generated component $V_{s,*A_y}$ leads to the diamagnetic
current responsible for reducing the magnetic field strength in high-pressure regions, through
Ampère's law. Using the companion identity
$\partial P_{s,xx}/\partial A_y=J_{s,y}$ \refeq{eq:species_A_identity},
together with $dA_y/dx=B_z$ and \refeq{eq:electrostatic_flow_exact},
\begin{equation}
V_{s,*A_y} = \frac{1}{\rho_sB_z}\frac{\partial P_{s,xx}}{\partial A_y}\frac{dA_y}{dx}
= \frac{J_{s,y}}{\rho_s} = V_{s,\perp},
\label{eq:magnetostatic_flow_exact}
\end{equation}
consistent with it being the only surviving piece of the total perpendicular flow. 

This cancellation is a consequence of the same structural property
in this symmetry-reduced model that led to the identity
\refeq{eq:species_phi_identity}. It is simply a restatement
in terms of species flows rather than the stress tensor's response to the
potentials. That structural property is not required by species force balance
in general; it is a consequence of this type of symmetry reduction, under which
any solution of the Vlasov equation takes the form \refeq{eq:general_F}, forcing
$P_{s,xx}$, $\rho_s$, and $J_{s,y}$ to depend on $x$ only through $\phi(x)$ and $A_y(x)$. The companion
identity \refeq{eq:magnetostatic_flow_exact}, itself a restatement of \refeq{eq:species_A_identity},
is a further, necessary consequence of \refeq{eq:electrostatic_flow_exact} together with the flow
definitions \refeq{eq:mean_flow_definition} and \refeq{eq:pressure_gradient_split}, requiring
$V_{s,*A_y}$ to reduce to the total flow $V_{s,\perp}$.

To demonstrate this cancellation in the full self-consistent equilibrium setting, properties particular to the
canonical bi-Maxwellian are exploited to design a test with specific verifiable predictions in the presence of
zeroth-order self-generated electric fields which can be controlled in specific ways by adjusting
canonical density parameters of individual species. Furthermore, choosing flat temperature profiles
with the canonical bi-Maxwellian allows for a situation in which a species flow is invariant to both
$\phi$ and $E_x$, a property that is not guaranteed in general by
\refeqs{eq:electrostatic_flow_exact}{eq:magnetostatic_flow_exact} alone, since
dependence on $\phi$ may be retained in $V_{s,*A_y}$.

At flat temperature, $\phi$ enters the species stress \refeq{eq:reduced_species_stress}
only through an overall factor $\exp[-q_s\phi/T_{\perp s}]$ common to
$\partial P_{s,xx}/\partial A_y$ and $\partial P_{s,xx}/\partial\phi$, which
cancels in their ratio leaving:
\begin{multline}
V_{s,\perp} = \frac{T_{\perp s}}{q_s}\,\frac{\partial}{\partial A_y}
\ln\Big[\int_{-\infty}^{\infty}N_s(X_s) \\
{}\times\exp\!\left(-\frac{m_sv_y^2}{2T_{\perp s}}\right)\,dv_y\Big],
\label{eq:flow_leading_order_flatT}
\end{multline}
with $X_s$ as in \refeq{eq:gcenter}.

Zeroth-order potentials, and the electric fields they generate, arise
from breaking charge balance in the canonical densities
\refeq{eq:V0_explicit}. Here, charge balance is broken by steepening
the canonical density profile of one species while keeping the other
species' profile fixed.

This construction gives a sharp prediction for the fixed species.
Since $V_{s,*A_y}$ depends only on that species' own profile, and
nothing about it has changed, its total flow $V_{s,\perp}$ should
remain invariant, even though a nonzero electric field has been
generated. Steepening $\sN_e$ relative to $\sN_i$ illustrates this: it
generates a potential set by \refeq{eq:V0_explicit}, changes
$V_{e,\perp}$ through the electron's own profile in the usual way, but
leaves $V_{i,\perp}$ completely unchanged. Steepening $\sN_i$ instead
gives the mirror statement.

The zeroth-order electric field also has a simple symmetry, giving a
second prediction to test alongside the flow invariance above:
exchanging $\frac{q_i}{e}\sN_i$ and $\sN_e$, while leaving each
species' own temperature profile fixed, reverses its sign. This follows directly
from \refeq{eq:V0_explicit}, since swapping the numerator and
denominator inside the logarithm flips its sign. The reversal is not
exact, since \refeq{eq:V0_explicit} is itself a zeroth-order
asymptotic result.

% TODO: matlab_implementation/scripts/species_steepen_scan.m generates
% the data/macros the figures below reference (figures/data/
% species_steepen_scan/); run it to populate that directory before
% compiling this subsection with real numbers. widthList there
% currently spans {0.2, 0.4, 0.6, 0.9}, ncell=1024, ngauss=32.

\input{figures/data/species_steepen_scan/species_steepen_scan_vars.tex}

Relative to the nominal parameters of
Table~\ref{tab:illustrative_parameters}, both species' canonical density
transition widths are set to the common reference value
$w_{sN}=1.2L_0$ (equal widths, before either study's steepening is
applied below), and temperatures are flattened, $T_s^+=T_s^-=1.0$ for
both species, so that the flat-temperature limit applies exactly. This isolates the
canonical density-width swap as the only difference between species.

The two results above are demonstrated together with a single pair of
scans: one steepening the electron canonical density profile relative to
the ion profile (held fixed at the reference width), and one steepening
the ion profile relative to the electron profile (held fixed).
\reffig{fig:sep_scan_potentials} and
\reffig{fig:sep_scan_fields} show the resulting potentials and fields
for both studies, together with the leading-order sign-reversal prediction of
\refeq{eq:V0_explicit}.

\begin{widefigure}
\centering
\scalebox{0.87}{\input{figures/compiled/sep_scan_potentials_trim.tex}}
\par\vspace{7pt}
\input{figures/data/species_steepen_scan/legend_width_mirror.tex}
\captionof{figure}{\label{fig:sep_scan_potentials} Electrostatic potential $\sV$
(a,c) and vector potential $\sA$ (b,d), steepen-electron study (top
row) and steepen-ion study (bottom row), each colored by the steepened
species' transition width $w_{sN}$. The steepen-ion row overlays
$-1\times$ the steepen-electron curve (dashed) at each matching width,
the leading-order sign-reversal check of \refeq{eq:V0_explicit}.}
\end{widefigure}

\begin{widefigure}
\centering
\scalebox{0.87}{\input{figures/compiled/sep_scan_fields_trim.tex}}
\par\vspace{7pt}
\input{figures/data/species_steepen_scan/legend_width_mirror.tex}
\captionof{figure}{\label{fig:sep_scan_fields} Electric field $E_x$ (a,c) and
magnetic field $B_z$ (b,d), same layout and sign-reversal overlay as
\reffig{fig:sep_scan_potentials}.}
\end{widefigure}

\reffig{fig:sep_scan_density} shows the prescribed canonical density
alongside the resulting true, self-consistent density for both species
in each study, the same comparison as \reffig{fig:representative_density},
now for the canonical density-width swap studies.

\begin{widefigure}
\centering
\scalebox{0.87}{\input{figures/compiled/sep_scan_density_trim.tex}}
\par\vspace{7pt}
\input{figures/data/species_steepen_scan/legend_density.tex}
\captionof{figure}{\label{fig:sep_scan_density} Prescribed canonical density
$N_s/N_0$ (a,c) compared with the true, self-consistent density
$n_s/N_0$ (b,d), steepen-electron study (top row) and steepen-ion
study (bottom row), colored by the steepened species' transition
width $w_{sN}$ as in \reffig{fig:sep_scan_potentials}, same comparison as
\reffig{fig:representative_density} with $w_{sN}$ varied per
\reffig{fig:sep_scan_potentials}'s caption.}
\end{widefigure}

\reffig{fig:sep_scan_flow} shows the accompanying species flow,
decomposed as in \reffig{fig:representative_flow}, confirming the
invariance of the total flow for the non-steepened species in each
study, despite the large, self-generated electric field
(\reffig{fig:sep_scan_fields}). This is a direct consequence of the
predicted cancellation between its $E\times B$ flow and the
electrostatic part of its pressure-gradient-driven flow, combined with
the potential independence special to the canonical bi-Maxwellian with
flat temperature profiles.

\begin{widefigure}
\centering
\scalebox{0.87}{\input{figures/compiled/sep_scan_flow_trim.tex}}
\par\vspace{7pt}
\input{figures/data/species_steepen_scan/legend_flow.tex}
\captionof{figure}{\label{fig:sep_scan_flow} Perpendicular flow $V_{s,\perp}$
\refeq{eq:mean_flow_definition} for each species, decomposed into
$V_{s,*}$ (dashed) and $V_E$ (dotted), for the electron- and
ion-steepened studies (top/bottom row). The non-steepened species'
total $V_{s,\perp}$ (solid) is predicted to be invariant to the
steepened species' width in each study
\refeq{eq:flow_leading_order_flatT}, even though its $V_{s,*}$ and $V_E$
components individually are not.}
\end{widefigure}

\reffig{fig:sep_scan_loading} shows the loaded distribution function of
both species at each study's most extreme steepening width,
illustrating how steepening one species' canonical density profile
reshapes the phase-space structure of both species.

\begin{widefigure}
\centering
\scalebox{0.87}{\input{figures/compiled/sep_scan_loading_trim.tex}}
\vspace{10pt}
\captionof{figure}{\label{fig:sep_scan_loading} Binned realization of $10^6$
particles loaded by inverse Rosenblatt sampling
(Sec.~\ref{sec:particle_loading}), shown by color (denser regions in
yellow), compared with contours of the tabulated target density
$g_s(\xi,u_{s,y})$ (black lines), for each species at each study's most extreme steepening
width, steepen-electron study (top row) and steepen-ion study (bottom
row).}
\end{widefigure}

%% file: figures/data/theta_mag_scan/theta_mag_scan_vars.tex
\newcommand{\magScanXmin}{-12}
\newcommand{\magScanXmax}{12}
\newcommand{\magScanLegendColumns}{3}
\newcommand{\magScanPhiYlim}{0.026883}
\newcommand{\magScanFieldYlim}{0.0370407}
\newcommand{\magScanFlowEYmin}{-0.03624}
\newcommand{\magScanFlowEYmax}{0.7952}
\newcommand{\magScanFlowIYmin}{-0.7952}
\newcommand{\magScanFlowIYmax}{0.03624}
\newcommand{\magScanUmin}{-3.5}
\newcommand{\magScanUmax}{3.5}
\newcommand{\magScanUDataMin}{-4}
\newcommand{\magScanUDataMax}{4}

%% file: figures/compiled/theta_mag_peak_scaling_trim.tex
\includegraphics[trim=0pt 586.13463pt 110.65811pt 0pt,clip]{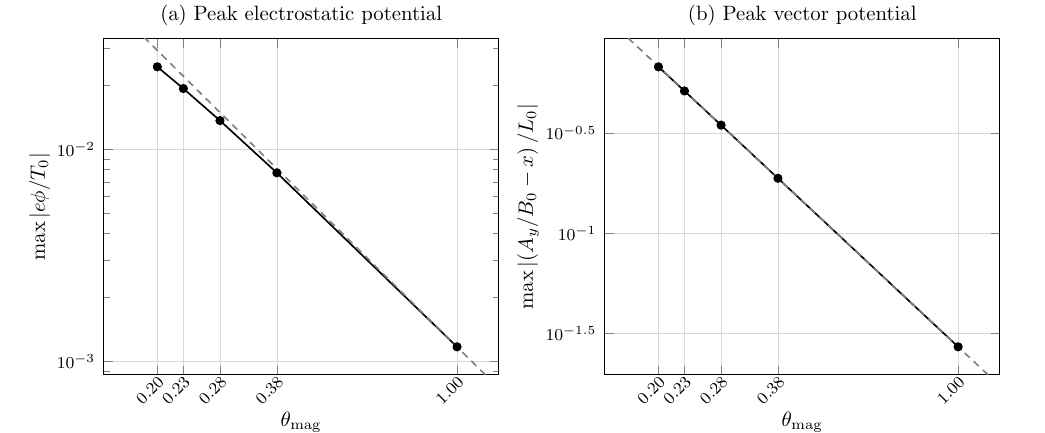}

%% file: figures/compiled/theta_mag_potentials_trim.tex
\includegraphics[trim=0pt 590.40236pt 110.65811pt 0pt,clip]{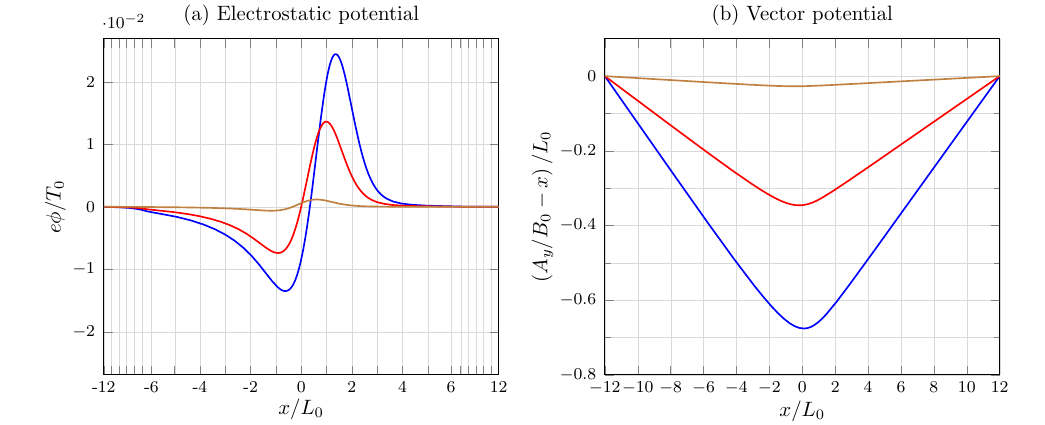}

%% file: figures/data/theta_mag_scan/legend_theta.tex
\begin{center}\fbox{\begin{minipage}{0.94\linewidth}\centering\footnotesize
\begin{tabular}{l@{\hspace{1.5em}}l@{\hspace{1.5em}}l}
\LegendSwatch{blue}{solid}~$\theta_{\mathrm{mag}}=0.2$ & \LegendSwatch{red}{solid}~$\theta_{\mathrm{mag}}=0.28$ & \LegendSwatch{brown}{solid}~$\theta_{\mathrm{mag}}=1$ \\
\end{tabular}
\end{minipage}}\end{center}

%% file: figures/compiled/theta_mag_fields_trim.tex
\includegraphics[trim=0pt 590.40236pt 110.65811pt 0pt,clip]{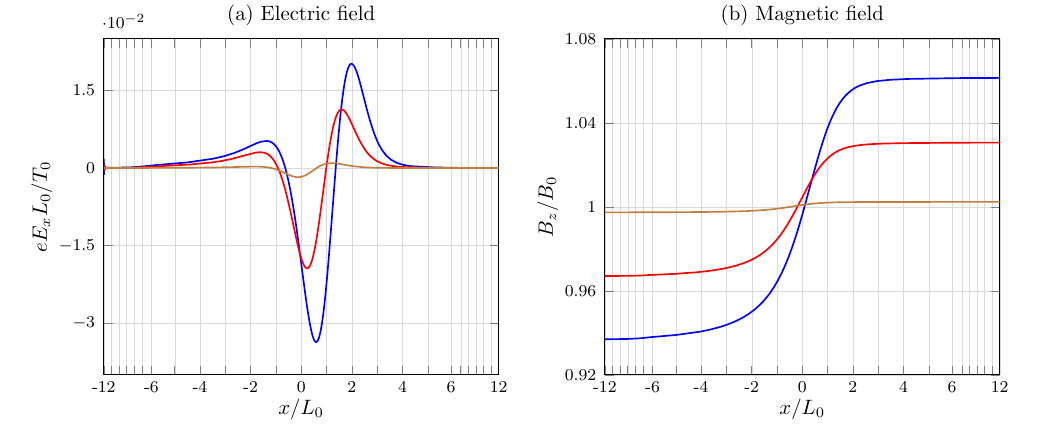}

%% file: figures/compiled/theta_mag_density_trim.tex
\includegraphics[trim=0pt 590.40236pt 110.65811pt 0pt,clip]{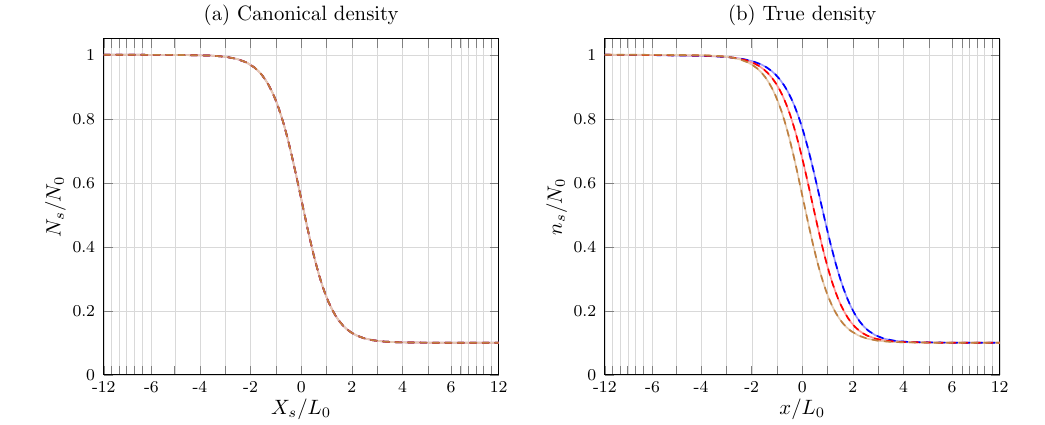}

%% file: figures/data/theta_mag_scan/legend_density_theta.tex
\begin{center}\fbox{\begin{minipage}{0.94\linewidth}\centering\footnotesize
\begin{tabular}{l@{\hspace{1.5em}}l@{\hspace{1.5em}}l}
\LegendSwatch{blue!40!white}{solid}~Electron, $\theta_{\mathrm{mag}}=0.2$ & \LegendSwatch{red!40!white}{solid}~Electron, $\theta_{\mathrm{mag}}=0.28$ & \LegendSwatch{brown!40!white}{solid}~Electron, $\theta_{\mathrm{mag}}=1$ \\
\LegendSwatch{blue}{dashed}~Ion, $\theta_{\mathrm{mag}}=0.2$ & \LegendSwatch{red}{dashed}~Ion, $\theta_{\mathrm{mag}}=0.28$ & \LegendSwatch{brown}{dashed}~Ion, $\theta_{\mathrm{mag}}=1$ \\
\end{tabular}
\end{minipage}}\end{center}

%% file: figures/compiled/theta_mag_flow_trim.tex
\includegraphics[trim=0pt 590.40236pt 110.65811pt 0pt,clip]{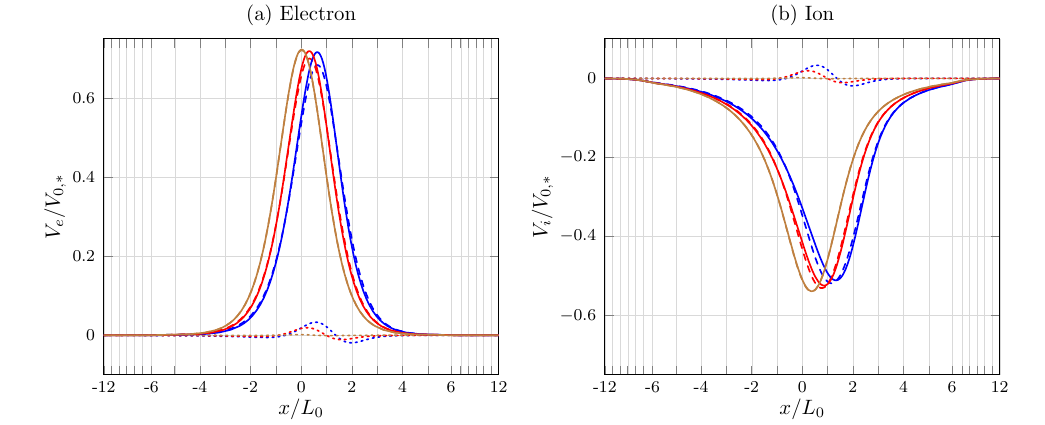}

%% file: figures/data/theta_mag_scan/legend_flow_theta.tex
\begin{center}\fbox{\begin{minipage}{0.94\linewidth}\centering\footnotesize
\begin{tabular}{l@{\hspace{1.5em}}l@{\hspace{1.5em}}l}
\LegendSwatch{blue}{dotted}~$\theta_{\mathrm{mag}}=0.2$, $V_E$ & \LegendSwatch{red}{dotted}~$\theta_{\mathrm{mag}}=0.28$, $V_E$ & \LegendSwatch{brown}{dotted}~$\theta_{\mathrm{mag}}=1$, $V_E$ \\
\LegendSwatch{blue}{dashed}~$\theta_{\mathrm{mag}}=0.2$, $V_{s,*}$ & \LegendSwatch{red}{dashed}~$\theta_{\mathrm{mag}}=0.28$, $V_{s,*}$ & \LegendSwatch{brown}{dashed}~$\theta_{\mathrm{mag}}=1$, $V_{s,*}$ \\
\LegendSwatch{blue}{solid}~$\theta_{\mathrm{mag}}=0.2$, $V_{s,\perp}$ & \LegendSwatch{red}{solid}~$\theta_{\mathrm{mag}}=0.28$, $V_{s,\perp}$ & \LegendSwatch{brown}{solid}~$\theta_{\mathrm{mag}}=1$, $V_{s,\perp}$ \\
\end{tabular}
\end{minipage}}\end{center}

%% file: figures/compiled/theta_mag_loading_trim.tex
\includegraphics[trim=0pt 204.02895pt 110.65811pt 0pt,clip]{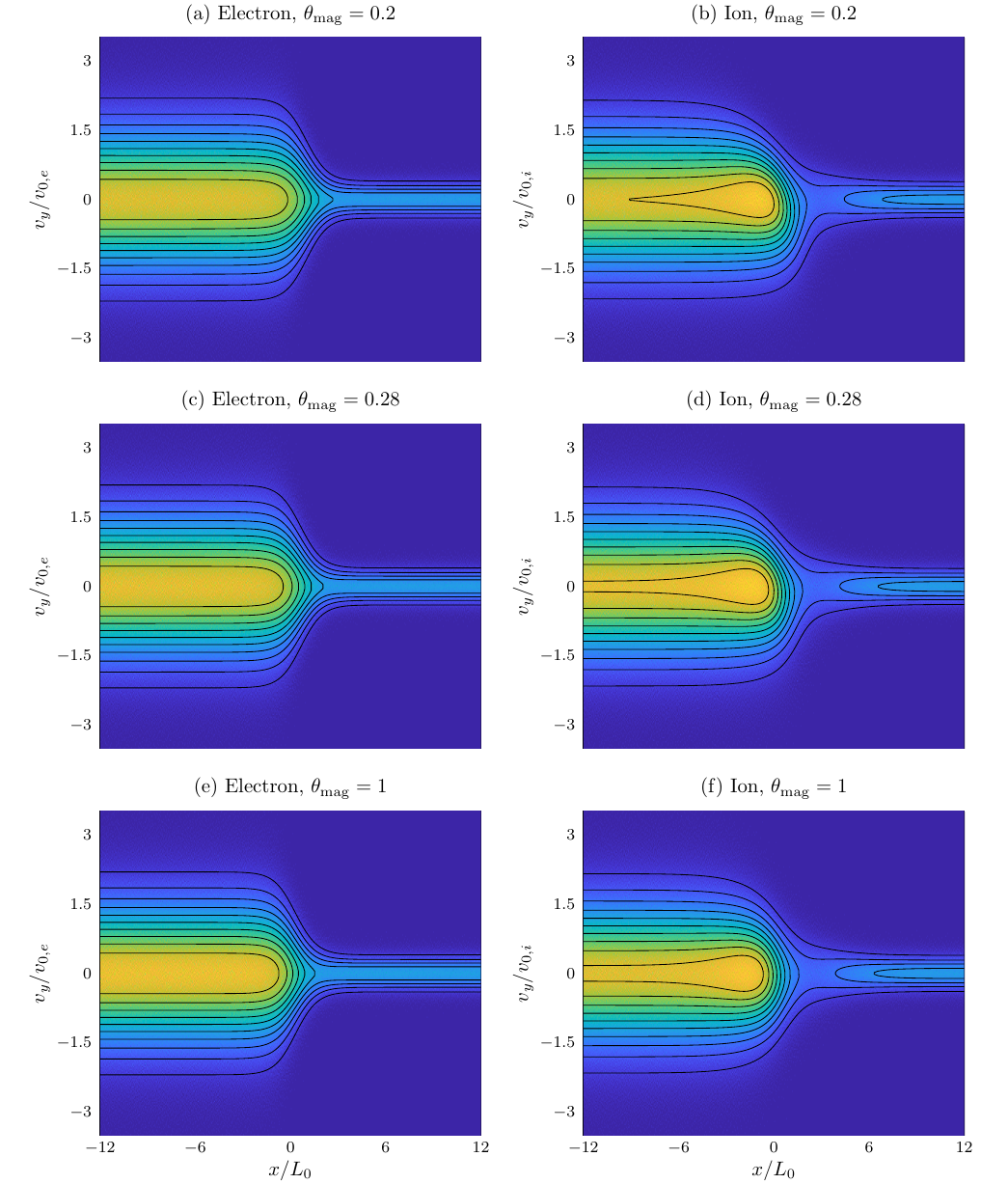}

%% file: figures/data/species_steepen_scan/species_steepen_scan_vars.tex
\newcommand{\stpScanXmin}{-12}
\newcommand{\stpScanXmax}{12}
\newcommand{\stpScanLegendColumns}{4}
\newcommand{\stpScanPhiYlim}{0.598177}
\newcommand{\stpScanFieldYlim}{1.49768}
\newcommand{\stpScanAYmin}{-0.0284565}
\newcommand{\stpScanAYmax}{0.00258695}
\newcommand{\stpScanBzYmin}{0.997297}
\newcommand{\stpScanBzYmax}{1.0027}
\newcommand{\stpScanFlowYmin}{-3.803}
\newcommand{\stpScanFlowYmax}{3.803}
\newcommand{\stpScanDenYmin}{0.0549741}
\newcommand{\stpScanDenYmax}{1.0455}
\newcommand{\stpScanPhiYtick}{-0.4,-0.2,0,0.2,0.4}
\newcommand{\stpScanFieldYtick}{-1,0,1}
\newcommand{\stpScanAYtick}{-0.02,-0.01,0}
\newcommand{\stpScanBzYtick}{0.998,0.999,1,1.001,1.002}
\newcommand{\stpScanFlowYtick}{-2,0,2}
\newcommand{\stpScanDenYtick}{0.2,0.4,0.6,0.8,1}
\newcommand{\stpScanUmin}{-3.5}
\newcommand{\stpScanUmax}{3.5}
\newcommand{\stpScanUDataMin}{-4}
\newcommand{\stpScanUDataMax}{4}

%% file: figures/compiled/sep_scan_potentials_trim.tex
\includegraphics[trim=0pt 397.21565pt 110.65811pt 0pt,clip]{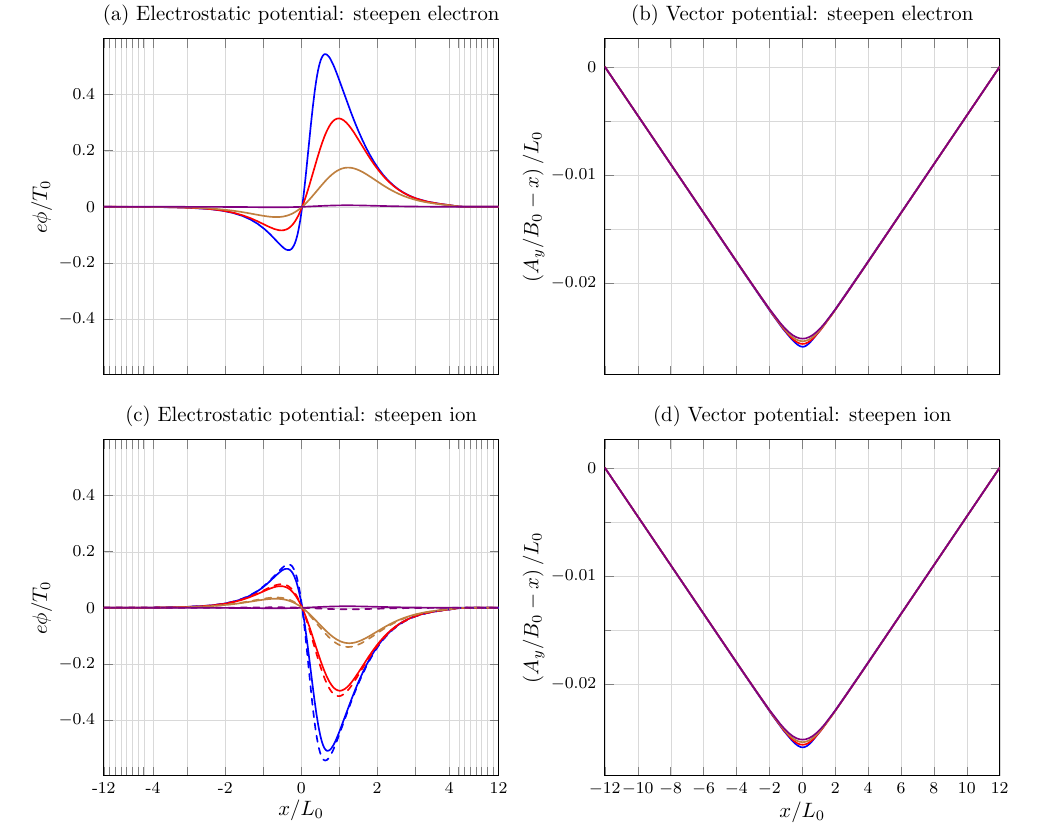}

%% file: figures/data/species_steepen_scan/legend_width_mirror.tex
\begin{center}\fbox{\begin{minipage}{0.94\linewidth}\centering\footnotesize
\begin{tabular}{l@{\hspace{1.5em}}l@{\hspace{1.5em}}l@{\hspace{1.5em}}l}
\LegendSwatch{blue}{solid}~$w_{sN}=0.3$ & \LegendSwatch{red}{solid}~$w_{sN}=0.6$ & \LegendSwatch{brown}{solid}~$w_{sN}=0.9$ & \LegendSwatch{violet}{solid}~$w_{sN}=1.2$ \\
\multicolumn{4}{c}{\LegendSwatch{black}{dashed}~$-1\times$ steepen electron (leading-order sign-reversal)} \\
\end{tabular}
\end{minipage}}\end{center}

%% file: figures/compiled/sep_scan_fields_trim.tex
\includegraphics[trim=0pt 397.21565pt 110.65811pt 0pt,clip]{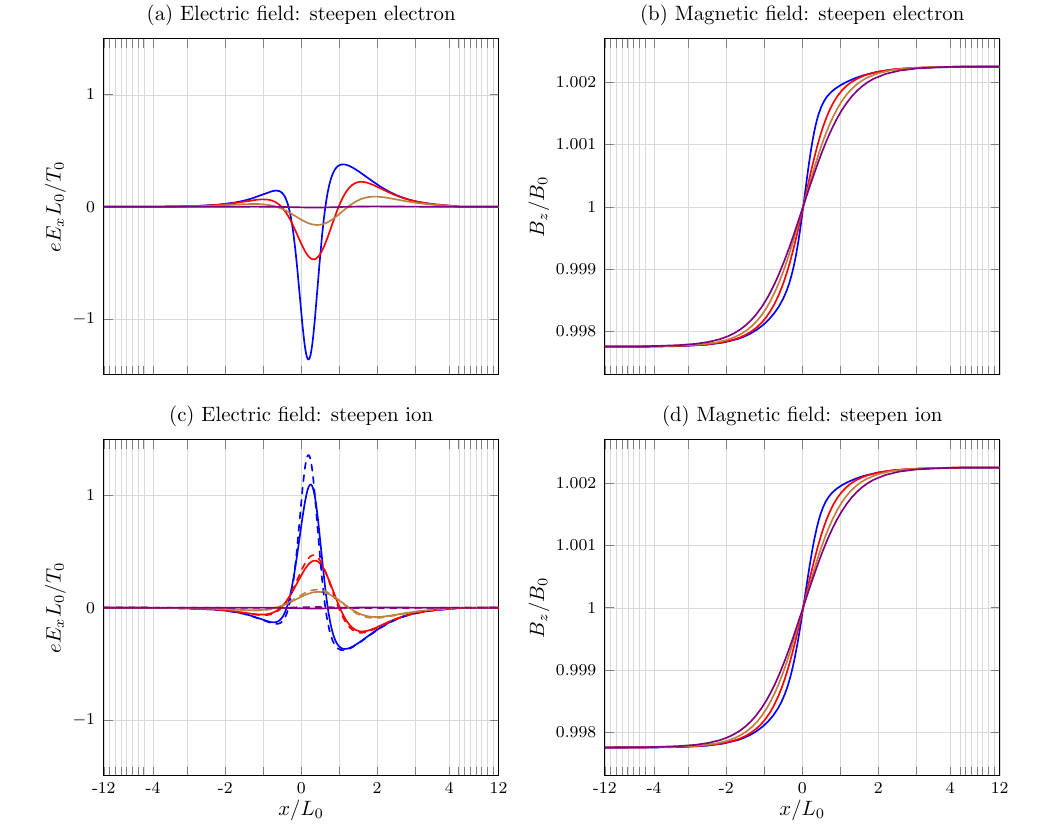}

%% file: figures/compiled/sep_scan_density_trim.tex
\includegraphics[trim=0pt 397.21565pt 110.65811pt 0pt,clip]{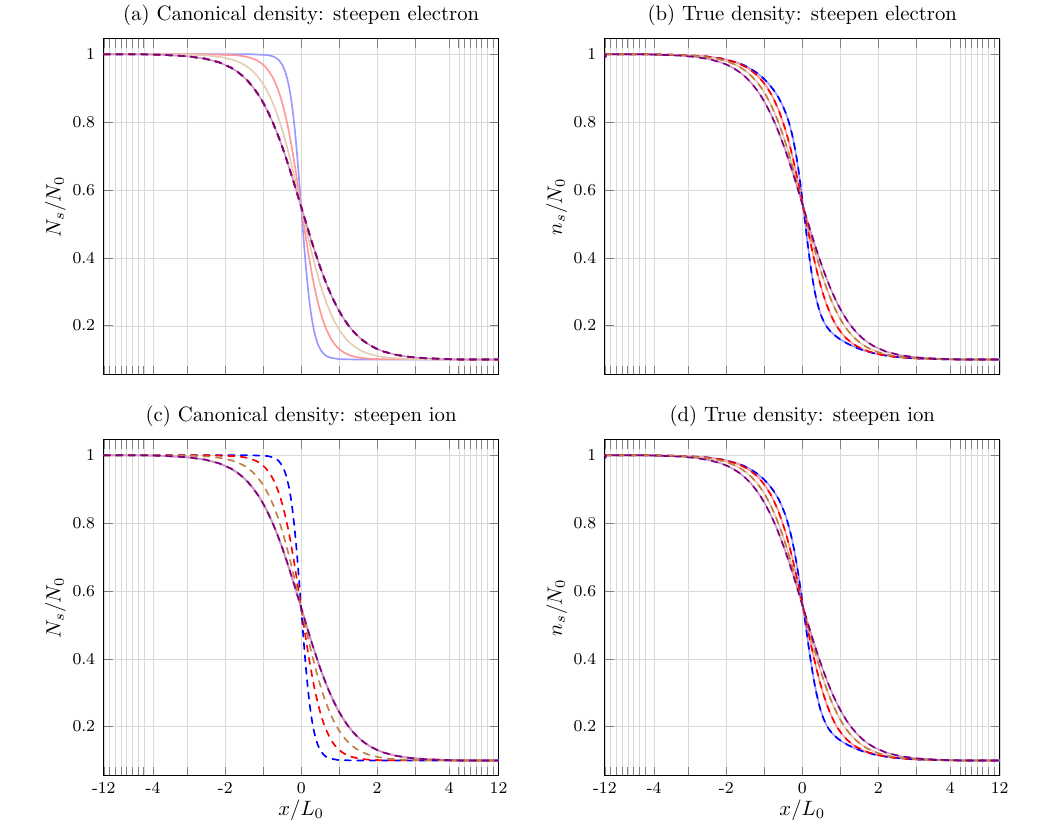}

%% file: figures/data/species_steepen_scan/legend_density.tex
\begin{center}\fbox{\begin{minipage}{0.94\linewidth}\centering\footnotesize
\begin{tabular}{l@{\hspace{1.5em}}l@{\hspace{1.5em}}l@{\hspace{1.5em}}l}
\LegendSwatch{blue!40!white}{solid}~Electron, $w_{sN}=0.3$ & \LegendSwatch{red!40!white}{solid}~Electron, $w_{sN}=0.6$ & \LegendSwatch{brown!40!white}{solid}~Electron, $w_{sN}=0.9$ & \LegendSwatch{violet!40!white}{solid}~Electron, $w_{sN}=1.2$ \\
\LegendSwatch{blue!100!white}{dashed}~Ion, $w_{sN}=0.3$ & \LegendSwatch{red!100!white}{dashed}~Ion, $w_{sN}=0.6$ & \LegendSwatch{brown!100!white}{dashed}~Ion, $w_{sN}=0.9$ & \LegendSwatch{violet!100!white}{dashed}~Ion, $w_{sN}=1.2$ \\
\end{tabular}
\end{minipage}}\end{center}

%% file: figures/compiled/sep_scan_flow_trim.tex
\includegraphics[trim=0pt 397.21565pt 110.65811pt 0pt,clip]{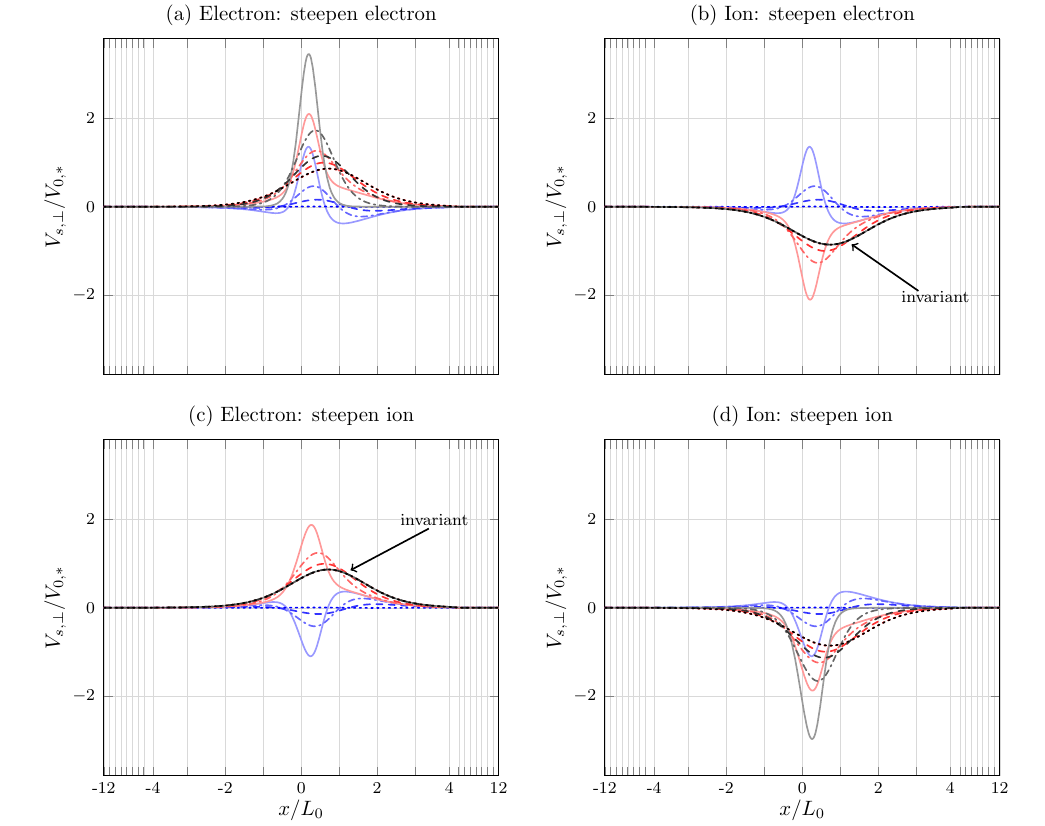}

%% file: figures/data/species_steepen_scan/legend_flow.tex
\begin{center}\fbox{\begin{minipage}{0.94\linewidth}\centering\footnotesize
\begin{tabular}{l@{\hspace{1.5em}}l@{\hspace{1.5em}}l@{\hspace{1.5em}}l}
\LegendSwatch{blue!40!white}{solid}~$w_{sN}=0.3$, $V_E$ & \LegendSwatch{blue!60!white}{dashdotted}~$w_{sN}=0.6$, $V_E$ & \LegendSwatch{blue!80!white}{dashed}~$w_{sN}=0.9$, $V_E$ & \LegendSwatch{blue!100!white}{dotted}~$w_{sN}=1.2$, $V_E$ \\
\LegendSwatch{red!40!white}{solid}~$w_{sN}=0.3$, $V_{s,*}$ & \LegendSwatch{red!60!white}{dashdotted}~$w_{sN}=0.6$, $V_{s,*}$ & \LegendSwatch{red!80!white}{dashed}~$w_{sN}=0.9$, $V_{s,*}$ & \LegendSwatch{red!100!white}{dotted}~$w_{sN}=1.2$, $V_{s,*}$ \\
\LegendSwatch{black!40!white}{solid}~$w_{sN}=0.3$, $V_{s,\perp}$ & \LegendSwatch{black!60!white}{dashdotted}~$w_{sN}=0.6$, $V_{s,\perp}$ & \LegendSwatch{black!80!white}{dashed}~$w_{sN}=0.9$, $V_{s,\perp}$ & \LegendSwatch{black!100!white}{dotted}~$w_{sN}=1.2$, $V_{s,\perp}$ \\
\end{tabular}
\end{minipage}}\end{center}

%% file: figures/compiled/sep_scan_loading_trim.tex
\includegraphics[trim=0pt 397.21565pt 110.65811pt 0pt,clip]{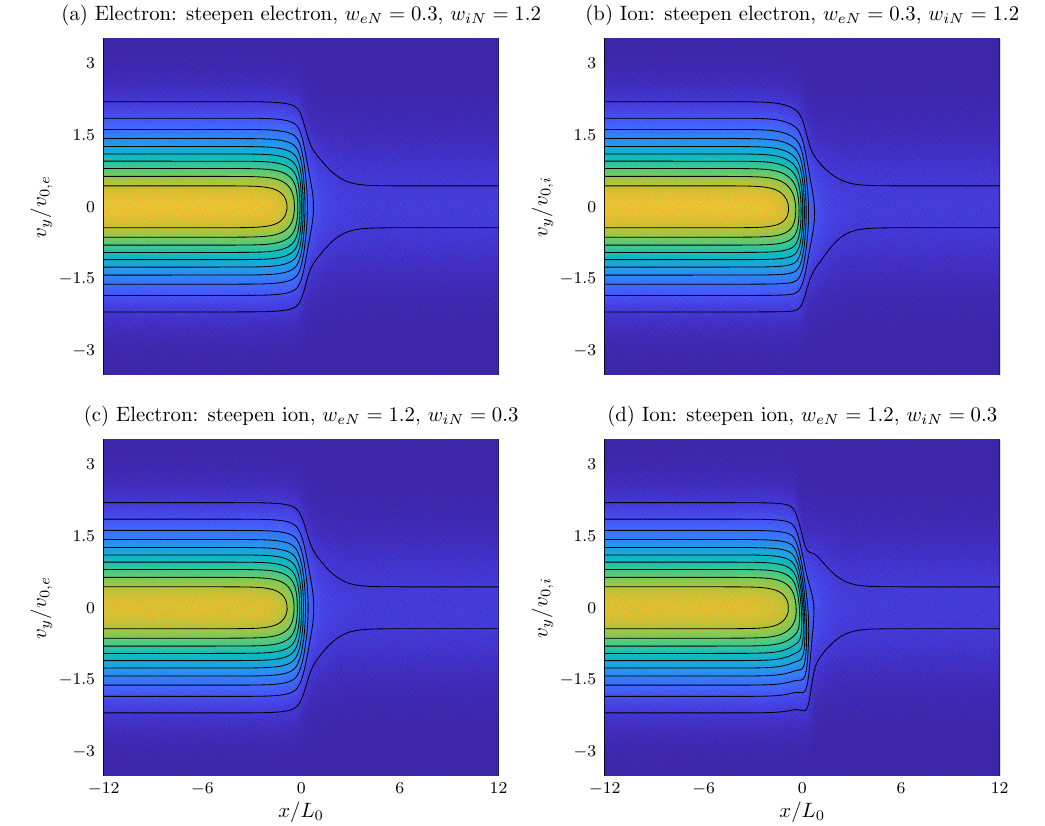}

%% file: v2/section_08.tex
\section{\label{sec:conclusions}Conclusions}

A method has been presented for constructing self-consistent,
symmetry-reduced Vlasov--Maxwell equilibria with nonuniform profiles and
equilibrium electric fields. With each species' distribution a
function of the constants of motion, the equilibrium problem reduces
to a pair of nonlinear ordinary differential equations for the
electrostatic and vector potentials, which are solved here as a
boundary-value problem with the full Poisson equation retained and the
magnetic field determined self-consistently. The
discretization preserves the associated energy invariant, the
pseudo-particle Hamiltonian, exactly at the discrete level, and the
velocity moments are evaluated by Gauss--Hermite quadrature, so the
distribution function is not restricted to forms with closed-form
moments.

The method was verified through discrete Hamiltonian conservation,
spatial and quadrature convergence, and agreement with an independent
asymptotic description of the interior in the well-magnetized,
quasi-neutral limit. A canonical bi-Maxwellian family, with profiles prescribed as
functions of a guiding-center-like first integral, illustrates the
construction end to end. It reduces to the familiar local-Maxwellian
prescription when orbits are small but remains an exact equilibrium
when they are not, and it comes with a procedure for loading particles
consistent with the equilibrium in particle-in-cell simulations.

Within this model, equilibrium electric fields arise at first order
from FLR separation between species and finite-$\beta$ effects, and at zeroth order when the
species' canonical densities are not charge-balanced. A key result
is an exact cancellation, for each species and for any
distribution function of the constants of motion, of the $E\times B$
flow by the electrostatically generated part of the
pressure-gradient-driven flow, so that each species' perpendicular
flow is carried entirely by the magnetostatically generated part. As a
consequence, a large self-consistent electric field need not produce a
correspondingly large species flow. For the canonical bi-Maxwellian
family with flat temperature profiles, a species' flow is invariant to
both the electric field and the potential, confirmed numerically in
Sec.~\ref{sec:flow_invariance}. The cancellation follows from a
structural property of the species force balance that arises from
this type of symmetry reduction, and does not depend on how the
potential is determined. A single-particle picture of it is given in
Appendix~\ref{sec:appendix_particle_picture}.

The present setting is deliberately simple. The geometry is a
one-dimensional slab with a unidirectional magnetic field, and
configurations in which the field passes through zero, although
admitted by the formulation, have not been examined. Although the full
Poisson equation allows Debye-scale charge separation to be treated,
the studies here focus on the quasi-neutral regime. The construction
also does not address the inverse problem of determining the
canonical density and canonical temperature profiles needed to reproduce
specified physical (true) density and temperature profiles; this remains a
recognized difficulty for canonical-Maxwellian-type equilibria more
broadly \cite{Ku2026}, and has been addressed by other means in recent
work, for uniform-temperature Vlasov--Poisson equilibria in a fixed
magnetic field \cite{Vogman2019} and for gyrokinetic distributions
\cite{Li2025}.

Nevertheless, the framework provides a controlled setting for
constructing and analyzing self-consistent kinetic equilibria with
electric fields. An immediate application is as initial states for
fully kinetic studies of gradient-driven instabilities in the presence
of equilibrium flows, where the absence of relaxation transients
allows the dynamics of interest to be isolated, as has been
demonstrated with Vlasov--Poisson equilibria in a fixed magnetic
field \cite{Vogman2020,Vogman2021}. The approach is also a
starting point for extending the formulation, nonlinear algebraic
solvers, quadrature, and particle loading to more general
symmetry-reduced configurations, and for examining the $E\times B$
cancellation property in that setting, including the cylindrical and
toroidal geometries in which kinetic and hybrid equilibria with
electric fields have recently been constructed
\cite{Tasso2014,Kuiroukidis2015,Kaltsas2024,Kaltsas2025}.
More broadly, equilibrium solvers
of this kind could serve as lightweight companions to kinetic codes,
analogous to the Grad--Shafranov solvers that supply equilibria for
MHD and gyrokinetic studies, so that kinetic simulations can start from
genuine Vlasov--Maxwell equilibrium data at little additional cost.

%% file: v2/appendix_a.tex
\section{\label{sec:asymptotic_approach}Asymptotic Solution Approach}

Here, an approach for obtaining asymptotic solutions to the ODE system \refeqs{eq:dl_phi_ode}{eq:dl_A_ode} is presented. Suppose for $\|\boldsymbol{\varrho}\|_\infty \ll 1$, the dimensionless stress can be approximated by
\begin{align*}
\sP(\sV,\sA;\boldsymbol{\varrho}) \approx \sP^{(0)}(\sV,\sA) + h \left( \|\boldsymbol{\varrho}\|_\infty \right) \sP^{(\varrho)}(\sV,\sA),
\end{align*}
where $h(\epsilon) \rightarrow 0$ as $\epsilon \rightarrow 0$.

Now, consider three smallness parameters defined as
\begin{align*}
\epsilon_\lambda = \lambda^2, \ \ \ \ \ \ \epsilon_\beta = \beta, \ \ \ \ \ \ \epsilon_\varrho = h\left(\|\boldsymbol{\varrho}\|_\infty\right).
\end{align*}
Substituting the stress expansion above into \refeqs{eq:dl_phi_ode}{eq:dl_A_ode}, multiplying the vector-potential equation by $\epsilon_\beta$ to express it in terms of positive powers of the smallness parameters, and retaining terms through first order gives:
\begin{align}
\label{eq:dl_phi_ode_expanded}
    \epsilon_\lambda \ppd{\sV}{\xi} &\approx \pd{\sP^{(0)}}{\sV} + \epsilon_\varrho \pd{\sP^{(\varrho)}}{\sV} \\
\label{eq:dl_A_ode_expanded}
    \ppd{\sA}{\xi} &\approx -\frac{\epsilon_\beta}{2} \pd{\sP^{(0)}}{\sA}.
\end{align}
The potentials are assumed to admit regular, multivariate expansions in the three smallness parameters:
\begin{align*}
\sV &= \sV^{(0)} + \epsilon_\lambda \sV^{(\lambda)} + \epsilon_\beta \sV^{(\beta)} + \epsilon_\varrho \sV^{(\varrho)} + ... \\
\sA &= \sA^{(0)} + \epsilon_\lambda \sA^{(\lambda)} + \epsilon_\beta \sA^{(\beta)} + \epsilon_\varrho \sA^{(\varrho)} + ...
\end{align*}
Inserting these assumed forms into \refeqs{eq:dl_phi_ode_expanded}{eq:dl_A_ode_expanded}, performing a multivariate expansion, and collecting terms of like products in $\epsilon_\lambda$, $\epsilon_\beta$, and $\epsilon_\varrho$ gives a hierarchy of equations for the expansion coefficients.
At $O(1)$, \refeq{eq:dl_A_ode_expanded} gives
\begin{align}
\ppd{\sA^{(0)}}{\xi} = 0. \label{eq:ELA_O1}
\end{align}
Imposing the boundary conditions of Sec.~\ref{sec:bcs} at this order gives the solution $\sA^{(0)} = \xi$, recovering the uniform background magnetic field.

Using $\sA^{(0)}=\xi$, the $O(1)$ terms of \refeq{eq:dl_phi_ode_expanded} give
\begin{align}
\left. \pd{}{\sV} \sP^{(0)} \right\vert_{(\sV^{(0)},\xi)} = 0.
\label{eq:V0_balance_general}
\end{align}
This is a statement of quasi-neutrality that determines $\sV^{(0)}$. It is important to note that $\sV^{(0)}$ is found through an algebraic equation and thus does not make use of boundary conditions. This is a result of the singular nature of the perturbation in the parameter $\epsilon_\lambda$. A comprehensive treatment of the problem across the entire domain would require matched asymptotic analysis, including a separate handling of the sheath physics near the boundary. However, that is not considered here. Instead, the focus is on the outer solution, characterized by a regular expansion and valid only away from the boundary.

The remaining terms in \refeq{eq:dl_A_ode_expanded} that are linear in the smallness parameters give
\begin{align}
\ppd{\sA^{(\beta)}}{\xi} &= -\frac{1}{2} \left. \pd{}{\sA} \sP^{(0)} \right\vert_{(\sV^{(0)},\xi)}, \\
\ppd{\sA^{(\lambda)}}{\xi} &= 0, \\ 
\ppd{\sA^{(\varrho)}}{\xi} &= 0.
\end{align}
Since the boundary conditions on $\sA$ have been imposed on the lower-order solution $\sA^{(0)}$, a consistent treatment requires homogeneous Dirichlet boundary conditions to be imposed for the higher-order terms in the expansion. This gives $\sA^{(\lambda)}=\sA^{(\varrho)}=0$, while the finite-$\beta$ correction to the vector potential, $\sA^{(\beta)}$, is determined by solving a one-dimensional Poisson equation for a known right-hand side, subject to homogeneous Dirichlet boundary conditions. The right-hand side here is the lowest-order current in the plasma, which may include diamagnetic and $E \times B$ effects.

The remaining terms in \refeq{eq:dl_phi_ode_expanded} that are linear in the smallness parameters give, after using $\sA^{(\lambda)}=\sA^{(\varrho)}=0$ above,
\begin{align}
\sV^{(\lambda)} \left. \ppd{}{\sV} \sP^{(0)}  \right\vert_{(\sV^{(0)},\xi)} = \ppd{\sV^{(0)}}{\xi},
\end{align}
which provides a small space-charge correction.

Similarly,
\begin{align}
\sV^{(\beta)} \left. \ppd{}{\sV} \sP^{(0)}  \right\vert_{(\sV^{(0)},\xi)} = - \sA^{(\beta)} \left. \pmd{}{\sA}{\sV} \sP^{(0)}  \right\vert_{(\sV^{(0)},\xi)},
\end{align}
describing a correction in the potential needed to maintain quasi-neutrality when the self-generated vector potential $\sA^{(\beta)}$ is accounted for.

Finally,
\begin{align}
\sV^{(\varrho)} \left. \ppd{}{\sV} \sP^{(0)}  \right\vert_{(\sV^{(0)},\xi)} = - \left. \pd{}{\sV} \sP^{(\varrho)}  \right\vert_{(\sV^{(0)},\xi)}.
\end{align}
The right-hand side represents a charge density that can occur due to FLR separation effects. In particular, in the presence of gradients, there can be a charge density arising from FLR separation between different species, due to differences in the size of their gyroradii. The potential on the left-hand side $\sV^{(\varrho)}$ is the potential correction required to counter this effect to maintain quasi-neutrality. Electric fields generated from FLR separation effects have been explored previously in the context of the $E_r$ well in H-mode \cite{Hazeltine1993,Shoucri2002,Shoucri2004,Shoucri2012,LeeWhite2017,LeeWhite2019}.

As a concrete illustration of the pressure expansion above, consider the canonical bi-Maxwellian species stress of \refeq{eq:reduced_species_stress}, written with $\sX_s=\sA+\varrho_s u$ as
\begin{multline*}
\sP_s(\sV,\sA;\varrho_s)
= \int_{-\infty}^{\infty}
\frac{\sN_s(\sA+\varrho_s u)}{\sqrt{2\pi}}
\left[\sT_s(\sA+\varrho_s u)\right]^{1/2}
\\
\times
\exp\left[-\frac{u^2/2+(q_s/e)\sV}{\sT_s(\sA+\varrho_s u)}\right] du.
\end{multline*}
Expanding $\sN_s$ and $\sT_s$ in a Taylor series about $\sX_s=\sA$ and integrating term by term over the resulting Gaussian moments in $u$ generates a series in powers of $\varrho_s$. The term linear in $\varrho_s$ carries a single explicit factor of $u$ multiplying a function that is otherwise even in $u$, so it integrates to zero identically over the symmetric domain; the leading correction to the $\varrho_s \to 0$ limit therefore enters at $O(\varrho_s^2)$. Let $s'$ denote the species attaining $\|\boldsymbol{\varrho}\|_\infty$, i.e., $|\varrho_{s'}|=\max_s|\varrho_s|$, and identify $\epsilon_\varrho=\varrho_{s'}^2=\|\boldsymbol{\varrho}\|_\infty^2$, consistent with $h(\epsilon)=\epsilon^2$ above. Since $\varrho_s=\sqrt{m_sT_0}/(q_sB_0L_0)$, the ratio
\begin{align}
\left(\frac{\varrho_s}{\varrho_{s'}}\right)^{2} = \left(\frac{q_{s'}}{q_s}\right)^{2}\frac{m_s}{m_{s'}}
\end{align}
is a fixed mass-and-charge ratio, independent of $B_0$, so the per-species expansion can be written directly in terms of $\epsilon_\varrho$ as
\begin{align*}
\sP_s(\sV,\sA;\varrho_s) \approx \sP_s^{(0)}(\sV,\sA) + \epsilon_\varrho\,\sP_s^{(\varrho)}(\sV,\sA).
\end{align*}
Explicitly,
\begin{align}
\sP_s^{(0)}(\sV,\sA) = \sN_s\sT_s\exp\left[-\frac{(q_s/e)\sV}{\sT_s}\right], \label{eq:P0_species_explicit}
\end{align}
recovering the usual Boltzmann relation for the canonical density and canonical temperature evaluated at $\sX_s=\sA$, and
\begin{widetext}
\begin{multline}
\sP_s^{(\varrho)}(\sV,\sA) = \left(\frac{q_{s'}}{q_s}\right)^{2}\frac{m_s}{m_{s'}}\,\sN_s\sT_s^2 \exp\left[-\frac{(q_s/e)\sV}{\sT_s}\right]\bigg\{\frac{\sN_s''}{2\sN_s}\left(1+\frac{(q_s/e)\sV}{\sT_s}\right) + \left(2+\frac{(q_s/e)\sV}{\sT_s}\right)\left[\frac{\sT_s''}{2\sT_s}+\frac{\sN_s'\sT_s'}{\sN_s\sT_s}\right] \\
+ \frac{1}{2}\left(\frac{\sT_s'}{\sT_s}\right)^{2}\left[2+\frac{(q_s/e)\sV}{\sT_s}\left(2+\frac{(q_s/e)\sV}{\sT_s}\right)\right]\bigg\}, \label{eq:P1_species_explicit}
\end{multline}
\end{widetext}
where $\sN_s$, $\sT_s$, and their primed derivatives with respect to $\sX_s$ are all evaluated at $\sX_s=\sA$. Summing over species gives $\sP^{(0)}=\sum_s\sP_s^{(0)}$ and $\sP^{(\varrho)}=\sum_s\sP_s^{(\varrho)}$, recovering the general form $\sP \approx \sP^{(0)} + \epsilon_\varrho \sP^{(\varrho)}$ assumed above.

%% file: v2/appendix_b.tex
\section{Single-Particle Illustration of the $E \times B$ Cancellation \label{sec:appendix_particle_picture}}

Section~\ref{sec:flow_invariance} showed, on a macroscopic level, the cancellation of the $E\times B$ flow with 
the electrostatic component of the pressure-gradient-driven flow. This appendix illustrates that cancellation at 
the particle level, in a case that can be worked out in full detail. In particular, uniform fields are considered:
\begin{align}
\vE &= E_0 \hat{x} \\
\vB &= B_0 \hat{z}.
\end{align}
The corresponding electrostatic and vector potentials are taken as:
\begin{align}
\phi &= - E_0 x \\
\vA &= B_0 x \hat{y}.
\end{align}

The motion of a charged particle in this case is exactly solvable. Consider the family of orbits passing 
through a given point $(x_0,y_0)$ at given time $t_0$. Solving the Newton-Lorentz equations under the specified conditions
yields for positions:
\begin{align}
x(t) &= x_0 - \frac{v_\perp}{\Omega_s}\Big[\sin(\varphi-\Omega_s(t-t_0))-\sin\varphi\Big], \label{eq:x_family}\\
y(t) &= y_0 + \frac{v_\perp}{\Omega_s}\Big[\cos(\varphi-\Omega_s(t-t_0))-\cos\varphi\Big] \notag\\
&\quad+ V_E (t-t_0) \label{eq:y_family}
\end{align}
and for velocities:
\begin{align}
v_x(t) &= v_\perp\cos(\varphi-\Omega_s(t-t_0)) \label{eq:vx_family} \\
v_y(t) &= v_\perp\sin(\varphi-\Omega_s(t-t_0)) + V_E, \label{eq:vy_family}
\end{align}
for constants $v_\perp\ge0$ and $\varphi$, where $\Omega_s\equiv q_sB_0/m_s$. All particles share the same
net drift, $V_E\equiv-E_0/B_0$.

Next, the constants $v_\perp$ and $\varphi$ are expressed in terms of the first integrals $W_{s,\perp}$ and $X_s$:
\begin{align}
v_\perp &= \sqrt{\frac{2 W_{s,\perp}}{m_s} - 2 V_E \Omega_s X_s + V_E^2} \label{eq:vperp_form} \\
\sin{\varphi} &= \frac{\Omega_s X_s - V_E}{\sqrt{\frac{2 W_{s,\perp}}{m_s} - 2 V_E \Omega_s X_s + V_E^2}}. \label{eq:vperp_phi}
\end{align}
Real $\varphi$ in \refeq{eq:vperp_phi} requires $|\sin\varphi|\le1$, which gives the region of $(X_s,W_{s,\perp})$ space for which a particle's orbit reaches $(x_0,y_0)$:
\begin{align}
-\sqrt{\frac{2 W_{s,\perp}}{m_s \Omega_s^2}} \le X_s \le \sqrt{\frac{2 W_{s,\perp}}{m_s \Omega_s^2}}. \label{eq:Xmax}
\end{align}
Writing \refeq{eq:vy_family} in terms of $(X_s,W_{s,\perp})$, gives:
\begin{multline}
v_y(t) = (\Omega_s X_s - V_E)\cos(\Omega_s(t-t_0)) \\
{}\mp\sqrt{\frac{2W_{s,\perp}}{m_s}-\Omega_s^2X_s^2}\,\sin(\Omega_s(t-t_0)) + V_E,
\end{multline}
with the upper sign ($-$) for the $v_x(t_0)>0$ branch and the lower sign ($+$) for $v_x(t_0)<0$. The two branches coincide when the square root vanishes, at the endpoints of \refeq{eq:Xmax}. Considering the value of $v_y$ as the orbit crosses through $(x_0,y_0)$ gives:
\begin{align}
v_y(t_0) &= \Omega_s X_s, \label{eq:cancellation_exact}
\end{align}
with no dependence on $V_E$. The collective contribution of the full family of orbits at $(x_0,y_0)$ then depends only on the distribution of
particles in $(X_s,W_{s,\perp})$ space over the region \refeq{eq:Xmax}. For example, if particles
were distributed uniformly over $X_s$, there would be no net flow, since the region is symmetric in $X_s$ for any given $W_{s,\perp}$.

To illustrate, \reffig{fig:orbits_through_origin} shows a collection of positively charged monoenergetic orbits uniformly distributed in $X_s$ passing through a common point $(x_0,y_0)$ at time $t_0$.
The $\hat{z}$ direction is taken to be out of the page. In the frame moving with the net drift $V_E$, the picture becomes that of circular orbits gyrating about fixed centers, since the $E\times B$ drift vanishes identically in this frame by construction; it reappears as a genuine nonzero drift in the lab frame, shown explicitly in \reffig{fig:orbits_through_origin}. For positively charged particles
with $E_0 > 0$, the potential energy decreases to the right. Therefore, particles on the same energy contour will have orbits with larger kinetic energies (and hence larger gyroradii) as their centers move to the right.
This gives rise to a familiar picture used to explain the Spitzer paradox \cite{Spitzer1952,Spitzer1962}, which distinguishes flows produced from
genuine net particle advection from those produced only by collective motions. Here, in the $E \times B$
frame, particles have no net drift, yet collectively there is a net upward drift since the larger upward velocities coming from particles to the right are not
cancelled out by the smaller downward velocities of particles coming from the left. When the population is distributed uniformly in $X_s$, this collective drift
becomes exactly $-V_E$ in the $E \times B$ frame. Moving back to the laboratory frame, this
is exactly the drift needed to counter the $V_E$ drift of the particles.

\begin{widefigure}
\centering
\includegraphics[width=0.48\linewidth]{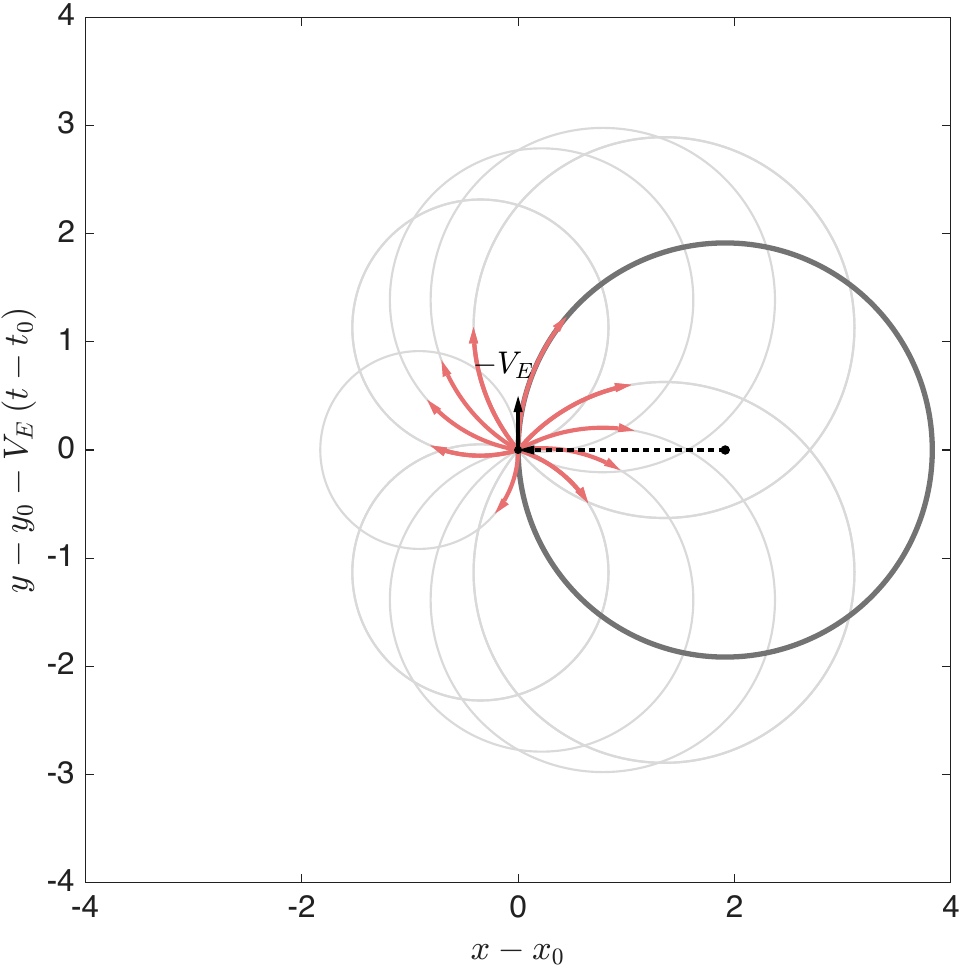}
\hfill
\includegraphics[width=0.48\linewidth]{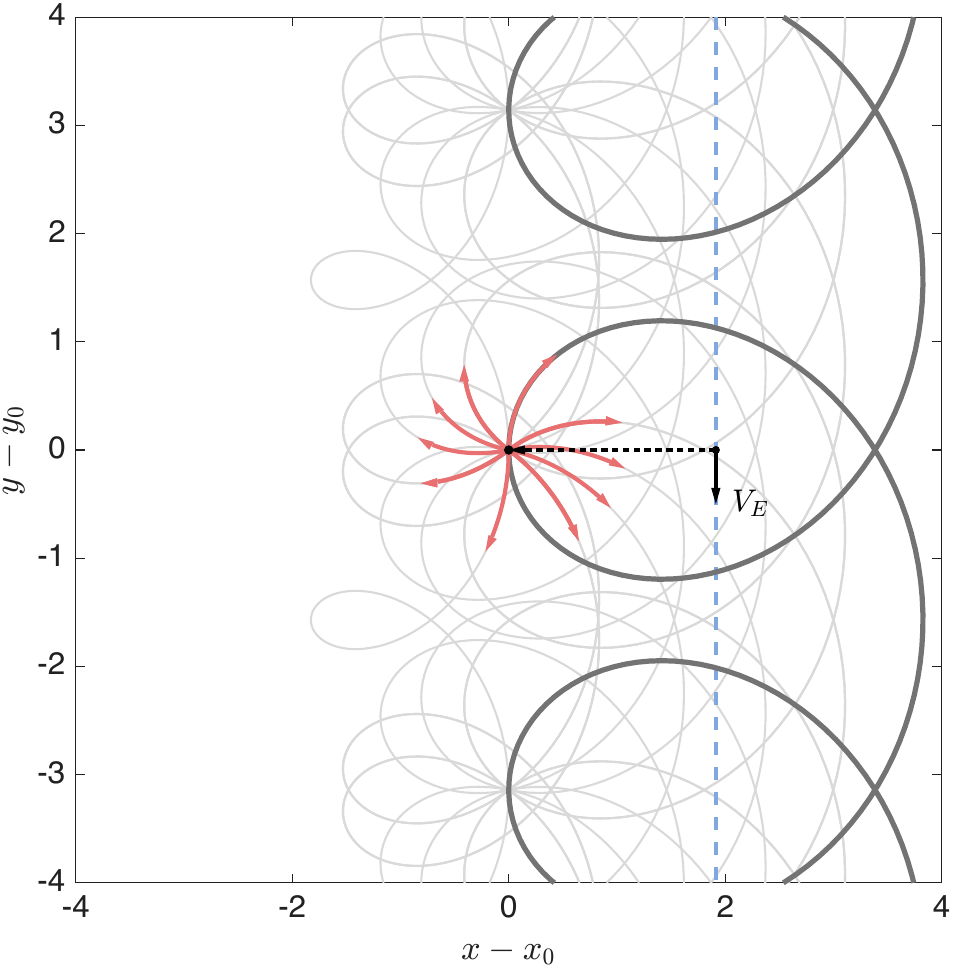}
\vspace{10pt}
\captionof{figure}{\label{fig:orbits_through_origin} The family of orbits \refeqs{eq:x_family}{eq:y_family} through a common point $(x_0,y_0)$ at time $t_0$, sampled over the full admissible range of $X_s$ \refeq{eq:Xmax}, in the $V_E$ (drift-comoving) frame (left) and the lab frame (right). The orbit at the upper bound of \refeq{eq:Xmax} is drawn bold, with its orbit center marked and a position-vector arrow to the crossing point; all others are light gray. Each orbit's velocity at the crossing is shown as a short tangent arc; the black arrow is their mean, averaging \refeq{eq:cancellation_exact} over $X_s$. In the lab frame, the bold orbit's center additionally traces a line showing its drift; this line is absent in the drift-comoving frame, where the center is fixed by construction. Here $V_E\sqrt{m_s/(2W_{s,\perp})}=-1/(2\sqrt{2})$.}
\end{widefigure}